\documentclass[twocolumn, times]{aastex63}

\usepackage{rotating}
\usepackage{booktabs}
\usepackage[version=4]{mhchem}
\usepackage{multirow}
\usepackage{graphicx}
\usepackage{color}
\usepackage{xcolor}
\usepackage{xcolor}

\begin{document}

\title{\large{A Next-Generation Exoplanet Atmospheric Retrieval Framework for Transmission Spectroscopy (\texttt{NEXOTRANS}): Comparative Characterization for WASP-39 b Using JWST NIRISS, NIRSpec PRISM, and MIRI Observations}}

\correspondingauthor{Liton Majumdar}
\email{liton@niser.ac.in, dr.liton.majumdar@gmail.com}

\author[0009-0007-7880-0250]{Tonmoy Deka}
\affiliation{Exoplanets and Planetary Formation Group, School of Earth and Planetary Sciences, National Institute of Science Education and Research, Jatni 752050, Odisha, India}
\affiliation{Homi Bhabha National Institute, Training School Complex, Anushaktinagar, Mumbai 400094, India}

\author[0009-0001-3342-3996]{Tasneem Basra Khan}
\affiliation{Exoplanets and Planetary Formation Group, School of Earth and Planetary Sciences, National Institute of Science Education and Research, Jatni 752050, Odisha, India}
\affiliation{Homi Bhabha National Institute, Training School Complex, Anushaktinagar, Mumbai 400094, India}
\altaffiliation{These authors contributed equally to this work.}

\author[0009-0002-6571-9406]{Swastik Dewan}
\affiliation{Exoplanets and Planetary Formation Group, School of Earth and Planetary Sciences, National Institute of Science Education and Research, Jatni 752050, Odisha, India}
\affiliation{Homi Bhabha National Institute, Training School Complex, Anushaktinagar, Mumbai 400094, India}
\altaffiliation{These authors contributed equally to this work.}

\author[0009-0002-4995-9346]{Priyankush Ghosh}
\affiliation{Exoplanets and Planetary Formation Group, School of Earth and Planetary Sciences, National Institute of Science Education and Research, Jatni 752050, Odisha, India}
\affiliation{Homi Bhabha National Institute, Training School Complex, Anushaktinagar, Mumbai 400094, India}

\author[0009-0001-4100-9218]{Debayan Das }
\affiliation{Exoplanets and Planetary Formation Group, School of Earth and Planetary Sciences, National Institute of Science Education and Research, Jatni 752050, Odisha, India}
\affiliation{Homi Bhabha National Institute, Training School Complex, Anushaktinagar, Mumbai 400094, India}
\affiliation{Department of Physical Sciences, Indian Institute of Science Education and Research, Kolkata, Mohanpur 741246, West Bengal, India }

\author[0000-0001-7031-8039]{Liton Majumdar}
\affiliation{Exoplanets and Planetary Formation Group, School of Earth and Planetary Sciences, National Institute of Science Education and Research, Jatni 752050, Odisha, India}
\affiliation{Homi Bhabha National Institute, Training School Complex, Anushaktinagar, Mumbai 400094, India}

\begin{abstract}

The advent of JWST has marked a new era in exoplanetary atmospheric studies, offering higher-resolution data and greater precision across a broader spectral range than previous space-based telescopes. Accurate analysis of these datasets requires advanced retrieval frameworks capable of navigating complex parameter spaces. We present \texttt{NEXOTRANS}, an atmospheric retrieval framework that integrates Bayesian inference using \texttt{UltraNest}/\texttt{PyMultiNest} with four machine learning algorithms: \texttt{Random Forest}, \texttt{Gradient Boosting}, \texttt{K-Nearest Neighbor}, and \texttt{Stacking Regressor}. This hybrid approach enables a comparison between traditional Bayesian methods and computationally efficient machine learning techniques. Additionally, \texttt{NEXOTRANS} incorporates \texttt{NEXOCHEM}, a module for solving equilibrium chemistry. We applied \texttt{NEXOTRANS} to JWST observations of the Saturn-mass exoplanet WASP-39 b, spanning wavelengths from 0.6 $\mu$m to 12.0 $\mu$m using NIRISS, NIRSpec PRISM, and MIRI. Four chemistry models -- free, equilibrium, modified hybrid equilibrium, and modified equilibrium-offset chemistry -- were explored to retrieve precise Volume Mixing Ratios (VMRs) for H$_2$O, CO$_2$, CO, H$_2$S, and SO$_2$. Absorption features in both NIRSpec PRISM and MIRI data constrained SO$_2$ log VMRs to values between $-6.25$ and $-5.73$ for all models except equilibrium chemistry. High-altitude aerosols, including ZnS and MgSiO$_3$, were inferred, with constraints on their VMRs, particle sizes, and terminator coverage fractions, providing insights into cloud composition. For the best-fit modified hybrid equilibrium model, we derived super-solar elemental abundances of  $\text{O/H} = 14.12^{+2.86}_{-1.82} \times$ solar,  $\text{C/H} = 21.37^{+4.93}_{-3.18} \times$ solar,  and $\text{S/H} = 5.37^{+0.79}_{-0.65} \times$ solar,  
along with a $\text{C/O}$ ratio of $1.35^{+0.05}_{-0.02} \times$ solar. These results demonstrate \texttt{NEXOTRANS}'s potential to enhance JWST data interpretation, advancing comparative exoplanetology efficiently.


\end{abstract}

\keywords{Exoplanets (498); Exoplanet atmospheres (487); Transmission Spectroscopy (2133); Exoplanet atmospheric structure (2310); Exoplanet atmospheric composition (2021)}

\section{Introduction} \label{sec:intro}

The Hubble Space Telescope (HST) has been instrumental in advancing our understanding of exoplanet atmospheres, particularly through its powerful spectroscopic tools \citep{seager2010exoplanet, sing2011hubble}. Observations using the Space Telescope Imaging Spectrograph (STIS) and the Wide Field Camera 3 (WFC3), with its optical and near-infrared coverage, enabled the first detection of water features in a giant planet's atmosphere and contributed to the understanding of chemical processes \citep{barman2007identification,Deming_2013}. Observations from HST also suggested the presence of high-altitude clouds or hazes. For example, studies of the super-Earth GJ 1214 b revealed a featureless spectrum, likely due to high-altitude clouds or haze, indicating the impact of aerosols on atmospheric opacity \citep{kreidberg2014clouds}. The influence of aerosols on observed spectra has provided critical insights into cloud formation and photochemistry. Studies of hot Jupiters have revealed that variations in atmospheric properties, such as cloud coverage and haze opacity significantly influence the observed spectra, leading to diverse atmospheric classifications ranging from clear to cloudy \citep{sing2013hst,sing2016continuum}. Additionally, Hubble has also facilitated comparative population studies, allowing different exoplanets to be examined within consistent observational frameworks, enabling more reliable insights into atmospheric diversity \citep{tsiaras2018population}. Some expected molecular features in the HST and Spitzer spectral data, in contrast to a few ground-based observations \citep{guilluy2019exoplanet,guilluy2022gaps, Giacobbe2021}, were found to be absent, giving rise to the ``Missing methane" problem in cooler planets \citep{Stevenson2010,Benneke2019}.
These discrepancies highlighted the need for further investigation and the necessity of a broader wavelength coverage to accurately constrain abundances and cloud properties. These factors, combined with the limitations of previous instruments, spurred the development and utilization of JWST, offering a more comprehensive and detailed approach to understanding exoplanet atmospheres across a wider range of wavelengths \citep{fortney2024characterizing}.

JWST, with its larger 6.5-meter mirror and resolving power of up to R$\sim$3000, marks a significant advancement over HST, providing unparalleled detail in studying exoplanet atmospheres. Unlike HST, which primarily observes in visible and ultraviolet light with limited near-infrared capabilities (up to 1.7 µm via the Wide Field Camera 3), JWST is particularly optimized for infrared wavelengths, allowing it to peer through dust clouds and study cooler objects in the universe with remarkable clarity \citep{gardner2005science}. This infrared focus makes JWST particularly effective for analyzing the atmospheric compositions of exoplanets, where molecular signatures such as water, methane, and carbon dioxide can be detected more distinctly than ever before \citep{beichman2014observations}. Instruments such as the Near Infrared Imager and Slitless Spectrograph (NIRISS), the Near Infrared Spectrograph (NIRSpec), and the Mid-Infrared Instrument (MIRI) enable simultaneous observations across a wide range of infrared wavelengths, providing comprehensive datasets for detailed analyses of planetary atmospheres.

Since its launch, the JWST has made several groundbreaking discoveries in exoplanetary atmospheres. Owing to the broader wavelength coverage and high sensitivity of JWST instruments, transmission spectra reveal several molecular absorption features, enabling highly confident detections of molecules such as H$_2$O. One significant finding is the detection of carbon dioxide (CO$_2$) and sulfur dioxide (SO$_2$) in the atmosphere of WASP-39 b, suggesting complex chemical processes at play, including photochemistry \citep{Tsai2023, alderson2023early, rustamkulov2023early}. Additionally, JWST has identified the presence of methane (CH$_4$) in the atmosphere of WASP-80 b, marking a critical achievement as it facilitates comparisons with methane levels in the gas giants of our own solar system \citep{bell2023methane}. JWST's capabilities extend to revealing the atmospheres of rocky exoplanets, such as 55 Cancri e, where evidence suggests a volatile-rich atmosphere rather than a bare, molten surface. The findings dismiss the possibility of the planet being a lava world enveloped by a thin atmosphere of vaporized rock. Instead, they indicate the presence of a volatile atmosphere, likely rich in CO$_2$ or CO, which may be sustained by outgassing from a magma ocean \citep{hu2024secondary, patel2024jwst}. Furthermore, JWST's NIRSpec phase curve observations for WASP-121 b have enabled detailed temperature mapping across its dayside, revealing that the eastern hemisphere generally exhibits higher temperatures than the western hemisphere \citep{mikal2023jwst}. These observations unveil the complexities of atmospheric circulation patterns on exoplanets, showing how temperature and pressure variations influence weather systems \citep{mikal2022diurnal}. Finally, JWST has detected the presence of methane (CH$_4$) and possible dimethyl sulfide (DMS) in the atmosphere of K2-18 b. These findings suggest the planet could represent an ocean world beneath a H$_2$-dominated atmosphere, potentially harboring conditions conducive to biological activity \citep{madhusudhan2023carbon}.

The extended wavelength coverage of JWST has also enabled the detailed identification and characterization of clouds and aerosols in exoplanetary atmospheres by probing a wider range of particle compositions and sizes across wavelengths than ever before \citep{Constantinou, Constantinou_2023}. This capability surpasses HST’s narrower range, allowing more detailed observations of how aerosol opacity changes with wavelength—critical for identifying particle composition and understanding scattering properties \citep{wakeford2015transmission, pinhas2017signatures}. The presence of aerosols typically suppresses the transmission features of certain species, such as Na and K, due to Mie scattering in the lower-wavelength regime. Accurate modeling of aerosols is therefore crucial for interpreting JWST observations, especially in systems where non-grey clouds significantly influence atmospheric properties. Failure to account for such effects can lead to misleading abundance estimates and atmospheric mischaracterization. By addressing these challenges, one can robustly infer chemical abundances with greater confidence \citep{lacy2020jwst, Constantinou_2023, ormel2019arcis, mai2019exploring}.

One of the essential tools for analyzing spectroscopic observations and characterizing exoplanetary atmospheres is a retrieval algorithm. These algorithms utilize parametric forward models to simulate spectra and employ probabilistic methods to determine the best-fit parameters. By exhaustively sampling the available parameter space, Bayesian retrievals enable robust constraints on a wide range of atmospheric properties. Numerous retrieval codes leveraging Bayesian inference already exist in the community, providing valuable insights into the chemical composition, temperature structure, and evidence of clouds both during the HST era and now in the JWST era \citep{macdonald2023catalog, madhusudhan2009temperature, lee2012optimal, line2013systematic, waldmann2015tau, lavie2017helios, gandhi2018retrieval, pinhas2018retrieval, molliere2019petitradtrans, min2020arcis, kitzmann2020helios, cubillos2021pyrat, kawahara2022exojax, niraula2022impending, robinson2023exploring}. However, as spectroscopic resolution and forward model complexity increase, the scalability and computational efficiency of these techniques face growing challenges. This issue is further exacerbated by the high-quality data from JWST and other upcoming missions such as Ariel \citep{ariel}, which will generate large datasets requiring faster analysis methods. Advances in Machine Learning (ML) now offer a promising alternative for performing posterior inferences. Several retrieval algorithms incorporating ML have already been developed, elevating these tools to new levels of sophistication and capability \citep{2018NatAs...2..719M, zingales2018exogan, cobb2019ensemble, fisher2020interpreting, nixon_madhusudhan_2020, martinez2022convolutional, yip2024sample}.

To fully harness the wealth of information provided by JWST's high-quality data, we introduce \texttt{NEXOTRANS} -- a Next-generation EXOplaneT Retrieval and ANalysiS tool, designed for modeling and comparative retrieval of exoplanet atmospheric data. \texttt{NEXOTRANS} leverages the capabilities of both Bayesian inference and computationally efficient machine learning methods to effectively constrain parameter space and identify robust solutions for observed spectra. The framework provides two nested sampling methods: \texttt{PyMultiNest} and \texttt{UltraNest}, along with four machine learning algorithms: \texttt{Random Forest}, \texttt{Gradient Boosting}, \texttt{k-Nearest Neighbour}, and \texttt{Stacking Regressor}. The \texttt{Stacking Regressor}, an ensemble of the first three algorithms, demonstrates enhanced performance compared to its individual components. Additionally, \texttt{NEXOTRANS} includes a built-in equilibrium chemistry module, \texttt{NEXOCHEM}, which offers grid-based retrieval utilizing equilibrium chemistry, as well as modern approaches such as the\textbf{ }modified hybrid and equilibrium offset chemistry. It also incorporates a comprehensive treatment of clouds, hazes, and aerosols, enabling comparative exoplanetology by evaluating different modeling approaches.

We demonstrate the capabilities of \texttt{NEXOTRANS} by performing a detailed and comprehensive analysis of JWST observations of WASP-39 b, a well-studied exoplanet known for its large radius and low density, which result in an extended atmosphere that enhances observable signals during transmission spectroscopy. Previous analyses using Hubble data \citep{wakeford2017complete}, along with recent Early Release Science (ERS) observations from JWST Cycle 1 \citep{rustamkulov2023early, alderson2023early, ahrer2023early}, have yielded robust atmospheric constraints and revealed chemical species such as H$_2$O, CO$_2$, CO, H$_2$S, and SO$_2$. In this work, we extend these analyses by incorporating the latest JWST MIRI observations \citep{powell2024sulfur} in the mid-infrared wavelength range (5–12 $\mu$m). Additionally, we combine these data with JWST NIRISS (0.6–2.8 $\mu$m) \citep{feinstein2023early} and NIRSpec PRISM (0.5–5.5 $\mu$m) \citep{rustamkulov2023early} observations. This comprehensive approach provides broader wavelength coverage and improved constraints on the atmospheric properties of WASP-39 b.
\begin{figure*}
    \centering
    \includegraphics[width=\linewidth]{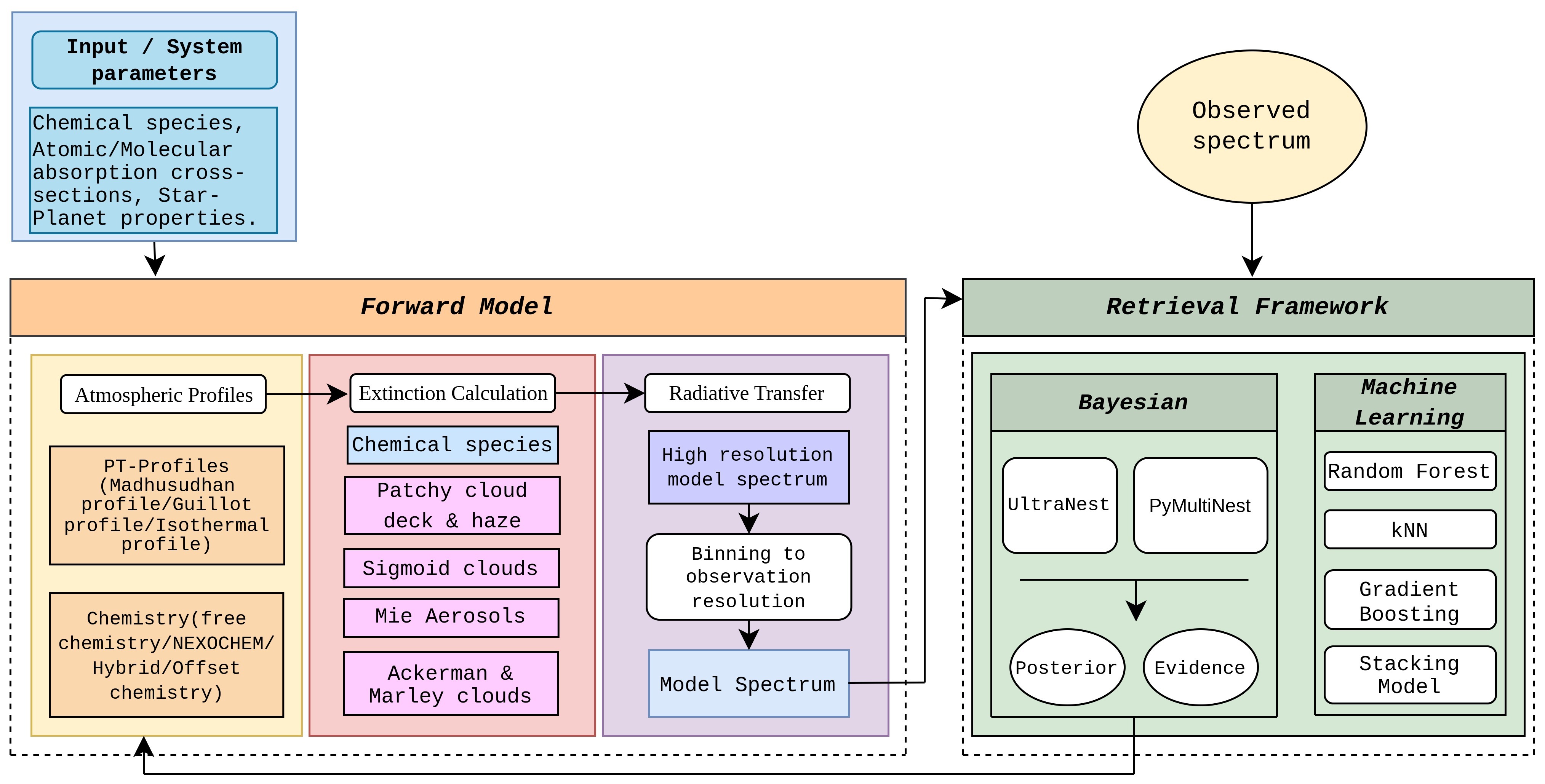}
    \caption{A schematic overview of the retrieval framework implemented in NEXOTRANS. This framework consists of two primary components: the Forward Model and the Retrieval Framework. The Forward Model simulates the exoplanetary atmosphere and produces a model transmission spectrum, while Bayesian inference and machine learning techniques are employed to perform robust parameter estimation.}
    \label{fig:flow}
\end{figure*}

This paper is organized as follows: Section \ref{subsec: forward model} explains the working principles of \texttt{NEXOTRANS}, providing an overview of the forward model and its capabilities. 
Section \ref{subsection : bayesian retrieval} focuses on the Bayesian nested sampling methods, while Section \ref{subsection : machine learning} describes the machine learning algorithms employed in this study. In Section \ref{retrieval_validation_section}, we validate the \texttt{NEXOTRANS} retrieval framework. Section \ref{subsec : JWST dataset} discusses the JWST datasets utilized. We then demonstrate the capabilities of \texttt{NEXOTRANS} through a comprehensive comparative retrieval analysis in Section \ref{sec:result}. In Section \ref{sec:dis}, we discuss the best-fit model. Section \ref{sec:conc} concludes with a summary of the inferences obtained about the atmosphere of WASP-39 b. Appendix \ref{sec: nexochem_exp} details our in-house developed equilibrium chemistry code, \texttt{NEXOCHEM}, and its benchmarking against \texttt{FastChem}, which was used to generate equilibrium chemistry grids. Appendix \ref{subsec:base} discusses the base models used in the machine learning retrievals and provides an intercomparison of four machine learning algorithms. Appendix \ref{retrieval_benchmark_results} presents the \texttt{NEXOTRANS} benchmark retrieved parameter values and their comparison with other retrieval frameworks from \citet{powell2024sulfur}. The best-fit retrieved spectra using hybrid equilibrium chemistry on individual datasets - NIRISS, NIRSpec PRISM and MIRI are presented in \ref{individual_section}. Additionally, the corner plots retrieved from \texttt{NEXOTRANS}, generated using PyMultiNest and the Stacking Regressor for hybrid and equilibrium-offset chemistry, are provided in Appendix \ref{corner_plots}.

\section{\textbf{THE NEXOTRANS RETRIEVAL FRAMEWORK AND ITS APPLICATION.}} \label{sec: retrieval framework} 

Like other retrieval algorithms, \texttt{NEXOTRANS} integrates a parametric forward model with a retrieval framework, as illustrated in Figure \ref{fig:flow}. The forward model simulates a transmission spectrum based on a given set of parameters, such as the P-T profile, chemical composition, and the presence of clouds, hazes, or aerosols in the planet's atmosphere. \texttt{NEXOTRANS} employs traditional Bayesian statistical retrieval techniques to sample the model's parameter space, alongside machine learning algorithms to identify the best-fit model.

\subsection{\textbf{THE FORWARD MODEL}} \label{subsec: forward model}

\subsubsection{\textbf{Radiative Transfer}} \label{subsubsec: radiative transfer}





\texttt{NEXOTRANS} computes the transmission spectrum of a transiting exoplanet by assuming a plane-parallel geometry. We employ the 1-D path distribution method developed by \citet{Robinson(2017)}.

The 1-D transit depth, as described by \citet{MacDonald(2022)}, is given by:

\begin{equation}\label{depth}
\delta_\lambda \approx \frac{R_{\mathrm{p}, \text{top}}^2 - 2 \int_0^{R_{\mathrm{p}, \text{top}}} b e^{-\tau_{\lambda}(b)} d b}{R_*^2}
\end{equation}

where \( R_{\mathrm{p}, \text{top}} \) is the radial distance from the center of the planet to the top of the atmosphere, and \( \tau_{\lambda}(b) \) is the optical depth, which depends on the wavelength \(\lambda\) and the impact parameter \(b\) of a light ray passing through the atmosphere. 

Following \citet{Robinson(2017)}, the optical depth \(\tau_{\lambda}(b)\) along the path of a ray can be computed using the path distribution \(\mathcal{P}\) as:

\begin{equation}
\tau_{\lambda}(b) = \int_0^{\infty} \kappa_\lambda(r) \mathcal{P}_b(r) \, dr
\end{equation}

If \(\kappa_\lambda(r) dr\) represents the differential vertical optical depth for a layer of thickness \(dr\), then:

\begin{equation}
\tau_{\lambda}(b) = \int_0^{\infty} \mathcal{P}_b(r) \, d\tau_{\lambda, \mathrm{vert}}
\end{equation}

which can also be expressed as:

\begin{equation}
\tau_{\lambda}(b) = \sum_{j=1}^{N_{\text{layers}}} \Delta \tau_{\lambda, j} \mathcal{P}_{i,j}
\end{equation}

where \(\mathcal{P}_{i,j}\) is the path distribution of a ray with an impact parameter \(b_i\) traversing through the atmospheric layer \(j\).

Thus, the transit depth in Equation (\ref{depth}) becomes:

{\small
\begin{equation}
\delta_\lambda \approx \frac{R_{\mathrm{p}, \text{top}}^2 - \frac{1}{\pi} \sum_{i=1}^{N_b} \exp \left( - \sum_{j=1}^{N_{\text{layers}}} \Delta \tau_{\lambda, j} \mathcal{P}_{i,j} \right) b_i \Delta b_i}{R_*^2}
\end{equation}
}


The term $b_i \Delta b_i$ represents the effective contribution of an atmospheric annular segment at an impact parameter $b_i$, where $\Delta b_i$ denotes its thickness. This expression approximates the actual annulus area, given by $2\pi b_i \Delta b_i$, but the factor $2\pi$ is omitted as it cancels out in the derivation of the transit depth.

The path distribution from Equation (3) can therefore be interpreted as a linear optical depth encountered between two layers with optical depths \(\tau_{\lambda}\) and \(\tau_{\lambda} + d\tau_{\lambda}\). Furthermore, from Equation (2), the path distribution is also defined as a matrix \(\mathcal{P}_{i,j}\), where \(\mathcal{P}_b(r) \, dr\) represents the path traversed through the atmospheric layer \(j\) with thickness \(dr_j\) by a ray with an impact parameter \(b_i\).


In this work, we consider only the geometric limit, where light rays travel in straight paths through the planetary atmosphere without any scattering. Therefore, for an atmospheric layer centered at $r_l$ having width $\Delta r_l$, and for a ray incident on the atmosphere with impact parameter $b_i$, the 1D path distribution is given by, 

\begin{equation}\label{path_dist}
\mathcal{P}_{1\mathrm{D}, il}(r) =
\begin{cases} 
    0, & \quad (i) \\[10pt] 
    \frac{2}{\Delta r_l} \left( \sqrt{r_{{up}, l}^2 - b_i^2} \right), & \quad (ii) \\[10pt]
    \frac{2}{\Delta r_l} \left( \sqrt{r_{{up}, l}^2 - b_i^2} - \sqrt{r_{{low}, l}^2 - b_i^2} \right), & \quad (iii)
\end{cases}
\end{equation}

where $r_{up,l}$ and $r_{low,l}$ denote the upper and lower boundaries of the atmospheric layer centered at $r_l$.

Equation \ref{path_dist} holds under the following conditions:

\begin{equation}
\begin{aligned}
    (i) &\quad r_{{up}, l} \leqslant b_i, \\
    (ii) &\quad r_{{low}, l} < b_i < r_{{up}, l}, \\
    (iii) &\quad r_{{low}, l} \geqslant b_i.
\end{aligned}
\end{equation}

\begin{figure*}
    \centering
    \begin{minipage}{0.48\textwidth}
        \centering
        \includegraphics[width=\textwidth]{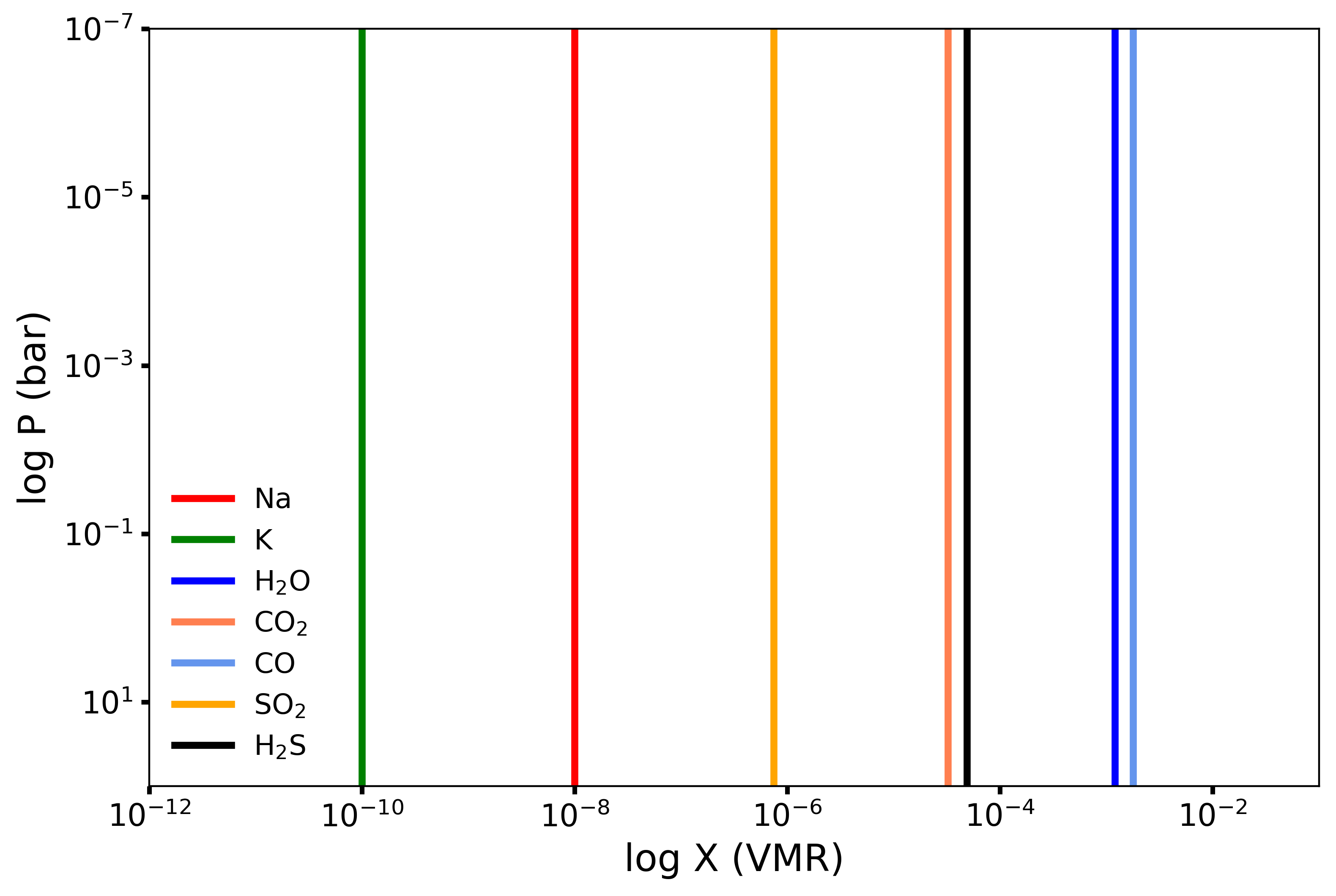}
        \parbox{0.8\linewidth}{\centering (a) Free chemistry}
    \end{minipage}
    \hfill
    \begin{minipage}{0.48\textwidth}
        \centering
        \includegraphics[width=\textwidth]{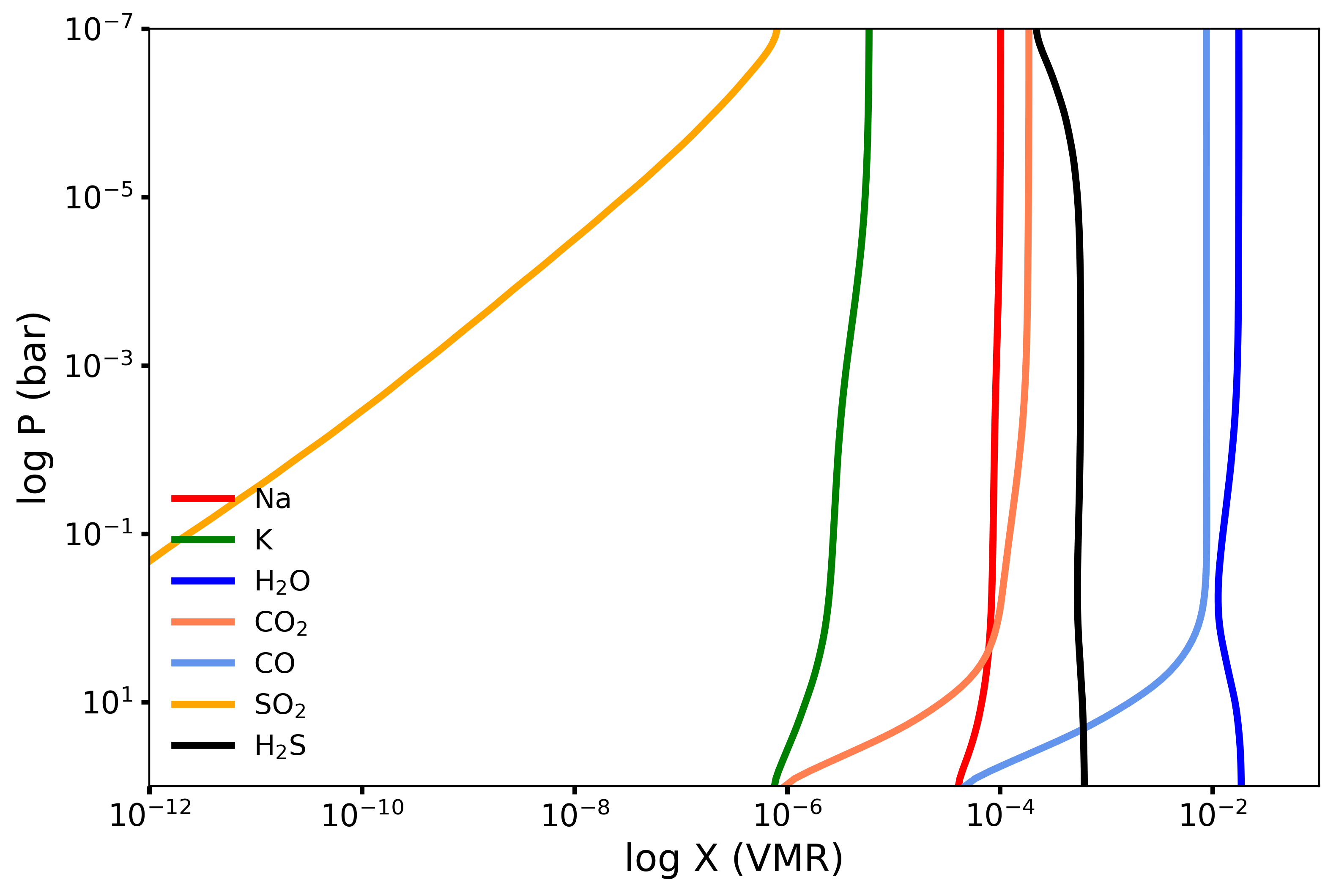}
        \parbox{0.8\linewidth}{\centering (b) Equilibrium chemistry}
        \label{fig:equ_chem}
    \end{minipage}

    \begin{minipage}{0.48\textwidth}
        \centering
        \includegraphics[width=\textwidth]{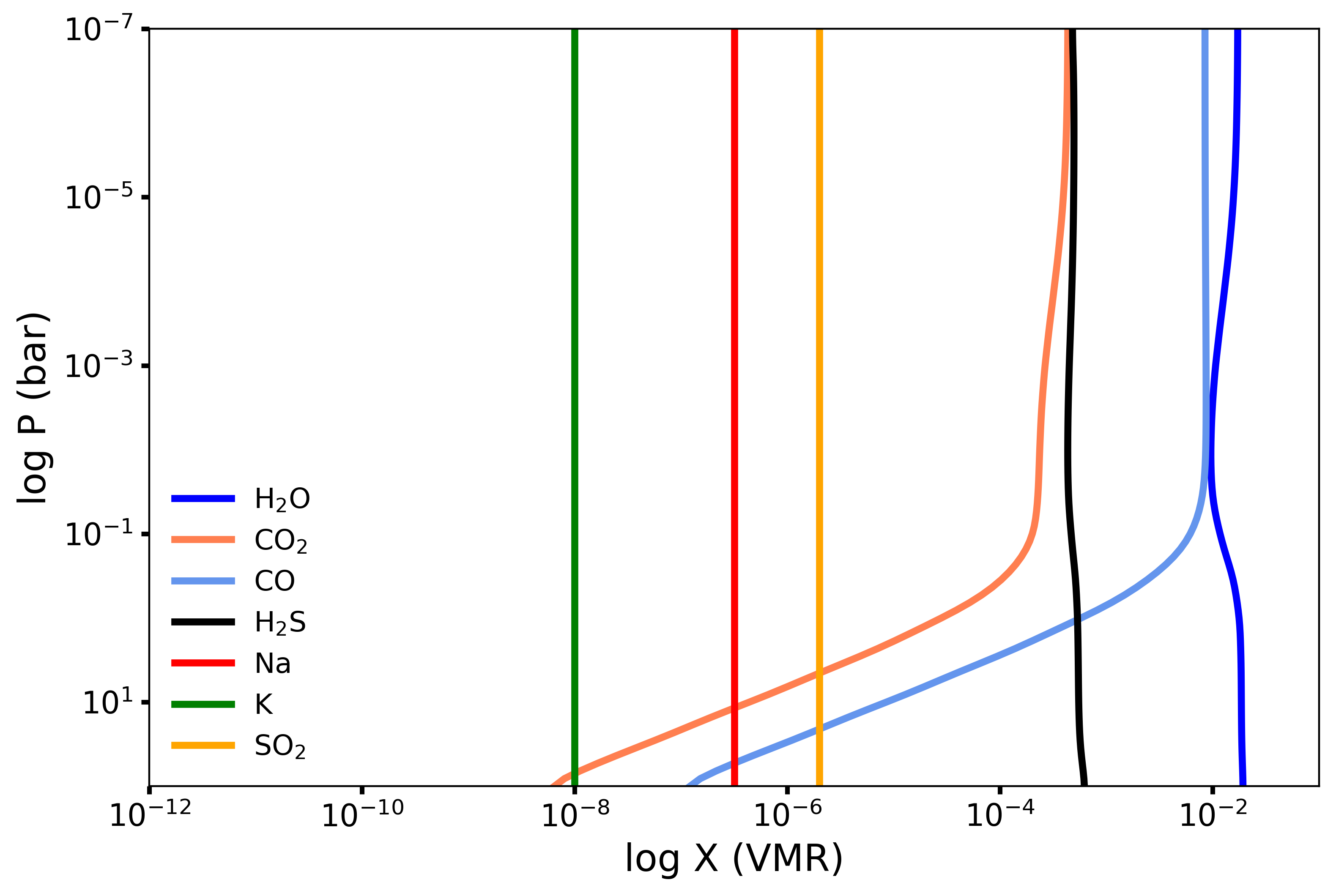}
        \parbox{0.8\linewidth}{\centering (c) Modified hybrid equilibrium chemistry}
    \end{minipage}
    \hfill
    \begin{minipage}{0.48\textwidth}
        \centering
        \includegraphics[width=\textwidth]{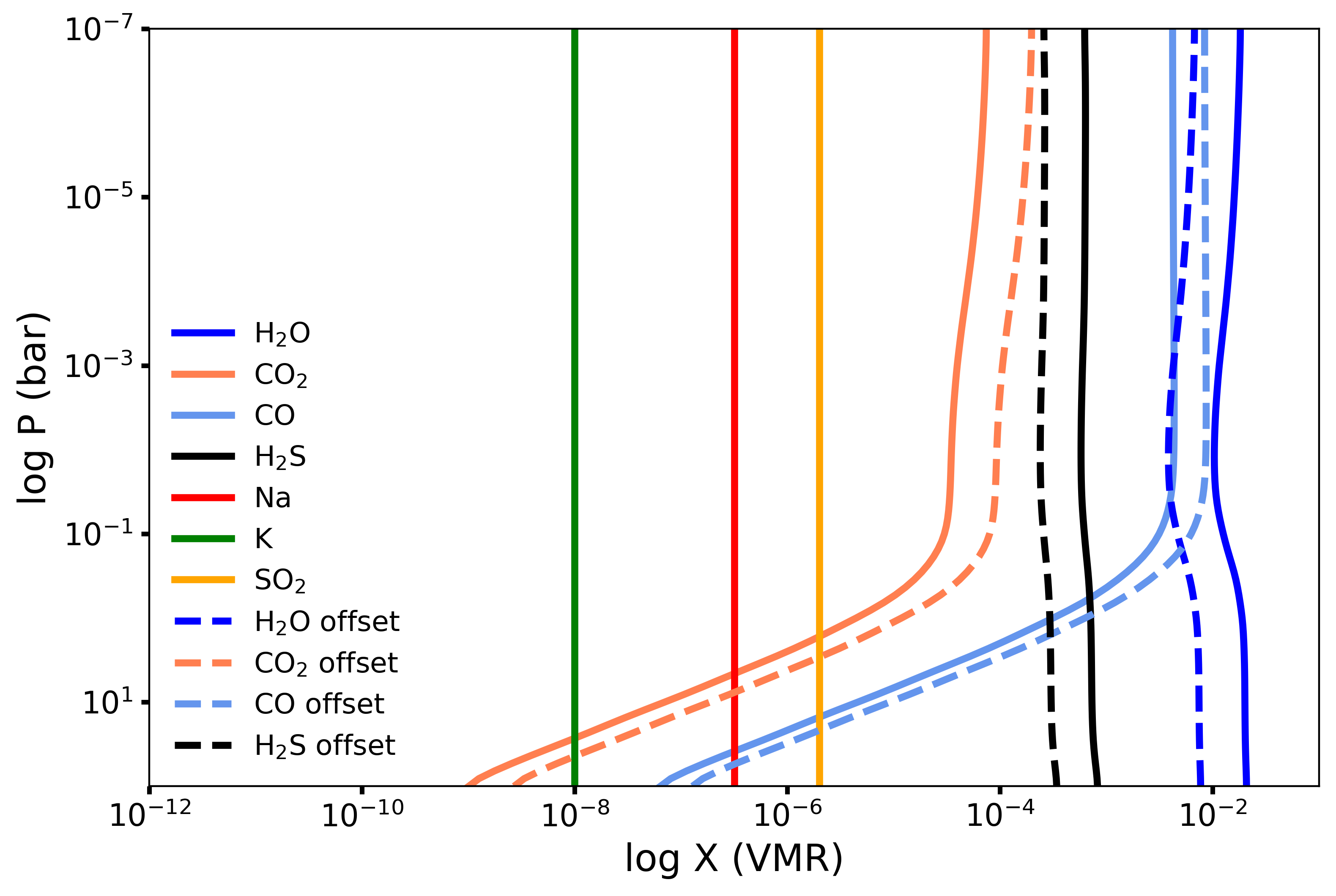}
        \parbox{0.8\linewidth}{\centering (d) Modified equilibrium offset chemistry}
        \label{fig:rqu_offset_chem}
    \end{minipage}
    
    \caption{An illustration of the four atmospheric chemistry methods implemented in \texttt{NEXOTRANS}. (a) shows the free chemistry approach, in which the mixing ratios of all species remain constant with altitude. (b) shows the equilibrium assumption, where the mixing ratio profiles are obtained from \texttt{NEXOCHEM} for a particular C/O and M/H value. (c) shows the modified hybrid chemistry approximation, where the mixing ratio profiles are a combination of free and equilibrium chemistry. Lastly, (d) shows the modified equilibrium offset chemistry approach, which, apart from being a combination of free and equilibrium chemistry, also involves scaling the profiles obtained from \texttt{NEXOCHEM} by an offset factor that shifts their position while maintaining the overall shape. }
    \label{fig:chemisty}
\end{figure*}

\subsubsection{\textbf{Pressure-Temperature Profile}} \label{subsubsection : pressure temperature profile}

The model atmosphere is divided into a series of layers, each characterized by a specific pressure, with the pressure being highest at the base of the atmosphere and decreasing towards the top. The temperature in each layer is computed using a parametric pressure-temperature (P-T) profile. \texttt{NEXOTRANS} provides three different P-T parameterizations: isothermal, the \cite{guillot2010radiative} profile, and the three-layer model of \cite{madhusudhan2009temperature}.

The Madhusudhan P-T parameterization divides the model atmosphere into three distinct regions:

\begin{equation}
    T = T_{0} + \left( \frac{\log \left( \frac{P}{P_{0}} \right)}{\alpha_{1}} \right)^{2}, \quad \mathrm{P_{0} < P < P_{1}}
\end{equation}

\begin{equation}
    T = T_{2} + \left( \frac{\log \left( \frac{P}{P_{2}} \right)}{\alpha_{2}} \right)^{2}, \quad \mathrm{P_{1} < P < P_{3}}
\end{equation}

\begin{equation}
    T = T_{2} + \left( \frac{\log \left( \frac{P_{3}}{P_{2}} \right)}{\alpha_{2}} \right)^{2}, \quad \mathrm{P > P_{3}}
\end{equation}

Here, \(\mathrm{P_{0}}\) and \(\mathrm{T_{0}}\) are the pressure and temperature at the top of the atmosphere, with \(\mathrm{P_1}\) and \(\mathrm{P_3}\) representing the pressures at the layer boundaries, and \(\mathrm{T_3}\) being the temperature at potential inversion points. The slopes of the P-T profile are determined by \(\alpha_1\) and \(\alpha_2\), with thermal inversions occurring when \(\mathrm{P_2 \leq P_1}\). For all cases, \(\mathrm{P_0 \leq P_1 \leq P_3}\).

The Guillot P-T profile \citep{guillot2010radiative, Barstow2020Outstanding} is described by:

\begin{align}
    T^{4}(\tau) &= \frac{3 T_{\text{int}}^{4}}{4} \left( \frac{2}{3} + \tau \right) 
    + \frac{3 T_{\text{irr}}^{4}}{4}(1-\alpha) \xi_{\gamma_{1}}(\tau) \notag \\
    &\quad + \frac{3 T_{\text{irr}}^{4}}{4} \alpha \, \xi_{\gamma_{2}}(\tau)
\end{align}

where

\begin{align}
    \xi_{\gamma_{i}} &= \frac{2}{3} + \frac{2}{3 \gamma_{i}} \left[ 1 + \left( \frac{\gamma_{i} \tau}{2} - 1 \right) e^{-\gamma_{i} \tau} \right] \notag \\
    &\quad + \frac{2 \gamma_{i}}{3} \left( 1 - \frac{\tau^{2}}{2} \right) \mathrm{E}_{2} \left( \gamma_{i} \tau \right)
\end{align}

and

\begin{align}
    T_{\text{irr}} &= \beta \left( \frac{R_*}{2a} \right)^{1/2} T_*
\end{align}

Here, the infrared opacity is represented by \( \kappa_{\text{IR}} \), while \( \gamma_1 = \frac{\kappa_{v1}}{\kappa_{\text{IR}}} \) and \( \gamma_2 = \frac{\kappa_{v2}}{\kappa_{\text{IR}}} \) denote the ratios of visible-band opacities in two specific bands to the infrared opacity. The parameter \( \alpha \) defines the ratio of fluxes between the two visible streams, and \( \beta \) represents the atmospheric recirculation efficiency. The planet’s internal temperature is denoted by \( T_{\text{int}} \), while \( T_{\text{irr}} \) corresponds to the temperature resulting from stellar irradiation. The parent star's radius and temperature are denoted by \( R_* \) and \( T_* \), respectively, and \( a \) represents the orbital semi-major axis. The infrared optical depth, \( \tau \), is expressed as \( \tau = \kappa_{\text{IR}} P / g \), where \( P \) is the atmospheric pressure and \( g \) is the gravitational acceleration. Additionally, \( \mathrm{E}_2 \) refers to the second-order exponential integral function.

Once the pressure and temperature are defined for each layer, the total number density and radial distance \(r\) of each layer are computed using the ideal gas law and the equation of hydrostatic equilibrium. This process requires specifying a reference pressure, \(\mathrm{P_{ref}}\), which corresponds to the pressure at \(r = \mathrm{R_p}\) (the observed planetary white-light radius).

\subsubsection{\textbf{Atmospheric Chemistry}} \label{subsubsection : chemistry} 

\texttt{NEXOTRANS} provides four methods to model the atmospheric chemistry (as shown in Figure \ref{fig:chemisty}):

The first method is free chemistry, where the volume mixing ratios of individual chemical species are treated as free parameters, and the mixing ratio profiles remain constant with altitude or pressure (Figure \ref{fig:chemisty}a).

The second method assumes chemical equilibrium. For this, we use an equilibrium chemistry module built into \texttt{NEXOTRANS} called \texttt{NEXOCHEM}, which has been thoroughly benchmarked with \texttt{FastChem} (see Appendix \ref{sec: nexochem_exp}), to obtain individual atomic and molecular volume mixing ratios. To minimize retrieval time, we precomputed a grid with \texttt{NEXOCHEM} across a range of possible combinations of temperature, pressure, C/O ratio, and metallicity (T = 300 -- 4000 K, P = 10$^{-7}$ -- 10$^{2} $ bar, C/O = 0.2 -- 2, and [Fe/H] = 10$^{-1}$ -- 10$^{3} $ times solar), as discussed in Appendix \ref{nexochem_grid}. Retrievals using the equilibrium chemistry assumption aim to constrain global compositional parameters, such as the atmospheric C/O ratio and metallicity, along with individual elemental ratios (Figure \ref{fig:chemisty}b).

Due to its high precision and ability to cover a large wavelength range, JWST can simultaneously detect numerous chemical species. With the detection of the photochemical product SO$_2$ \citep{crossfield2023photochemically} for the first time in an exoplanetary atmosphere, it is clear that JWST has the ability to detect chemical species resulting from disequilibrium processes. To account for these processes, we implement two approximate methods to model disequilibrium processes motivated by the approaches of \citet{Constantinou}. These methods are the hybrid equilibrium and equilibrium offset approaches (Figures \ref{fig:chemisty}c and \ref{fig:chemisty}d show how the vertical mixing ratios vary in these two methods). 

In the modified hybrid equilibrium approach of \texttt{NEXOTRANS}, the C/O ratio and metallicity are treated as free parameters, which are used to perform an equilibrium calculation with \texttt{NEXOCHEM} to determine the mixing ratios of chemical species. Additionally, to account for contributions from species such as SO$_2$, which are produced through disequilibrium processes like photochemistry, and species such as Na and K, whose mixing ratios largely remain constant with altitude, the hybrid method assumes vertically constant mixing ratios for these species. This approach offers greater flexibility and relaxation compared to strict chemical equilibrium assumptions.

On the other hand, the modified equilibrium offset approach combines equilibrium and vertically constant mixing ratios while allowing the mixing ratios of each equilibrium chemical species to be adjusted by a constant multiplicative factor. This method approximately includes the effects of disequilibrium chemistry, which can either enhance or deplete specific molecules, while preserving physically realistic vertical mixing ratio profiles.

It is important to note that our modified hybrid equilibrium and equilibrium-offset approaches differ slightly from the prescription outlined by \citet{Constantinou}. In their method of hybrid equilibrium, the free parameters consist of the C/H, O/H, N/H, and S/H ratios relative to solar values, along with free VMRs for species of interest (e.g., SO$_2$). In contrast, in \texttt{NEXOTRANS}, the free parameters are the metallicity and C/O ratio, along with the free VMRs of species of interest. Additionally, in the equilibrium offset method of \citet{Constantinou}, the elemental abundances remain fixed. However, in \texttt{NEXOTRANS}, the metallicity and C/O ratio remain free parameters alongside the multiplicative offset factors and free VMRs for species of interest, providing full flexibility for the retrieval to determine the best-fit solution. Throughout this text, the terms ``hybrid equilibrium" and ``equilibrium offset" specifically refer to our modified implementations rather than those of \citet{Constantinou}.

After establishing a suitable P-T profile parameterization and atmospheric chemistry, as discussed, we proceed to calculate the extinction coefficients, $\kappa_{\lambda}$, for each atmospheric layer.

\subsubsection{\textbf{Opacity Sources}} \label{subsubsection : opacity sources}

The primary sources of extinction presented by an exoplanet's atmosphere to the stellar rays can be summarised as:

\begin{align}
\kappa_\lambda(r) &= \kappa_{\text{chem}, \lambda}(r) + \kappa_{\text{Rayleigh}, \lambda}(r) \notag \\
&\quad + \kappa_{\mathrm{CIA}, \lambda}(r) + \kappa_{\text{cloud}, \lambda}(r)
\end{align}

where $\kappa_{\text{chem}, \lambda}$ is extinction due to photons absorbed by atoms and molecules, $\kappa_{\text{Rayleigh}, \lambda}$ is due to Rayleigh scattering, $\kappa_{\mathrm{CIA}, \lambda}$ is extinction from collision-induced absorption (CIA), and $\kappa_{\text{cloud}, \lambda}$ is due to absorption and scattering by clouds, hazes, or aerosol particles.

The extinction coefficient $\kappa$ can generally be calculated from the following equation:

\begin{equation} \label{equation12}
    \kappa(P,T) = n(P,T) \sigma(\lambda, P, T)
\end{equation}

where $n(P, T)$ is the number density of the atmosphere, which is a function of pressure and temperature, and $\sigma (\lambda, P, T)$ is the absorption cross-section, which is also a function of wavelength, in addition to being dependent on pressure and temperature.

In general, the absorption cross-sections of atoms and molecules are calculated from line lists. However, since absorption and scattering are distinct phenomena, the cross-sections for Rayleigh scattering must be treated differently. These depend on the polarizability of the molecule and the wavelength of the incident light, but are independent of pressure and temperature. Therefore, to model the effects of Rayleigh scattering, we follow \citet{Sneep} to calculate the Rayleigh cross-sections of H$_2$ and He:

\begin{equation} \label{equation 13}
\sigma_{\text {scat }, \lambda}=\frac{24 \pi^3}{n_{\text {ref }}^2 \lambda^4}\left(\frac{\eta_\lambda^2-1}{\eta_\lambda^2+2}\right)^2 K_\lambda,
\end{equation}

where $n_{\text{ref}}$ is the number density at a reference pressure and temperature, $\eta$ is the wavelength-dependent refractive index, and $K_{\lambda}$ is the King factor accounting for polarization corrections.

In this work, we consider chemical species such as Na, K, H$_2$O, CO$_2$, CO, H$_2$S, SO$_2$, CH$_4$ and HCN, which have previously been inferred in the atmosphere of WASP-39 b. Therefore, we make use of the publicly available absorption cross-sections of the following species provided in the \texttt{POSEIDON} opacity database\footnote{\url{https://poseidon-retrievals.readthedocs.io/en/latest/content/opacity_database.html}}: $\mathrm{H_2O}$ \citep{Poly},  $\mathrm{CO_2}$ \citep{tashkun2011cdsd}, CO \citep{li2015rovibrational}, $\mathrm{SO_2}$ \citep{underwood2016exomol}, $\mathrm{H_2S}$ \citep{azzam2016exomol}, HCN \citep{Barber}, CH$_4$ \citep{yurchenko2017hybrid}, Na \citep{kurucz1975table, wiese1966atomic, lindgaard1977transition, ralchenko2011nist}, and K \citep{wiese1966atomic, kurucz1975table}. We also use CIA (Collision-Induced Absorption) cross-sections of $\mathrm{H_2-H_2}$ and $\mathrm{H_2-He}$ \citep{richard2012new} contributing to the continuum. 



To include the opacity contributions due to clouds, aerosols or hazes, \texttt{NEXOTRANS} implements several methods. These are discussed briefly in Section \ref{subsubsection : cloud}\\

\subsubsection{\textbf{Cloud/Haze/Aerosol Models}} \label{subsubsection : cloud}

We include four different cloud implementations in \texttt{NEXOTRANS}: uniform/patchy grey clouds (with or without haze), sigmoid clouds, Mie scattering aerosols, and Ackerman-Marley clouds. These implementations are selected based on the distinct behaviors they exhibit due to variations in particle sizes.

We incorporate the patchy grey cloud model discussed in \citet{line2016influence}, in which, along with assuming an opaque grey cloud deck at pressures $\mathrm{P \geq P_{cloud}}$, we also include scattering due to hazes above this cloud deck level \citep{macdonald_2019}. Therefore, the combined extinction due to this cloud parameterization is given by:

\begin{equation} \label{haze}
\kappa_{\text{cloud}}(h) = \begin{cases} 
a \sigma_0 \left(\lambda / \lambda_0\right)^\gamma & \left(P < P_{\text{cloud}}\right) \\
\infty & \left(P \geq P_{\text{cloud}}\right)
\end{cases}
\end{equation}

where $a$ is the Rayleigh enhancement factor, $\sigma_0$ is the $\mathrm{H_2}$ Rayleigh scattering cross-section at the reference wavelength $\lambda_0$ (350 nm).

In this case, the transmission spectrum is given by:

\begin{equation} \label{equation15} \delta_{\lambda} = \phi \delta_{\lambda, \text{cloudy}} + (1-\phi)\delta_{\lambda, \text{clear}} \end{equation}

where $\phi$ is the fraction of the atmosphere covered by clouds, with values ranging from 0 to 1. A value of 0 represents a clear atmosphere without any clouds, while a value of 1 represents a uniformly distributed grey cloud deck.

The above parameterization corresponds to the large particle size limit of Mie scattering and the grey cloud assumption. However, to model aerosol scattering caused by particles smaller than $\sim$ 10 $\mathrm{\mu m}$ and non-grey clouds, we employ the parametric sigmoid cloud model mentioned in \citet{Constantinou}. The optical depth due to this cloud implementation varies as a sigmoid, given by the equation:

\begin{equation} \label{sigmoid}
    \tau_{\text{cloud}}(\lambda) = \frac{100}{1 + \exp(w(\lambda - \lambda_c))}
\end{equation}

where $\lambda_c$ is the center of the sigmoid and also the wavelength at which the cloud opacity diminishes. The parameter $w$ controls the slope of the sigmoid around $\lambda_c$, and $\lambda$ represents the model wavelengths. The sigmoid cloud model can also be combined with the haze model above the cloud deck level, as described in Equation (\ref{haze}). The sigmoid cloud model behaves in a grey-like manner only until the wavelength $\lambda_c$, after which the sigmoid function subsides.

In addition, \texttt{NEXOTRANS} includes Mie scattering aerosol models to incorporate the spectral contributions from species like MgSiO$_3$, ZnS, MnS, SiO$_2$, and Fe$_2$O$_3$. For this, we utilize the Mie scattering aerosol model as described in \citet{Constantinou_2023} and \citet{pinhas2017signatures}. This was demonstrated in \citet{Constantinou} and \citet{Constantinou_2023} by constraining aerosol properties based on the first observations of WASP-39 b obtained with JWST \citep{jwst2023identification, rustamkulov2023early, feinstein2023early}.

Additionally, we also implement the \citet{ackerman2001precipitating} cloud model to calculate vertical profiles of particle size distributions in condensation clouds. This model specializes in treating condensation clouds by balancing the upward turbulent diffusion of condensate and vapor with the sedimentation of condensates (i.e., particle settling), assuming that all condensation clouds are horizontally homogeneous.

\begin{equation} \label{ackerman1}
-K \frac{\partial q_t}{\partial z} - f_{\text{rain}} w_* q_c = 0
\end{equation}

where $K$ is the vertical eddy diffusion coefficient, $f_{\text{rain}}$ is defined as the ratio of the mass-weighted droplet sedimentation velocity to the convective velocity scale $w_*$, and $q_t = q_c + q_v$ is the total mole fraction of condensates and vapors.

The model dynamically couples the cloud structure with the pressure-temperature profile of the atmosphere. This is essential because the condensation curve of various species determines where clouds will form based on the local temperature and pressure conditions. Following this model, we can calculate the opacity contributions in the geometric limit due to condensate clouds, such as $\mathrm{MgSiO_3}$ or Fe clouds, for an atmospheric layer of thickness $\Delta z$, given by:

\begin{equation} \label{ackerman2}
\Delta \tau = \frac{3}{2} \frac{\epsilon \rho_a q_c}{\rho_p r_{\mathrm{eff}}} \Delta z
\end{equation}

where $\epsilon$ is the ratio of condensate to atmospheric molecular weights, $r_{\text{eff}}$ is the effective (area-weighted) droplet radius, $\rho_p$ is the density of a condensed particle, and $\rho_a$ is the atmospheric density.

\subsection{\textbf{BAYESIAN RETRIEVAL}} \label{subsection : bayesian retrieval}

The primary aim of a retrieval algorithm is to provide robust statistical estimates of the assumed parameters in the model and to offer model assessment metrics. This allows for the assessment of the validity and capability of a model to explain exoplanetary atmospheric observations. Recently, Nested Sampling \citep{Skilling;2006} has become an efficient method for Bayesian inference in exoplanetary retrievals due to its potential to better sample high-dimensional parameter spaces and manage complex posterior distributions that may contain several high-likelihood regions. Conventional Markov Chain Monte Carlo (MCMC) strategies may also struggle with such distributions, while Nested Sampling systematically explores the entire parameter space.

\texttt{NEXOTRANS} uses two implementations of Nested Sampling. One is MultiNest \citep{10.1111/j.1365-2966.2007.12353.x, 10.1111/j.1365-2966.2009.14548.x, feroz2013importance}, implemented via the Python wrapper \texttt{PyMultiNest} \citep{buchner2014x}, and the other is \texttt{UltraNest} \citep{buchner2021ultranest}.

The likelihood of observing the data $\mathcal{D}$, given a set of model parameters $\boldsymbol{\theta}$ for a model $\mathcal{M}$, is expressed as \citep{welbanks2021aurora}:

\begin{equation} \label{liklihood eqn}
\mathcal{L} = P(\mathcal{D}|\boldsymbol{\theta},\mathcal{M})
\end{equation} 

Incorporating the prior distribution $\pi(\boldsymbol{\theta})$, the marginalized likelihood (or evidence) is obtained by integrating the likelihood over the entire parameter space:

\begin{equation} \label{evidence eqtn}
\mathcal{Z} = P(\mathcal{D}|\mathcal{M}) = \int P(\mathcal{D}|\boldsymbol{\theta},\mathcal{M}) P(\boldsymbol{\theta}|\mathcal{M}) d\boldsymbol{\theta} 
\end{equation}

Moreover, given the data, the posterior probability distribution of each parameter is obtained using Bayes' theorem:

\begin{equation} \label{bayes equation} 
P(\boldsymbol{\theta}|\mathcal{D},\mathcal{M}) = \frac{P(\mathcal{D}|\boldsymbol{\theta},\mathcal{M}) P(\boldsymbol{\theta}|\mathcal{M})}{P(\mathcal{D}|\mathcal{M})}
\end{equation}

The likelihood function used by \texttt{NEXOTRANS} for data with independently distributed Gaussian errors is given by:

\begin{equation} \label{liklihood eqtn2}
\mathcal{L}(\mathcal{D}|\boldsymbol{\theta},\mathcal{M}_i) = \prod_{j=1}^N \frac{1}{\sqrt{2\pi\sigma_j^2}} \exp\left(-\frac{[\mathcal{D}_j - \mathcal{M}_{i,j}]^2}{2\sigma_j^2}\right)
\end{equation}

Hence, Nested Sampling serves as an integral tool, providing the evidence necessary for model comparison and evaluation, along with the posterior distributions.

\subsubsection{\textbf{MultiNest}} \label{subsubsection : multinest}

Nested Sampling iteratively samples live points from the prior distribution, finding the lowest-probability point each time and replacing it with a higher-probability point. Through this repetitive process, the entire parameter space is sampled by shrinking contours of equal likelihood, which gradually migrate to the region of highest likelihood \citep{10.1111/j.1365-2966.2009.14548.x}.

\texttt{MultiNest}, a widely used Nested Sampling algorithm, has been praised for its efficient exploration of parameter space using ellipsoidal rejection sampling, its ability to handle multimodal posteriors, and its simultaneous calculation of Bayesian evidence alongside parameter estimation. However, \texttt{MultiNest} and similar Nested Sampling implementations have been found to suffer from biases when likelihood contours are non-ellipsoidal \citep{dittmann2024notes}.

\subsubsection{\textbf{UltraNest}} \label{subsubsection : ultranest}

To address these limitations, \texttt{UltraNest} was developed as a more recent Nested Sampling implementation \citep{buchner2021ultranest}. While trading some computational efficiency for increased robustness, \texttt{UltraNest} incorporates several advanced features. These include the MLFriends algorithm for constrained-likelihood sampling, which uses Mahalanobis distance-based ellipsoids around live points, and a nested sampler capable of dynamically managing the live point population \citep{buchner2023nested}. \texttt{UltraNest} also employs bootstrapping for uncertainty estimation and a stopping criterion that takes into account the weights of the live points, using a tolerance-based approach in the case of noisy likelihoods. This enables better handling of multidimensional parameter spaces than other Bayesian inference methods.

\subsection{\textbf{MACHINE LEARNING RETRIEVAL}} \label{subsection : machine learning}

Bayesian methods, while yielding promising results in exoplanetary atmospheric retrievals, are computationally expensive, often requiring hours or days for a single retrieval depending on the availability of computational resources. This limitation has led to the exploration of machine learning (ML) methods for exoplanetary atmospheric retrievals \citep{2018NatAs...2..719M,fisher2020interpreting,nixon_madhusudhan_2020,MacDonald_2023}.

Machine learning methods rely heavily on data, and datasets are a crucial aspect of their application. \texttt{NEXOTRANS} addresses this challenge by integrating an in-house forward model, as discussed in Section \ref{subsec: forward model}, with data generation capabilities specifically designed for machine learning applications. \texttt{NEXOTRANS} offers a robust machine learning retrieval scheme that is adaptable to diverse exoplanetary atmospheres and provides enhanced flexibility in parameter selection.

\begin{figure}
    \centering
    \includegraphics[width=\linewidth]{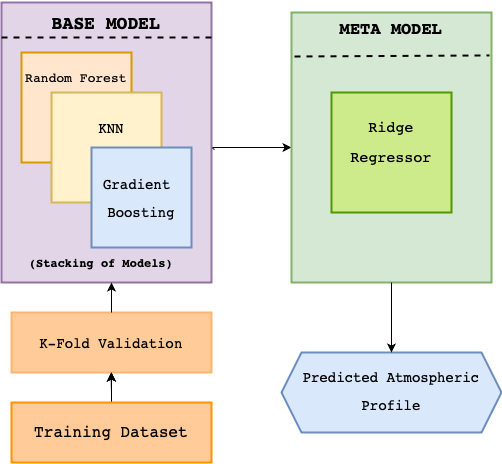}
    \caption{Schematic representation of the machine learning training structure used in \texttt{NEXOTRANS}. Training data, obtained after data generation, is used to train the base models individually, after which the \texttt{Ridge Regressor} is employed for the final prediction. This method is called supervised ensemble learning (\texttt{Stacking Regressor}).}.
    \label{fig:train}
\end{figure}

The machine learning component of \texttt{NEXOTRANS}, illustrated in Figure \ref{fig:train}, employs a supervised ensemble learning approach utilizing a \texttt{Stacking Regressor} model. This model incorporates \texttt{Random Forest}, \texttt{Gradient Boosting}, and \texttt{K-Nearest Neighbour} algorithms as base models, with a \texttt{Ridge Regressor} serving as the meta-model for final predictions. By leveraging the strengths of multiple algorithms, this architecture enhances overall performance and reliability. The \texttt{Stacking Regressor} model serves as the default configuration. A key feature of \texttt{NEXOTRANS} is its ability to perform retrievals with different base models (\texttt{Random Forest}, \texttt{Gradient Boosting}, \texttt{K-Nearest Neighbour}), enabling performance comparisons between these individual methods as well as the default model. This functionality provides valuable insights into the results generated by different approaches.

\subsubsection{\textbf{Data Generation}}\label{subsubsection : data generation}

\texttt{NEXOTRANS} offers a comprehensive suite for customizing parameters, enabling us to tailor the data generation process to their specific needs. Data can be generated by defining the wavelength range that matches the observational spectrum, with the bin size determined by the resolution. We can also define chemical species (e.g., H$_2$O, CO$_2$, CO), aerosol species (e.g., MgSiO$_3$, ZnS), disequilibrium species (e.g., SO$_2$), select the appropriate P-T profiles as described in Section \ref{subsubsection : pressure temperature profile}, choose the chemistry as outlined in Section \ref{subsubsection : chemistry}, and provide cloud, haze, or aerosol as discussed in Section \ref{subsubsection : cloud}. The generated data is stored as a 2D array, as shown below:

\begin{equation}
\begin{pmatrix}
  T_{S_1,\lambda_1} & T_{S_1,\lambda_2} & \cdots & T_{S_1,\lambda_n} & | & L_{S_1,1} & L_{S_1,2} & \cdots & L_{S_1,m} \\
  T_{S_2,\lambda_1} & T_{S_2,\lambda_2} & \cdots & T_{S_2,\lambda_n} & | & L_{S_2,1} & L_{S_2,2} & \cdots & L_{S_2,m} \\
  T_{S_3,\lambda_1} & T_{S_3,\lambda_2} & \cdots & T_{S_3,\lambda_n} & | & L_{S_3,1} & L_{S_3,2} & \cdots & L_{S_3,m} \\
  \vdots & \vdots & \ddots & \vdots & | & \vdots & \vdots & \ddots & \vdots \\
  T_{S_r,\lambda_1} & T_{S_r,\lambda_2} & \cdots & T_{S_r,\lambda_n} & | & L_{S_r,1} & L_{S_r,2} & \cdots & L_{S_r,m}
\end{pmatrix}
\label{eqtn}
\end{equation}

In this matrix, each row corresponds to a single spectrum $S_r$, where $r$ is the number of spectra generated. $T_{S_r,\lambda_n}$ represents the transit depth value corresponding to a particular wavelength index, ranging from 1 to $n$ for each spectrum. $L_{S_r,m}$ represents the values of the parameters, where $m$ is the number of parameters. The columns containing transit depth values are referred to as features, while those containing parameter values are called labels. Once the data generation is complete, it undergoes feature reduction as discussed in the next section.

 \subsubsection{\textbf{Feature Reduction}} \label{subsubsection : Feature engineering}


It is important to note that the high-dimensional nature of spectral data often necessitates feature reduction techniques prior to model training. For feature reduction, we aim to reduce the number of transit depths $T_{S_{r},\lambda_{n}}$, i.e., the number of columns in the dataset. We employ a method based on wavelength selection. For a given wavelength index $n$, corresponding to $\lambda_{n}$, we select only those $\lambda_{n}$ that show the maximum variation in transit depth across the spectrum. Thus, $\lambda_{x}$ represents the wavelengths chosen, where $x$ is the index of the wavelength at which the maximum variation occurs, where $x \subseteq n$. The number of features will then depend on the selected wavelengths. The transit depth corresponding to the selected wavelength ( $\lambda_{x}$) is chosen for training, as a result, the columns of the generated data are reduced, addressing the curse of dimensionality.

We must ensure that the number of features in the observational dataset matches the number of features used to train the model. To achieve this, we select the transit depths in the observational dataset corresponding to the chosen wavelength value $\lambda_{x}$. If the resolution does not match, we use the transit depth corresponding to the wavelength that is closest to the one used for training.

For the JWST datasets used in this study (NIRISS, NIRSpec PRISM, and MIRI), the average wavelength mismatch between the feature-reduced training datasets and the feature-reduced observational datasets is less than 0.0562 $\mu$m across the various generated datasets. The method employed here provides a straightforward way to ensure that the number of features used for training the model matches the number of features in the observational dataset without introducing any noise. The choice of $\lambda_{n}$ depends on the desired data generation approach. Data can be generated to match the resolution of the observational dataset, thereby minimizing the mismatch.

\subsubsection{\textbf{Training}} \label{subsubsection:training}

For the machine learning component of \texttt{NEXOTRANS}, we implement a supervised ensemble learning approach utilizing a \texttt{Stacking Regressor}. This technique combines multiple base models to improve predictive performance and robustness \citep{wolpert1992stacked, breiman1996bagging}. Our \texttt{Stacking Regressor} comprises three diverse base models: \texttt{Random Forest} \citep{breiman2001random}, \texttt{Gradient Boosting} \citep{friedman2001greedy}, and \texttt{k-Nearest Neighbor} (KNN) \citep{cover1967nearest}.

For training, the dataset is partitioned into $k$ folds, with $k-1$ folds used for training each base model and the $k$-th fold reserved for validation. This process is iterated $k$ times, ensuring that each fold serves as the validation set exactly once, thus training all base models comprehensively. The features in the $k$-th fold are then used to generate a prediction matrix using the trained base models.

This prediction matrix serves as the input feature set for a meta-model, which, in our case, is a \texttt{Ridge Regressor} \citep{hoerl1970ridge}. The \texttt{Ridge Regressor} is trained to optimally combine the predictions from the base models, with the true value labels of the $k$-th fold serving as the target variable. The loss function for the \texttt{Ridge Regressor} is defined as:

\begin{equation} \label{ridge regressor etn} L(\beta) = ||y - X\beta||^2_2 + \alpha ||\beta||^2_2 \end{equation}

where $y$ is the vector of true values, $X$ is the prediction matrix from the base models, $\beta$ is the coefficient vector, and $\alpha$ is the regularization parameter. The first term represents the ordinary least squares error, while the second term is the L2 regularization that helps prevent overfitting \citep{tibshirani1996regression}.

This ensemble learning approach allows the exploitation of the strengths of each base model while mitigating their individual weaknesses. The \texttt{Random Forest} provides robustness to outliers and captures non-linear relationships, \texttt{Gradient Boosting} excels at handling complex interactions between features, and \texttt{k-Nearest Neighbor} can effectively model local patterns in the data. The \texttt{Ridge Regressor}, as the meta-model, learns to optimally weight the predictions from these diverse base models, potentially leading to superior predictive performance compared to any single model approach. More about the \texttt{Random Forest}, \texttt{k-Nearest Neighbor}, and \texttt{Gradient Boosting} has been discussed in Appendix \ref{subsec:base}.

\begin{figure*}[]
\centering
\includegraphics[width=1.0\linewidth]{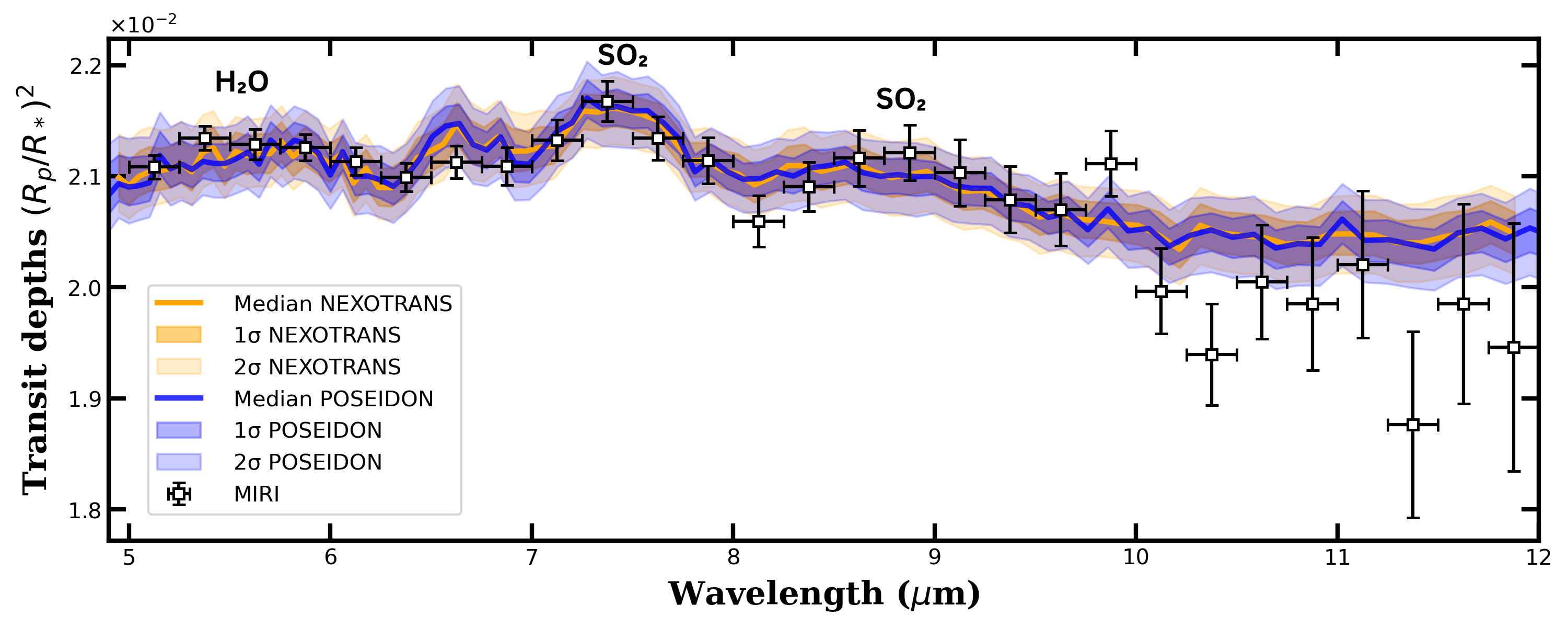}
\caption{Best-fit retrieved spectrum on Eureka! reduced data with \texttt{NEXOTRANS} and \texttt{POSEIDON}, both with the median and 1$\sigma$ error envelope. The model corresponds to the patchy grey cloud and haze assumption run with a resolution of 15,000 and binned to 100 for plotting. The \texttt{NEXOTRANS} spectrum is in orange, and the \texttt{POSEIDON} spectrum is in blue. The MIRI data points are plotted with black error bars.
}
\label{fig:benchmark-spectrum}
\end{figure*}

\begin{figure*}[]
\centering
\includegraphics[width=\linewidth]{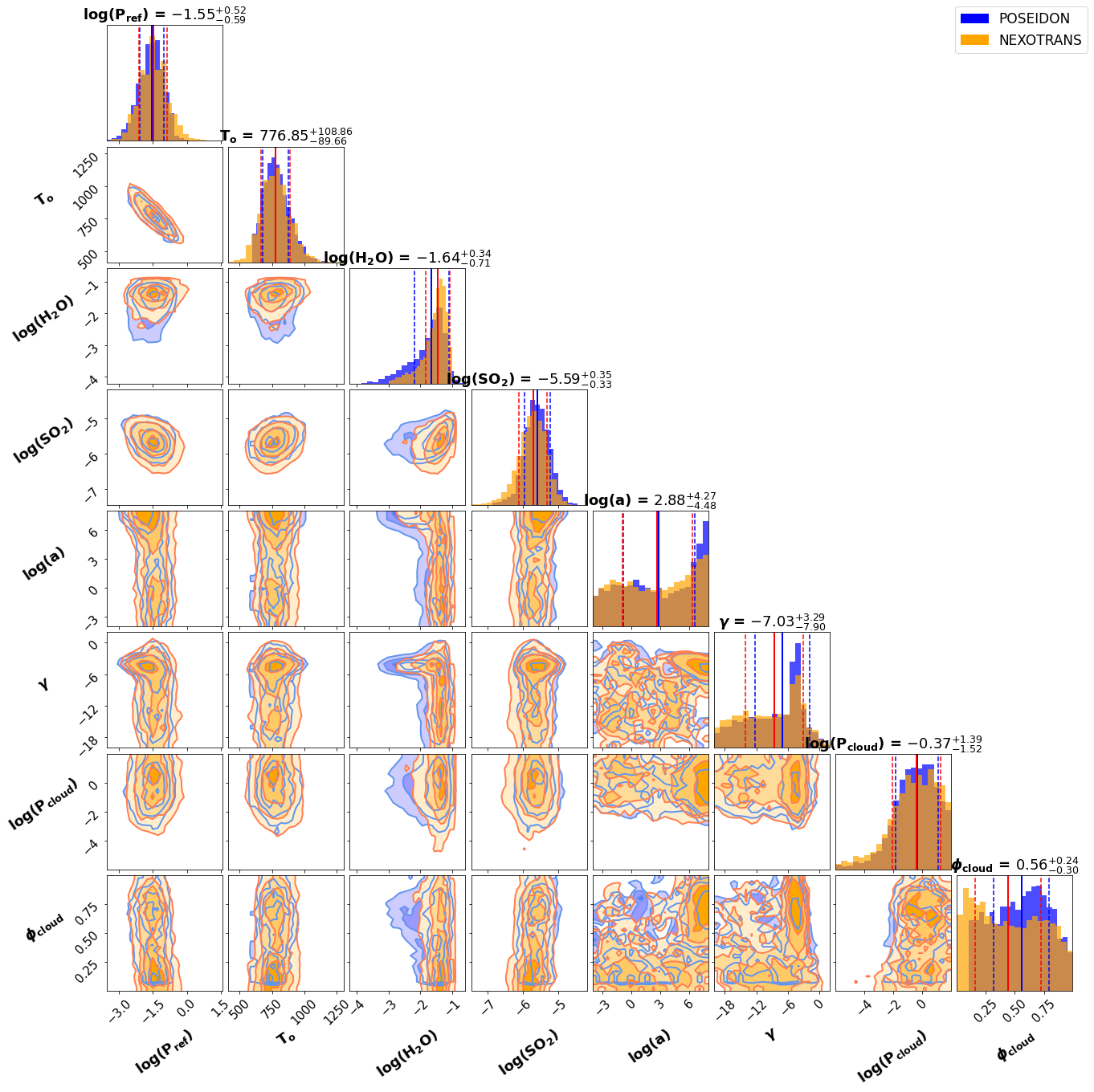}
\caption{Corner plots comparing the posterior distributions of parameters retrieved with \texttt{NEXOTRANS} and \texttt{POSEIDON} for the Eureka! reduced MIRI data. The reference abundance constraints stated above each histogram correspond to results from the \texttt{POSEIDON} framework.
}
\label{fig:overplot_corner}
\end{figure*}

To visualize the predictions, we examine the posterior distribution in the form of corner plots. In the case of Bayesian retrievals, these often yield a Gaussian distribution of parameter ranges, making it straightforward to generate corner plots. However, since machine learning does not provide a distribution, we need to use alternative methods to create these plots. Our methodology involves sampling the local prediction space around the observed data points. Specifically, we apply uniform random perturbations within a $\pm10\%$ range to the measured transit depths, generating an ensemble of 100 slightly varied spectral inputs. This approach creates a distribution of predictions that captures both the model's sensitivity to observational uncertainties and the intrinsic variance in its predictions. 
While this method is not fully Bayesian, it offers a practical solution for quantifying uncertainty in machine learning-based atmospheric retrievals. This approach provides insights into parameter correlations and model sensitivity that would be lacking in single-point predictions.

\subsection{\textbf{\texttt{NEXOTRANS} VALIDATION}} \label{retrieval_validation_section}

We benchmark \texttt{NEXOTRANS} by conducting retrievals on the JWST MIRI data covering the 5.0 - 12.0 $\mu$m range. We follow \citet{powell2024sulfur}, where retrievals were performed using seven different algorithms on data reduced by three distinct pipelines. In our analysis, we include \texttt{POSEIDON} retrievals as well. The retrievals are carried out under the same assumptions outlined in \citet{powell2024sulfur}, with a model resolution of R = 15,000 and 1000 live points for \texttt{PyMultiNest} retrievals. We assume a free chemistry model and explore three cloud parametrizations: grey, patchy grey, and patchy grey with haze.

The \texttt{NEXOTRANS} retrievals demonstrate strong agreement with the other results, as shown in Table \ref{benchmark_table} in the Appendix \ref{retrieval_benchmark_results}. Figure \ref{fig:benchmark-spectrum} shows the best-fit retrieved spectra obtained with \texttt{NEXOTRANS} and \texttt{POSEIDON}. Both retrievals demonstrate excellent agreement with the observed data, accurately capturing the absorption features attributed to H$_2$O and SO$_2$. The retrieved log(VMR) for SO$_2$ is consistent across all data reductions and retrieval frameworks. The spread of log(SO$_2$) values, from the lowest -1$\sigma$ bound to the highest +1$\sigma$ bound, ranges from -6.2 to -5.0. We also obtained log(H$_2$O) median values that are consistent with those from all other retrieval algorithms, with the -1$\sigma$ and +1$\sigma$ bounds spanning from -1.9 to -1.0.

\begin{table*}
\centering
\caption{Free parameters in the retrieval models.}
\label{tab:retrieval_models}
\renewcommand{\arraystretch}{1.2} 

\begin{tabular}{lc}
\hline
\hline
\textbf{Model}                        & \textbf{Free Parameters}  \\
\hline
\hline
Common Parameters* & log(P$_{ref}$), $\alpha_1$, $\alpha_2$, log(P$_1$), log(P$_2$), log(P$_3$),  
                                    T$_0$, log(MgSiO$_3$),  
                                   log(ZnS), r(MgSiO$_3$), r(ZnS), h$_c$, f$_c$  \\
\hline
Free Chemistry     & log(H$_2$O), log(CO$_2$), log(CO), log(SO$_2$),  
                                   log(H$_2$S),   
                                  log(CH$_4$), log(HCN), log(Na), log(K)   \\
\hline
Equilibrium Chemistry         &  C/O, Metallicity   \\
\hline
Hybrid Equilibrium Chemistry  & C/O, Metallicity, log(SO$_2$), log(Na), log(K)   \\
\hline
Equilibrium Offset Chemistry  & C/O, Metallicity,  $\delta$(H$_2$O),  
                                      $\delta$(CO$_2$), $\delta$(CO),  
                                   $\delta$(H$_2$S),   
                                  $\delta$(CH$_4$), $\delta$(HCN), log(SO$_2$), log(Na), log(K)   \\
\hline
\hline
\textbf{Number of Datapoints} & NIRISS: 331, PRISM: 105, MIRI: 28 \\
\hline
\end{tabular}
\label{Table:free_parameter}
* These parameters are included in all retrieval models.
\end{table*}


The retrieved cloud parameters are consistent with the findings of \citet{powell2024sulfur} as well as \texttt{POSEIDON} retrievals, with the median log cloud top pressure ranging from -0.38 to -2.26 bar and terminator coverage spanning from 0.35 to 0.45, all within the 1$\sigma$ intervals of the previous results. The retrieved \texttt{NEXOTRANS} haze parameters are as follows: log(a) = $2.72_{-4.52}^{+4.05}$, $2.10_{-3.53}^{+3.59}$, $1.52^{+4.16}_{-3.62}$ for \texttt{Eureka}, \texttt{Tiberius}, and \texttt{Sparta} reductions, respectively; $\gamma$ = $-8.55_{-7.35}^{+4.89}$, $-10.94_{-5.78}^{+6.35}$, $-10.71^{+6.87}_{-5.99}$ for \texttt{Eureka!}, \texttt{Tiberius}, and \texttt{Sparta} reductions, respectively.

Figure \ref{fig:overplot_corner} presents the overplotted corner plots of selected parameters constrained by both \texttt{NEXOTRANS} and \texttt{POSEIDON}, allowing for a direct comparison. The credible regions in the corner plot show strong agreement between the two frameworks. Notably, all the retrieved parameter values remain consistent within 1$\sigma$ across both frameworks. 

\subsection{\textbf{APPLICATION OF \texttt{NEXOTRANS} ON FULL JWST DATASETS FOR WASP 39 b}} \label{subsec : JWST dataset}

To demonstrate the capabilities and applications of \texttt{NEXOTRANS}, we used the full set of JWST observations available for WASP-39 b, spanning a wavelength range from 0.6 $\mu$m to 12.0 $\mu$m. These observations were obtained with the NIRISS, NIRSpec PRISM, and MIRI LRS instruments onboard JWST. The 0.6-2.8 $\mu$m observations \citep{feinstein2023early} were obtained as part of the JWST ERS program, using the SOSS mode of the NIRISS instrument onboard JWST. We utilized the atmospheric spectrum, reduced with the Supreme-SPOON pipeline, provided in the NASA Exoplanet Archive's atmospheric spectroscopy table\footnote{\url{https://exoplanetarchive.ipac.caltech.edu/cgi-bin/atmospheres/nph-firefly?atmospheres}}. We also used the 0.5-5.5 $\mu$m transmission spectrum observed \citep{rustamkulov2023early} with the JWST NIRSpec's PRISM mode and reduced with the FIREFLy pipeline. This data was obtained as part of the JWST Transiting Community Early Release Science Team program. Once again, we downloaded the data from the \href{https://exoplanetarchive.ipac.caltech.edu/cgi-bin/atmospheres/nph-firefly?atmospheres}{NASA Exoplanet Archive}. However, note that we did not use the PRISM data below 2.0 $\mu$m, since it was reported that those observations might be affected by detector saturation \citep{carter2024benchmark, Constantinou}. Additionally, we also include the Eureka! reduced latest mid-infrared transmission spectrum observations from \citet{powell2024sulfur}, measured by the JWST Mid-Infrared Instrument (MIRI) Low-Resolution Spectrometer (LRS) in the 5-12 $\mu$m range (\href{https://zenodo.org/records/10055845}{MIRI data}). This enabled us to perform the retrievals on the full spectrum of WASP-39 b obtained by JWST.

 \section{\textbf{Results}} \label{sec:result} 
Now, we discuss the results of the \texttt{NEXOTRANS} framework through a comprehensive series of retrievals, informed by previous atmospheric studies of WASP-39 b \citep{wakeford2017complete, Constantinou_2023, Constantinou}. By combining data from the NIRISS, NIRSpec PRISM, and MIRI LRS instruments, we provide robust constraints on the atmospheric parameters. We performed retrievals of WASP-39 b using four types of chemistry models, incorporating both a patchy cloud-haze model and an alternative approach that independently assumes aerosol species. This methodology is motivated by the inferred impact of non-grey cloud opacities in the lower wavelength regime \citep{Constantinou_2023, Constantinou}.

\begin{figure*} []
    \centering
    \begin{minipage}{0.50\textwidth}
        \centering
        \includegraphics[width=\textwidth]{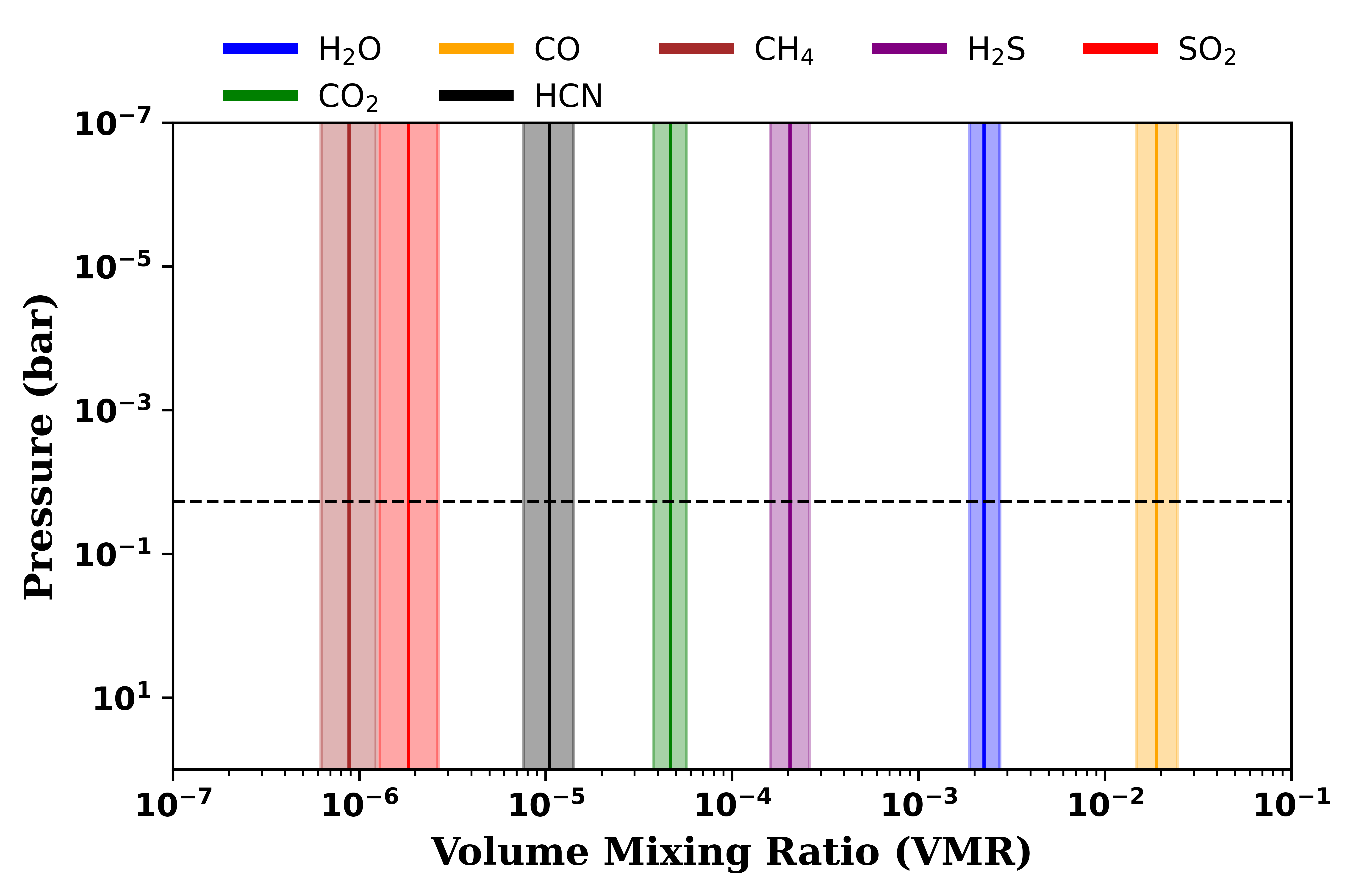}
        \parbox{0.8\linewidth}{\centering (a) Free chemistry retrieved VMR.}
        \label{fig:vmr_free}
        
    \end{minipage}
    \hfill
    \begin{minipage}{0.49\textwidth}
        \centering
        \vspace{0.7cm}
        \includegraphics[width=\textwidth]{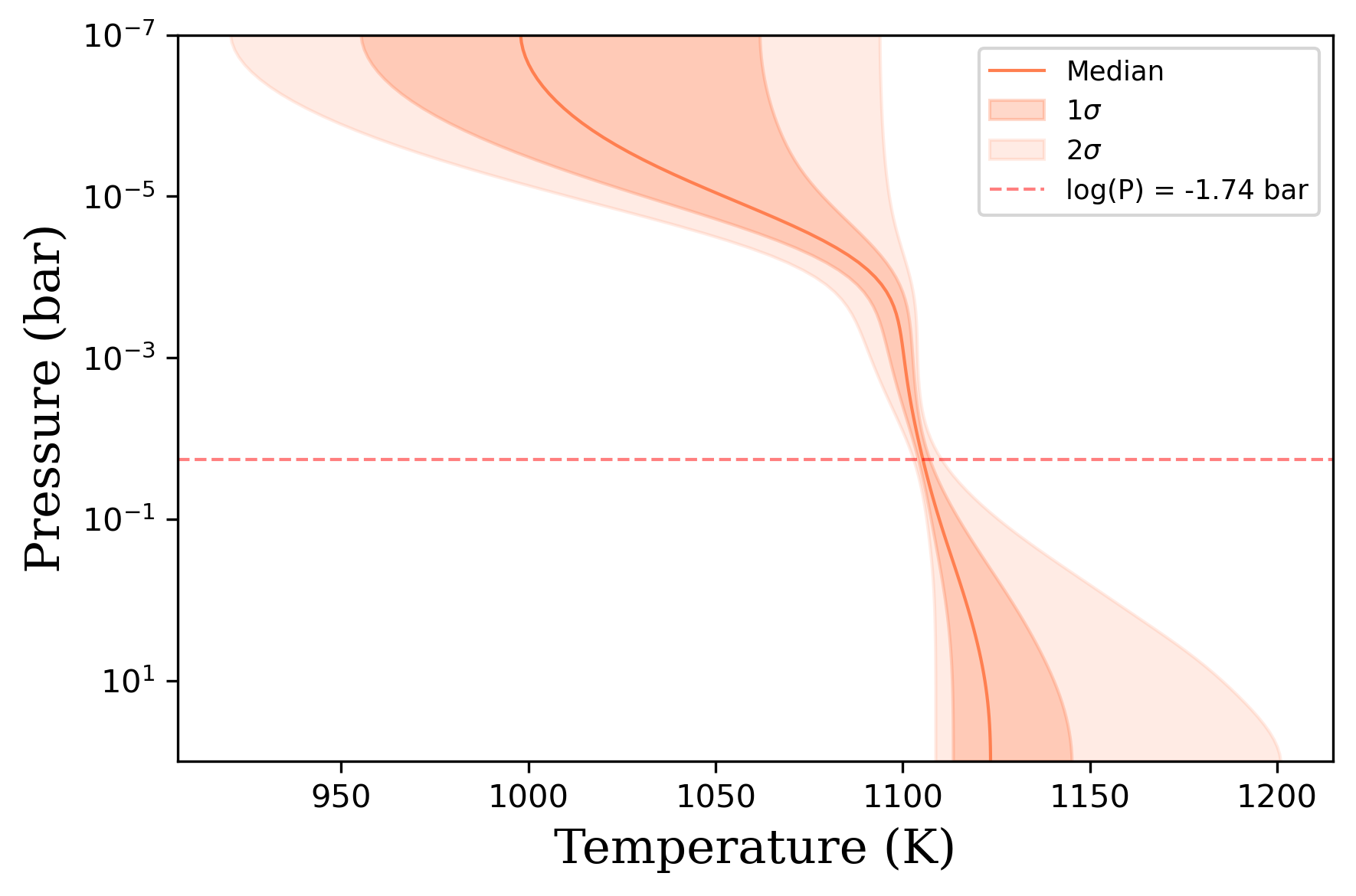}
        \parbox{0.8\linewidth}{\centering (b) Free Chemistry retrieved P-T profile.}
        \label{fig:pt_free}
    \end{minipage}
    
    \hspace{-0.1cm}
    \begin{minipage}{0.485\textwidth}
        \centering
        \includegraphics[width=\textwidth]{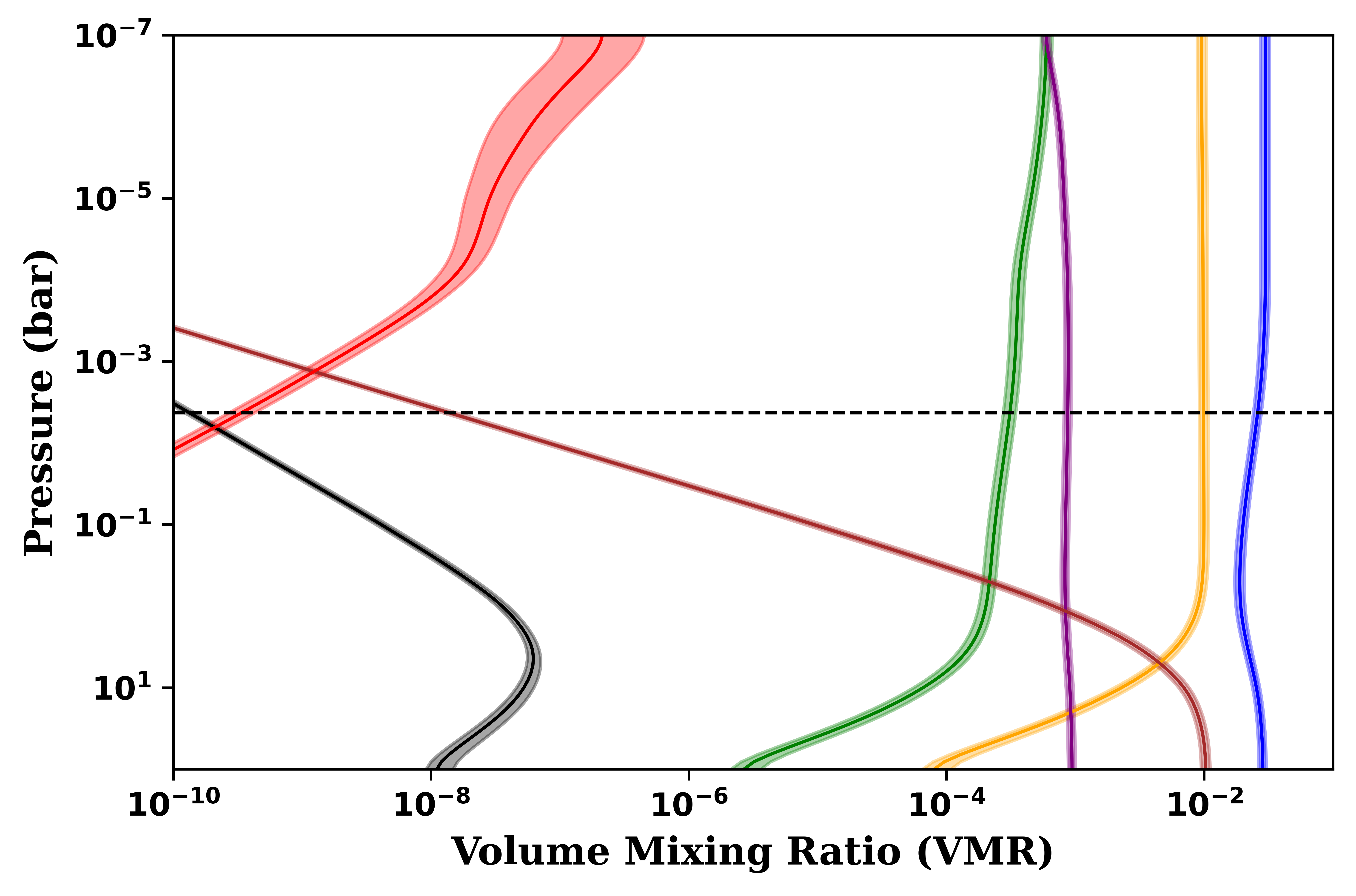}
        \parbox{0.8\linewidth}{\centering (c) Equilibrium chemistry retrieved VMR. }
        \label{fig:vmr_equ}
    \end{minipage}
    \hfill
    \hspace{0.1cm}
    \begin{minipage}{0.49\textwidth}
        \centering
        \hspace{0.5cm}
        \includegraphics[width=\textwidth]{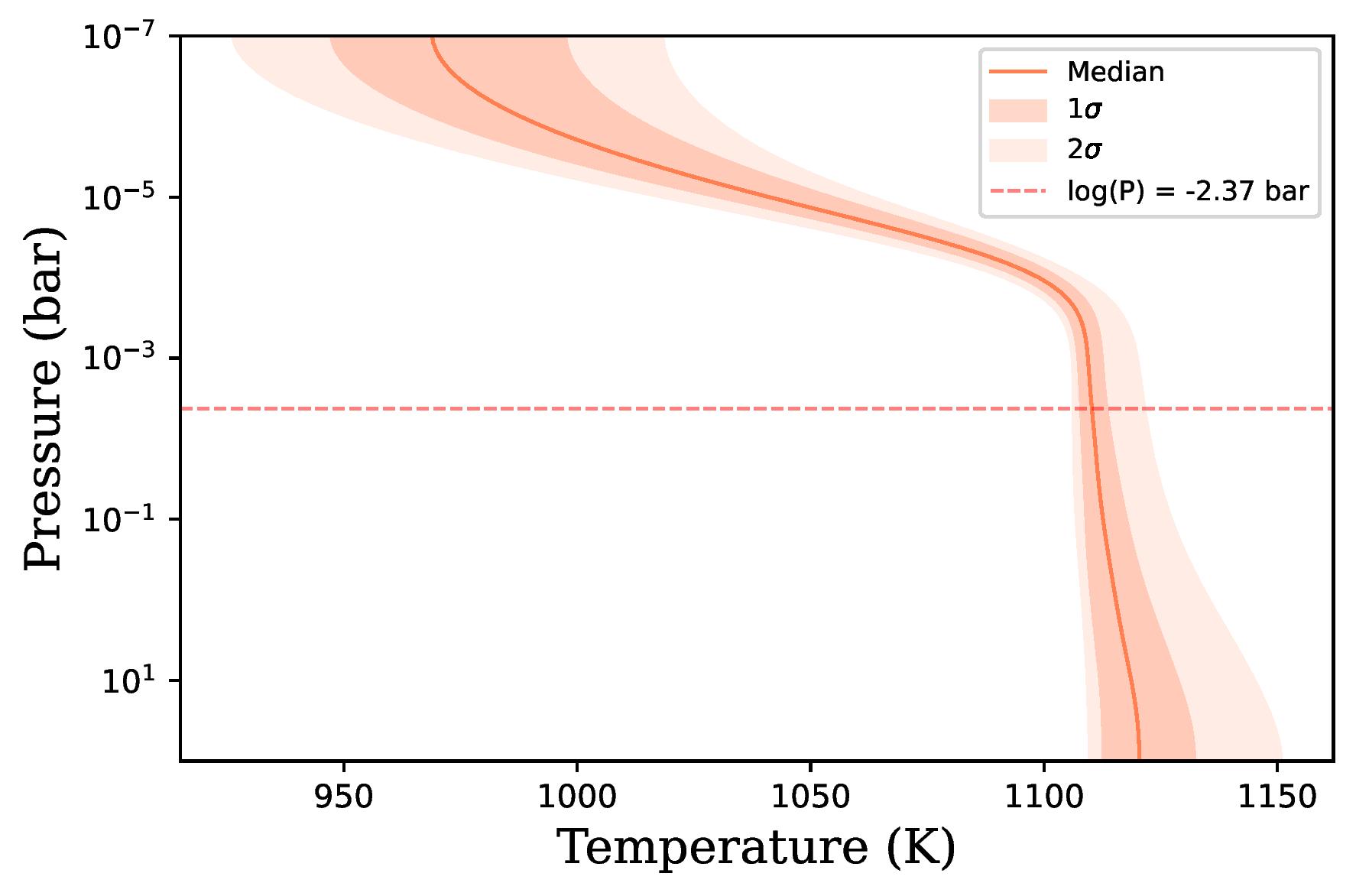}
        \parbox{0.8\linewidth}{\centering (d) Equilibrium Chemistry retrieved P-T profile.  }
        \label{fig:pt_equ}
    \end{minipage}
    
    \caption{Retrieved Volume Mixing Ratio profiles for: a) free chemistry and c) equilibrium chemistry, along with the corresponding $P$-$T$ profiles retrieved. The dotted horizontal lines represent the retrieved median reference pressures for each model. The 1$\sigma$ region for the VMR profiles are also indicated by the light colored bands.}
\label{fig:chemistry_vmr_part1}
\end{figure*}
The atmospheric model spans a pressure range from $10^{-7}$ to $100$ bar, with 100 layers uniformly distributed in logarithmic pressure. We assume a $\mathrm{H_2}$ + $\mathrm{He}$ dominated atmosphere with $\mathrm{He}/\mathrm{H_2} = 0.17$. For constraining the P-T profile of the atmosphere of WASP-39 b, we adopt the three region model from \citet{madhusudhan2009temperature}. The PyMultiNest Bayesian retrieval framework utilizes 2000 live points to sample the parameter space. At each point in the parameter space, \texttt{NEXOTRANS} computes the transmission spectrum of WASP-39 b at a resolution of $R = 20,000$, covering the wavelength range from $0.60 \,\mu\mathrm{m}$ to $12.0\,\mu\mathrm{m}$. The free parameters for the different model configurations are presented in Table \ref{Table:free_parameter}.

The machine learning (ML) model is trained using 60,000 spectra, each generated in the wavelength range from $0.60 \,\mu\mathrm{m}$ to $12.0\,\mu\mathrm{m}$ with a resolution of 158. Low resolution is chosen for data generation so that feature reduction spans across global peaks and does not get stuck in local peaks due to high resolution.

It is also worth mentioning that during our initial retrieval, assuming a patchy cloud deck and haze model with free chemistry, we retrieved an offset of 57 ppm for the NIRISS data and 311.13 ppm for the MIRI data, both with respect to NIRSpec PRISM data. Therefore, we corrected the respective datasets for these offsets and performed all our subsequent retrievals on the new datasets.

\begin{table*}
\centering
\caption{Retrieved abundances at log(P) = -2 bar, for different species under various chemistry models when assuming mie scattering aerosols. The best-fit factor, reduced $\chi^2$, for the different models is also added in the last column.}
\resizebox{1\textwidth}{!}{
\hspace{-1.9cm}
\begin{tabular}{lccccccccc}
\toprule
\hline
\multirow{2}{*}{} & \multicolumn{1}{c}{Na} & \multicolumn{1}{c}{K} & \multicolumn{1}{c}{H$_2$O} & \multicolumn{1}{c}{CO$_2$} & \multicolumn{1}{c}{CO} & \multicolumn{1}{c}{SO$_2$} & \multicolumn{1}{c}{H$_2$S} & \multicolumn{1}{c}{HCN} & \multicolumn{1}{c}{reduced $\chi^2$}\\
\cmidrule{1-10} 
 & \multicolumn{9}{c}{\textit{\textbf{Free Chemistry}}} \\
 
Bayesian& $-8.64^{+0.75}_{-0.76}$ & $-8.67^{+0.21}_{-0.23}$ & $-2.64^{+0.08}_{-0.08}$ & $-4.33^{+0.09}_{-0.09}$ & $-1.72^{+0.11}_{-0.10}$ & $-5.73^{+0.15}_{-0.16}$ & $-3.69^{+0.11}_{-0.11}$ & $-4.98^{+0.14}_{-0.14}$ & $2.99$\\
[0.2cm]
ML& $-6.51^{+0.02}_{-0.01}$ & $-8.01^{+0.02}_{-0.01}$ & $-2.47^{+0.52}_{-0.46}$ & $-3.79^{+0.33}_{-0.28}$ & $-1.50^{+0.35}_{-0.17}$ & $-6.25^{+0.01}_{-0.01}$ & $-3.39^{+0.08}_{-0.07}$ & $-4.70^{+0.05}_{-0.05}$    \\
[0.2cm]
\hline

 & \multicolumn{9}{c}{\textit{\textbf{Equilibrium}}} \\
Bayesian & $-3.87^{+0.03}_{-0.03}$ & $-5.36^{+0.02}_{-0.03}$ & $-1.61^{+0.02}_{-0.04}$ & $-3.54^{+0.05}_{-0.05}$ & $-2.00^{+0.04}_{-0.03}$ & $-10.03^{+0.10}_{-0.10}$ & $-3.06^{+0.03}_{-0.03}$ & $-9.49^{+0.05}_{-0.03}$ & $3.35$\\
[0.2cm]
ML& $-4.44^{+0.75}_{-0.22}$ & $-5.89^{+0.53}_{-0.26}$ & $-2.15^{+0.77}_{-0.24}$ & $-4.46^{+0.80}_{-0.34}$ & $-2.50^{+0.49}_{-0.23}$ & $-11.69^{+2.63}_{-1.48}$ & $-3.64^{+0.48}_{-0.23}$ & $-10.64^{+0.20}_{-0.10}$ \\
[0.2cm]
\hline

 & \multicolumn{9}{c}{\textit{\textbf{Hybrid Equilibrium}}} \\
Bayesian & $-7.97^{+0.44}_{-0.50}$ & $-8.74^{+0.15}_{-0.16}$ & $-2.89^{+0.01}_{-0.03}$ & $-4.59^{+0.05}_{-0.05}$ & $-1.99^{+0.06}_{-0.04}$ & $-5.80^{+0.11}_{-0.13}$ & $-3.93^{+0.03}_{-0.02}$ & $-8.42^{+0.09}_{-0.05}$ & $2.97$ \\
[0.2cm]
ML  & $-6.20^{+0.20}_{-1.30}$ & $-8.25^{+0.01}_{-0.01}$ & $-2.60^{+0.08}_{-0.08}$ & $-4.42^{+0.17}_{-0.55}$ & $-2.25^{+0.27}_{-0.49}$ & $-6.24^{+1.14}_{-0.32}$ & $-3.81^{+0.05}_{-0.20}$ & $-8.22^{+0.42}_{-0.57}$  \\ 
[0.2cm]
\hline

& \multicolumn{9}{c}{\textit{\textbf{Equilibrium Offset}}} \\
Bayesian & $-7.21^{+0.38}_{-0.68}$ & $-8.35^{+0.16}_{-0.18}$ & $-2.84^{+0.04}_{-0.06}$ & $-4.64^{+0.09}_{-0.06}$ & $-1.51^{+0.03}_{-0.05}$ & $-5.73^{+0.15}_{-0.16}$ & $-3.77^{+0.08}_{-0.08}$ & $-7.63^{+0.23}_{-0.45}$ & $2.98$\\
[0.2cm]
ML& $-6.80^{+1.00}_{-0.01}$ & $-8.30^{+0.01}_{-0.01}$ & $-2.57^{+0.08}_{-0.25}$ & $-4.17^{+0.07}_{-0.70}$ & $-1.89^{+0.07}_{-0.53}$ & $-6.00^{+1.00}_{-0.001}$ & $-3.61^{+0.03}_{-0.26}$ & $-8.23^{+0.09}_{-0.56}$  \\

\bottomrule
\end{tabular}
}
\label{tab:species_abundances}
\end{table*}

\subsection[\textbf{0.60 µm - 12.0 µm RETRIEVAL OF WASP-39 b WITH NEXOTRANS}]
{\textbf{0.60 $\mu$m - 12.0 $\mu$m RETRIEVAL OF WASP-39 b WITH \texttt{NEXOTRANS}}}


\subsubsection{\textbf{Free Chemistry Model Retrievals}}

We first perform a retrieval of the entire 0.6–12.0 $\mu$m dataset, assuming the commonly used patchy cloud deck and haze model discussed in Section \ref{subsubsection : cloud}. This model yielded a reduced $\chi^2$ of 3.30. For H$_2$O, we derived a log volume mixing ratio (VMR) of $-2.08^{+0.09}_{-0.09}$, dominated by many absorption peaks in the 0.6–4.0 $\mu$m range. A significant peak at $\sim4.50$ $\mu$m, attributed to CO$_2$, corresponds to a log VMR of $-3.74^{+0.10}_{-0.10}$. CO also contributes to absorption in the wavelength region redward of the CO$_2$ absorption region. The photo-chemical product SO$_2$ is also inferred at around 4.05 $\mu$m with a VMR of $-5.75^{+0.15}_{-0.16}$. H$_2$S is retrieved with a log VMR of $-4.51^{+0.30}_{-0.67}$ and contributes at around 3.5 $\mu$m. Na, K, and HCN are retrieved with log VMRs of $-8.97^{+1.97}_{-1.85}$, $-6.98^{+0.17}_{-0.19}$, and $-4.26^{+0.12}_{-0.12}$ respectively.

The retrieved cloud parameters are log(a) = $0.73^{+0.08}_{-0.08}$, $\gamma$ = $-0.71^{+0.09}_{-0.09}$, log(P$_{cloud}$) = $-0.52^{+1.28}_{-0.95}$, and $f_c$ = $0.99^{+0.0028}_{-0.01}$. The retrieved values suggest a largely homogeneous cloud coverage, with a high-altitude cloud deck located at 0.3 bar in terminator region. There is little evidence of Rayleigh enhancement from hazes, indicated by the absence of a steep slope at shorter wavelengths and the small value of log(a). However, such enhancements may become more prominent at shorter wavelengths when considering additional observations. The negative value of the scattering slope, $\gamma$, points to a scattering signature, suggesting the presence of small-grained haze particles with large scattering cross-sections. In later stages of our analysis, we identify these particles as Mie scattering aerosols. 

Indicated by this initial retrieval about the presence of haze particles, along with evidence provided by prior work on opacity contributions from non-grey clouds such as aerosols \citep{Constantinou, Constantinou_2023}, we perform all subsequent retrievals assuming opacity contributions due to Mie scattering aerosols, namely, ZnS and MgSiO$_3$. Therefore, we performed a free chemistry retrieval assuming these as the primary aerosol opacity contributors. We find that this retrieval achieves a better fit to the observations than the former in the lower wavelength regions, statistically with a lower reduced $\chi^2$ value of 2.99. The log VMR estimates of chemical species are presented in Table \ref{tab:species_abundances}. We obtain a log mixing ratio of $-11.55^{+5.30}_{-4.97}$ for MgSiO$_3$, with a corresponding modal particle size log$(r_c/\mu$m) = $-2.13^{+0.60}_{-0.51}$. ZnS is retrieved with a log mixing ratio of $-2.61^{+0.98}_{-1.27}$ and with a modal particle size of $-1.29^{+0.01}_{-0.01}$. The effective scale height factor was retrieved at a value of H$_c$ = 0.78, hinting at almost a vertically constant profile, covering 67\% of the terminator region.

The retrieved VMR profiles are shown in Figure \ref{fig:chemistry_vmr_part1}(a), and the P-T profile is shown in Figure \ref{fig:chemistry_vmr_part1}(b), which shows a gradient with temperatures increasing in the deep atmosphere. The overall temperatures lie between 950-1200 K.

The parameters retrieved using \texttt{Stacking Regressor} are shown in Table \ref{tab:species_abundances}. We obtained comparable results with \texttt{PyMultiNest}, thus validating the results of the machine learning retrieval.

\subsubsection{\textbf{\texttt{NEXOCHEM} Equilibrium Chemistry Retrievals}}

Equilibrium analysis in \citet{ahrer2023early} found a 1-100$\times$ solar metallicity and a sub-solar C/O ratio of $\leq$0.35 to explain the data. Our retrievals also obtained a similar C/O ratio of $0.23^{+0.02}_{-0.02}$ and a $54.95^{+3.93}_{-3.67}\times$ solar metallicity. \citet{Tsai2023} mentions that the equilibrium mixing ratio of SO$_2$ is less than $10^{-12}$ for 10$\times$ solar metallicity and less than $10^{-9}$ for as high as 100$\times$ solar metallicity, which is insufficient for any spectral features, contrary to what is seen in the JWST observations. \texttt{NEXOTRANS} retrievals also suggest the same conclusion while assuming chemical equilibrium. This is evident from the mixing ratios retrieved in our analysis using the \texttt{NEXOCHEM} equilibrium chemistry grids, as shown in Figure \ref{fig:chemistry_vmr_part1}(c), where SO$_2$ is seen to have a very low abundance in the photospheric region being probed. Due to this low abundance of SO$_2$, the retrieved model spectrum could not produce the spectral feature due to SO$_2$.

Based on these findings, it can be inferred that disequilibrium processes prevail in the atmosphere of WASP-39 b. Therefore, we assume more flexible chemistry methods, such as hybrid and equilibrium offset methods as discussed in Section \ref{subsubsection : chemistry}, in our next set of retrievals.

The molecular abundances are also retrieved using the machine learning model, and the results obtained are shown in Table \ref{tab:species_abundances}.

\begin{figure*} []
    \centering
    \begin{minipage}{0.49\textwidth}
        \centering
        \includegraphics[width=\textwidth]{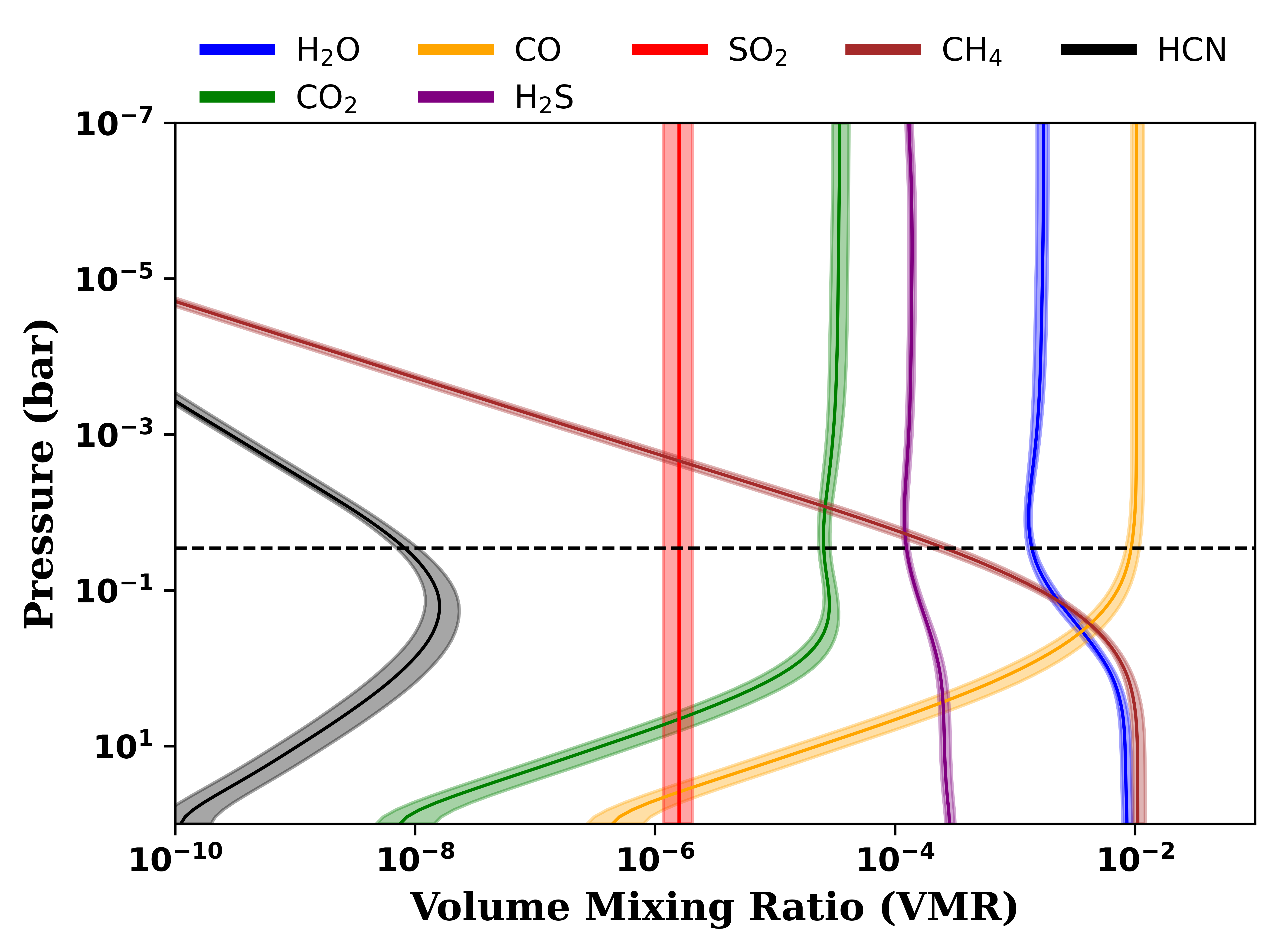}
        \parbox{0.8\linewidth}{\centering (a) Hybrid chemistry retrieved VMR.}

        \label{fig:vmr_hybrid}
    \end{minipage}
    \hfill
    \begin{minipage}{0.5\textwidth}
        \centering
        \vspace{0.6cm}
        \includegraphics[width=\textwidth]{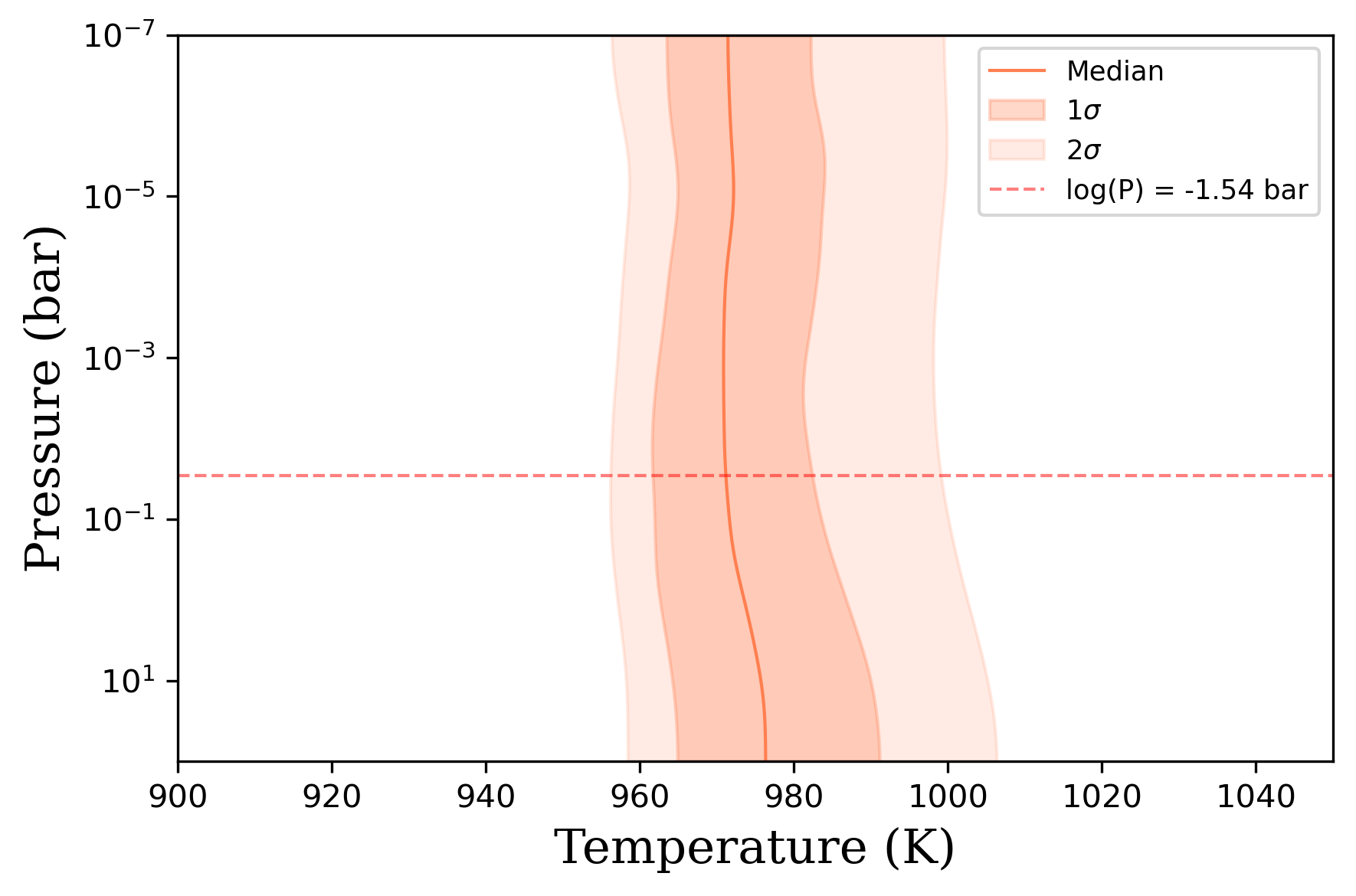}
        \parbox{0.8\linewidth}{\centering (b) Hybrid chemistry retrieved P-T profile. }
        \label{fig:pt_hybrid}
    \end{minipage}
    
    \vspace{0.5cm}
    
    \begin{minipage}{0.49\textwidth}
        \centering
        \includegraphics[width=\textwidth]{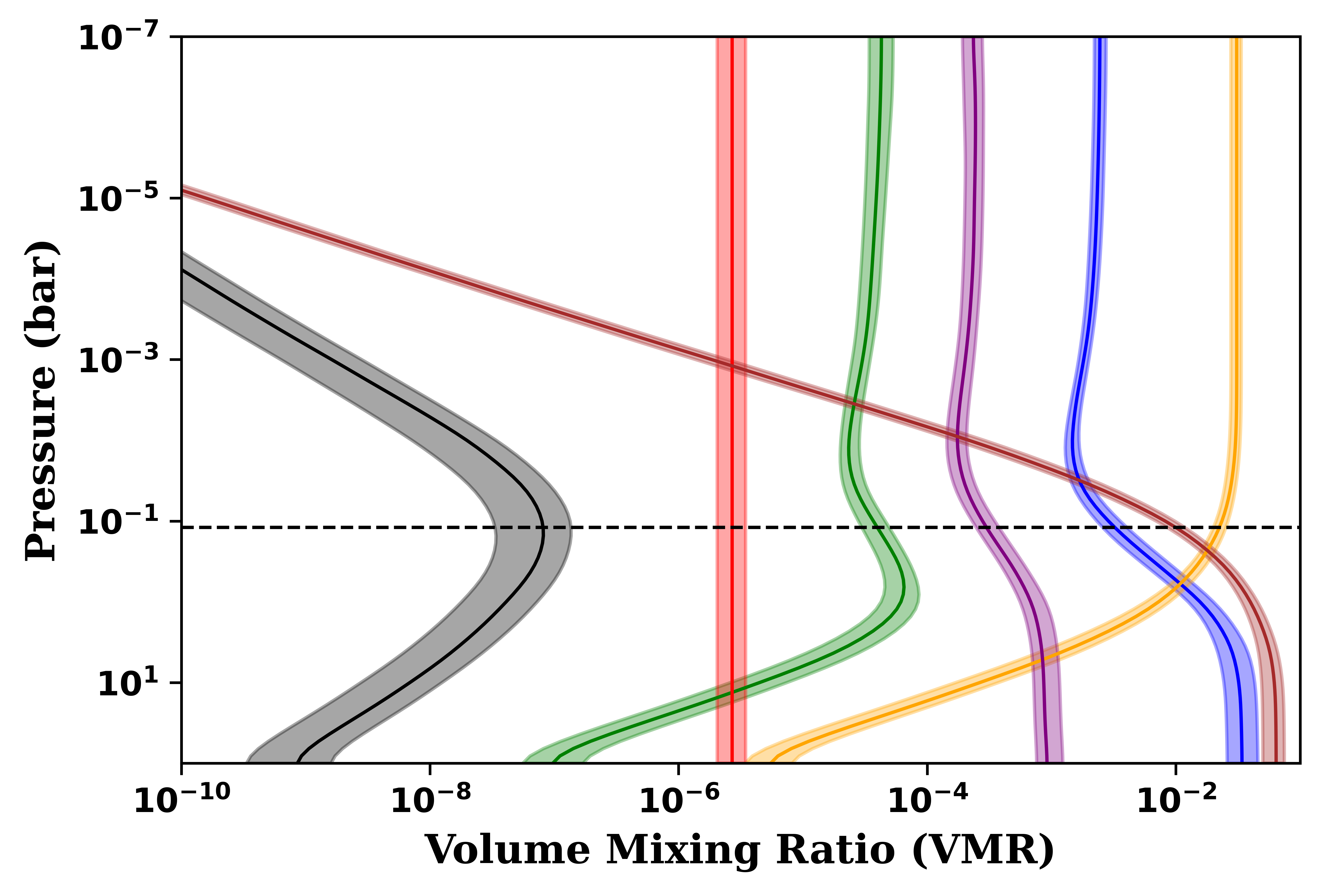}
        \parbox{0.8\linewidth}{\centering (c) Equilibrium offset chemistry retrieved VMR.  }
        \label{fig:vmr_offset}
    \end{minipage}
    \hfill
    \begin{minipage}{0.5\textwidth}
        \centering
        \includegraphics[width=\textwidth]{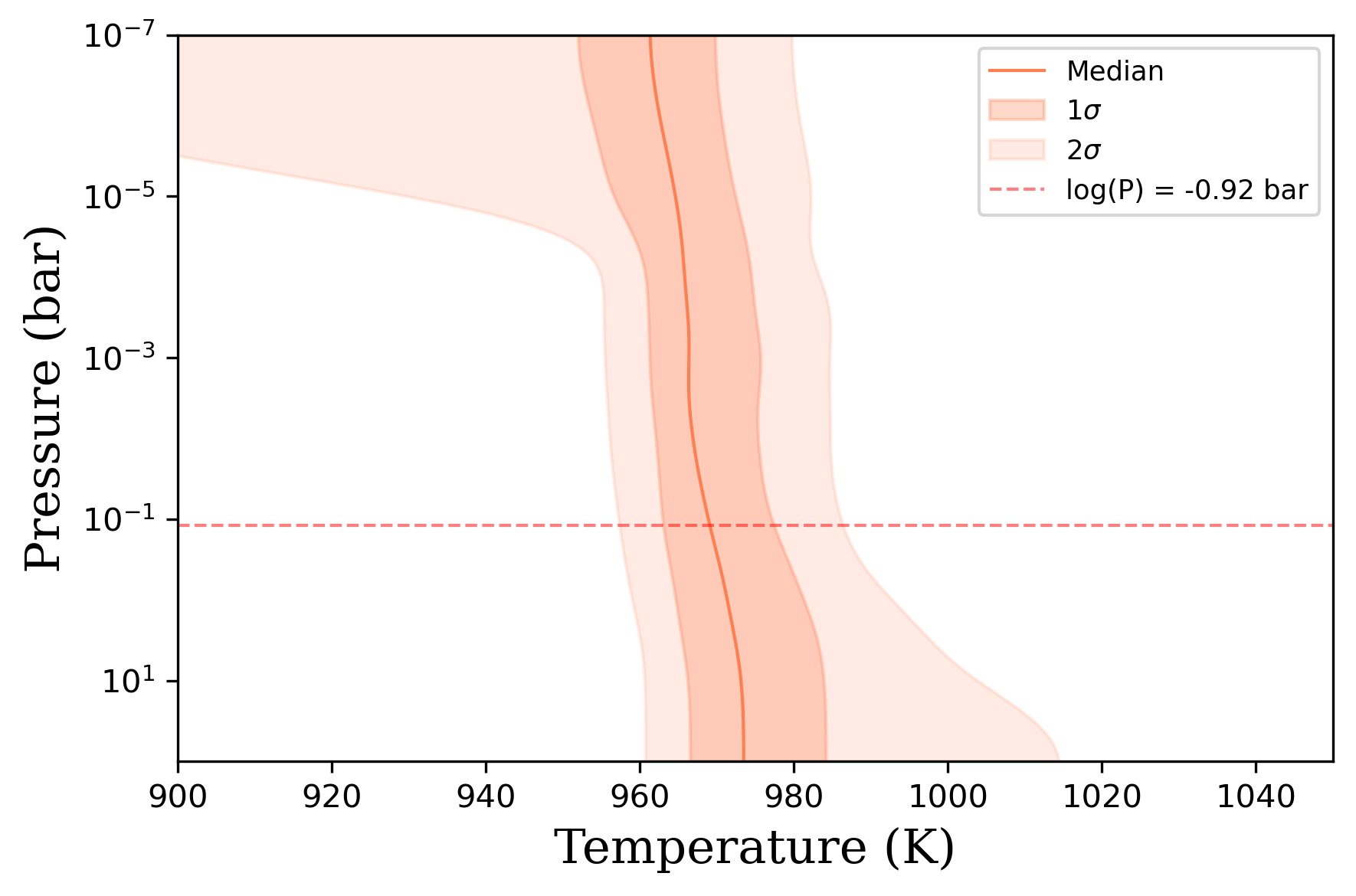}
        \parbox{0.8\linewidth}{\centering (d) Equilibrium offset chemistry retrieved P-T profile. }
        \label{fig:pt_offset}
    \end{minipage}
    
    \caption{Retrieved Volume Mixing Ratio Profiles for: a) Hybrid Equilibrium Chemistry and c) Equilibrium Offset Chemistry, along with the corresponding PT profiles retrieved. The dotted horizontal lines represent the retrieved median reference pressures for that model. The 1$\sigma$ region for the VMR profiles are also indicated by the light colored bands.}
    \label{chemistry_vmr_part2}
\label{fig:retrieved_vmr2}
\end{figure*}

\subsubsection{\textbf{Hybrid Equilibrium Chemistry Retrievals}} \label{subsection : hybd_result}

Here, we conduct the retrievals using the hybrid equilibrium method as outlined in Section \ref{subsubsection : chemistry}. In comparison to equilibrium chemistry retrievals, here the mixing ratios of SO$_2$, Na, and K are allowed to vary freely since these species don't participate in equilibrium chemistry calculations, with SO$_2$ being a photochemical product, allowing their volume mixing ratios (VMR) to remain constant with altitude. 

From this retrieval, we find a super-solar carbon-to-oxygen ratio of $0.80^{+0.03}_{-0.01}$, higher than the solar value of 0.59, which is consistent with previously published values. Additionally, we obtain a metallicity of approximately 15.49$\times$ solar. The retrieved vertical mixing ratio profiles are shown in Figure \ref{fig:retrieved_vmr2}(a), and the corner plot is shown in the Appendix, Figure \ref{fig:corner_hybrid_ultra}.

We note that the retrieved abundances for SO$_2$, Na, and K are: $-5.80^{+0.11}_{-0.13}$, $-7.97^{+0.44}_{-0.50}$, and $-8.74^{+0.15}_{-0.16}$, respectively, all within 1$\sigma$ intervals of previously published values \citep{Constantinou}. 

We also obtained constraints for the mixing ratio of ZnS and MgSiO$_3$ aerosols. Specifically, we retrieved a log-mixing ratio of $-2.54^{+0.74}_{-0.85}$, with a corresponding modal particle radius $\log(r_c/\mu \text{m}) = \-1.29^{+0.005}_{-0.005}$ for ZnS, and a log VMR of $-9.42^{+3.71}_{-4.87}$ and modal particle radius $\log(r_c/\mu \text{m}) = -1.74^{+0.40}_{-0.51}$ for MgSiO$_3$, an effective scale factor $H_c = 0.78^{+0.01}_{-0.01}$, and a terminator coverage fraction $f_c = 0.59^{+0.01}_{-0.01}$.

The retrieved P-T profile is
shown in Figure \ref{fig:retrieved_vmr2}(b)
with an overall temperature ranging between $\sim$ 960 K and 1000 K.

The retrievals, as shown in Table \ref{tab:species_abundances}, performed by the machine learning model are found to be consistent with the Bayesian retrieval.

\subsubsection{\textbf{Equilibrium Offset Chemistry Retrievals}}

Now, we discuss the more flexible method of equilibrium offset chemistry, as mentioned in Section \ref{subsubsection : chemistry}. This method allows the mixing ratios to deviate from equilibrium profiles using a multiplicative factor, which is allowed to vary freely to determine the best-fitting mixing ratio profiles, while preserving their overall shape. This approach can provide insights into processes that deviate from equilibrium conditions.

In this method, along with retrieving the C/O ratio and log(metallicity), the values of which are $0.89^{+0.01}_{-0.02}$ and $1.66^{+0.09}_{-0.10}$ respectively, we also retrieve an offset value that is multiplied with the mixing ratios coming out of \texttt{NEXOCHEM} equilibrium grids to fit the observations. We obtain equilibrium offsets of $1.38^{+0.30}_{-0.30}$, $0.94^{+0.22}_{-0.17}$, and $0.33^{+0.13}_{-0.09}$ for H$_2$O, CO, and CO$_2$, respectively. This shows that H$_2$O and CO are consistent with no significant offsets, whereas CO$_2$ is slightly depleted. The H$_2$S offset is also constrained at $1.23^{+0.32}_{-0.24}$, showing consistency without a greater offset. However, it should be noted that we obtained an enhanced CH$_4$ profile with a multiplicative offset of $1.73^{+0.18}_{-0.22}$ compared to what was seen  
\begin{sidewaystable}[]
\vspace{9cm}
\hspace{-1cm}
\rotatebox{0}{
    \begin{minipage}{\textwidth}
        \centering
        \caption{Retrieved elemental ratios at log(P) = -2 bar, under various chemistry models when assuming Mie scattering aerosols. The corresponding solar values \citep{asplund2021653} for the ratios are included in the last row for reference.}
        \label{tab:elemental_abundances}
        \begin{tabular}{lcccccccccc}
        \toprule
        \hline
        \multirow{2}{*}{} & \multicolumn{1}{c}{C/O} & \multicolumn{1}{c}{log[M/H]} & \multicolumn{1}{c}{log(O/H)} & \multicolumn{1}{c}{log(C/H)} & \multicolumn{1}{c}{log(S/H)} & \multicolumn{1}{c}{log(Na/H)} & \multicolumn{1}{c}{log(K/H)} & \multicolumn{1}{c}{log(S/O)} & \multicolumn{1}{c}{log(Na/O)} & \multicolumn{1}{c}{log(K/O)} \\
        \cmidrule{1-11} 
        & \multicolumn{10}{c}{\textit{\textbf{Free Chemistry}}} \\
        Bayesian & $0.89^{+0.34}_{-0.34}$ & $1.62^{+0.07}_{-0.06}$ & $-1.90^{+0.12}_{-0.11}$ & $-1.95^{+0.13}_{-0.12}$ & $-3.91^{+0.13}_{-0.13}$ & $-8.87^{+0.47}_{-0.47}$ & $-8.90^{+0.20}_{-0.21}$ & $-2.02^{+0.18}_{-0.17}$ & $-6.97^{+0.84}_{-0.78}$ & $-7.00^{+0.29}_{-0.25}$ \\
        [0.2cm]
        ML & $0.90^{+0.07}_{-0.17}$ & $1.37^{+0.25}_{-0.23}$ & $-1.68^{+0.24}_{-0.14}$ & $-1.72^{+0.26}_{-0.14}$ & $-3.62^{+0.07}_{-0.06}$ & $-6.74^{+0.02}_{-0.01}$ & $-8.24^{+0.02}_{-0.01}$ & $-1.94^{+0.18}_{-0.33}$ & $-5.06^{+0.15}_{-0.33}$ & $-6.56^{+0.15}_{-0.33}$ \\
        \hline
        
        & \multicolumn{10}{c}{\textit{\textbf{Equilibrium}}} \\
        Bayesian & $0.23^{+0.02}_{-0.02}$ & $1.74^{+0.03}_{-0.03}$ & $-1.61^{+0.05}_{-0.06}$ & $-2.21^{+0.07}_{-0.06}$ & $-3.29^{+0.06}_{-0.06}$ & $-4.10^{+0.06}_{-0.06}$ & $-5.59^{+0.05}_{-0.06}$ & $-1.61^{+0.07}_{-0.06}$ & $-2.42^{+0.07}_{-0.06}$ & $-3.91^{+0.07}_{-0.06}$ \\
        [0.2cm]
        ML & $0.28^{+0.01}_{-0.01}$ & $1.14^{+0.48}_{-0.22}$ & $-2.21^{+0.36}_{-0.15}$ & $-2.72^{+0.33}_{-0.18}$ & $-3.87^{+0.32}_{-0.30}$ & $-4.67^{+0.44}_{-0.18}$ & $-6.12^{+0.35}_{-0.20}$ & $-1.65^{+0.32}_{-0.70}$ & $-2.45^{+0.31}_{-0.83}$ & $-3.90^{+0.35}_{-0.71}$ \\ 
        [0.2cm]
        \hline
        
        & \multicolumn{10}{c}{\textit{\textbf{Hybrid Equilibrium}}} \\
        Bayesian & $0.80^{+0.03}_{-0.01}$ & $1.19^{+0.05}_{-0.04}$ & $-2.16^{+0.08}_{-0.06}$ & $-2.21^{+0.09}_{-0.07}$ & $-4.15^{+0.06}_{-0.05}$ & $-8.20^{+0.33}_{-0.36}$ & $-8.97^{+0.16}_{-0.17}$ & $-1.99^{+0.07}_{-0.09}$ & $-6.03^{+0.54}_{-0.47}$ & $-6.80^{+0.20}_{-0.19}$ \\
        [0.2cm]
        ML & $0.72^{+0.14}_{-0.13}$ & $1.18^{+0.17}_{-0.30}$ & $-2.31^{+0.16}_{-0.25}$ & $-2.31^{+0.16}_{-0.25}$ & $-4.04^{+0.05}_{-0.16}$ & $-6.43^{+0.16}_{-0.60}$ & $-8.48^{+0.01}_{-0.01}$ & $-1.72^{+0.39}_{-0.21}$ & $-4.11^{+0.26}_{-0.26}$ & $-6.16^{+0.27}_{-0.20}$ \\
        [0.2cm]
        \hline
        
        & \multicolumn{10}{c}{\textit{\textbf{Equilibrium Offset}}} \\
        Bayesian & $0.89^{+0.01}_{-0.02}$ & $1.66^{+0.09}_{-0.10}$ & $-1.72^{+0.06}_{-0.08}$ & $-1.73^{+0.06}_{-0.08}$ & $-3.99^{+0.10}_{-0.10}$ & $-7.44^{+0.30}_{-0.44}$ & $-8.58^{+0.17}_{-0.18}$ & $-2.28^{+0.12}_{-0.11}$ & $-5.72^{+0.72}_{-0.41}$ & $-6.86^{+0.22}_{-0.19}$ \\
        [0.2cm]
        ML & $0.75^{+0.01}_{-0.05}$ & $1.48^{+0.02}_{-0.38}$ & $-2.03^{+0.06}_{-0.30}$ & $-1.97^{+0.08}_{-0.27}$ & $-3.84^{+0.03}_{-0.20}$ & $-7.03^{+0.52}_{-0.01}$ & $-8.53^{+0.01}_{-0.01}$ & $-1.80^{+0.58}_{-0.07}$ & $-5.00^{+0.38}_{-1.00}$ & $-6.50^{+0.38}_{-0.06}$ \\
        & \multicolumn{10}{c}{\textit{\textbf{Solar Values}}} \\
        & $0.59$ &  & $-3.31$ & $-3.54$ & $-4.88$ & $-5.78$ & $-6.93$ & $-1.57$ & $-2.47$ & $-3.62$ \\
        \bottomrule
        \end{tabular}
    \end{minipage}
}
\end{sidewaystable}
\clearpage 

in previous studies \citep{Constantinou}. We retrieved the free chemistry mixing ratios for SO$_2$, Na, and K as $-5.57^{+0.11}_{-0.22}$, $-7.21^{+0.38}_{-0.68}$, and $-8.35^{+0.16}_{-0.18}$, respectively. These retrieved values lie within the ranges of those determined by previous studies \citep{Constantinou}.

The overall equilibrium offset chemical mixing ratio profiles are shown in Figure \ref{fig:retrieved_vmr2}(c), and the retrieved corner plot is shown in Figure \ref{fig:corner_eq_ultra}. The most contrasting difference in this retrieval is that of the aerosol parameters. Unlike the free-chemistry and hybrid equilibrium cases, we retrieve a lower log-mixing ratio for ZnS but a higher value for MgSiO$_3$. The retrieved parameters are log(ZnS) = $-14.58^{+2.73}_{-2.05}$, with a corresponding modal particle radius log($r_c/\mu$ m) = $-1.57^{+0.12}_{-0.13}$, and a log VMR of $-13.77^{+5.48}_{-4.10}$ with modal particle radius log($r_c/\mu$ m) = $-2.53^{+0.47}_{-0.27}$ for MgSiO$_3$. As shown in Figure \ref{fig:retrieved_vmr2}(d), we retrieved an almost identical P-T profile to that of the hybrid equilibrium case.

Chemical abundances under equilibrium-offset chemistry were also retrieved using machine learning algorithms. Table \ref{tab:species_abundances} summarizes the results obtained from the \texttt{Stacking Regressor}. As shown, the retrievals are consistent with those obtained via the Bayesian method, confirming the robustness of the ML approach. For this specific case, we also demonstrated how the parameters retrieved using \texttt{PyMultiNest} compare not only with the default ML model, \texttt{Stacking Regressor}, but also with the outcomes from individual base models (\texttt{Random Forest}, \texttt{Gradient Boosting}, and \texttt{k-Nearest Neighbors}), as presented in Table \ref{tab:ml-models-parameters} and discussed in Appendix \ref{subsec:base}.


\begin{figure*}[]
    \centering
    \hspace{-0.1cm}
    \includegraphics[width=0.858\linewidth]{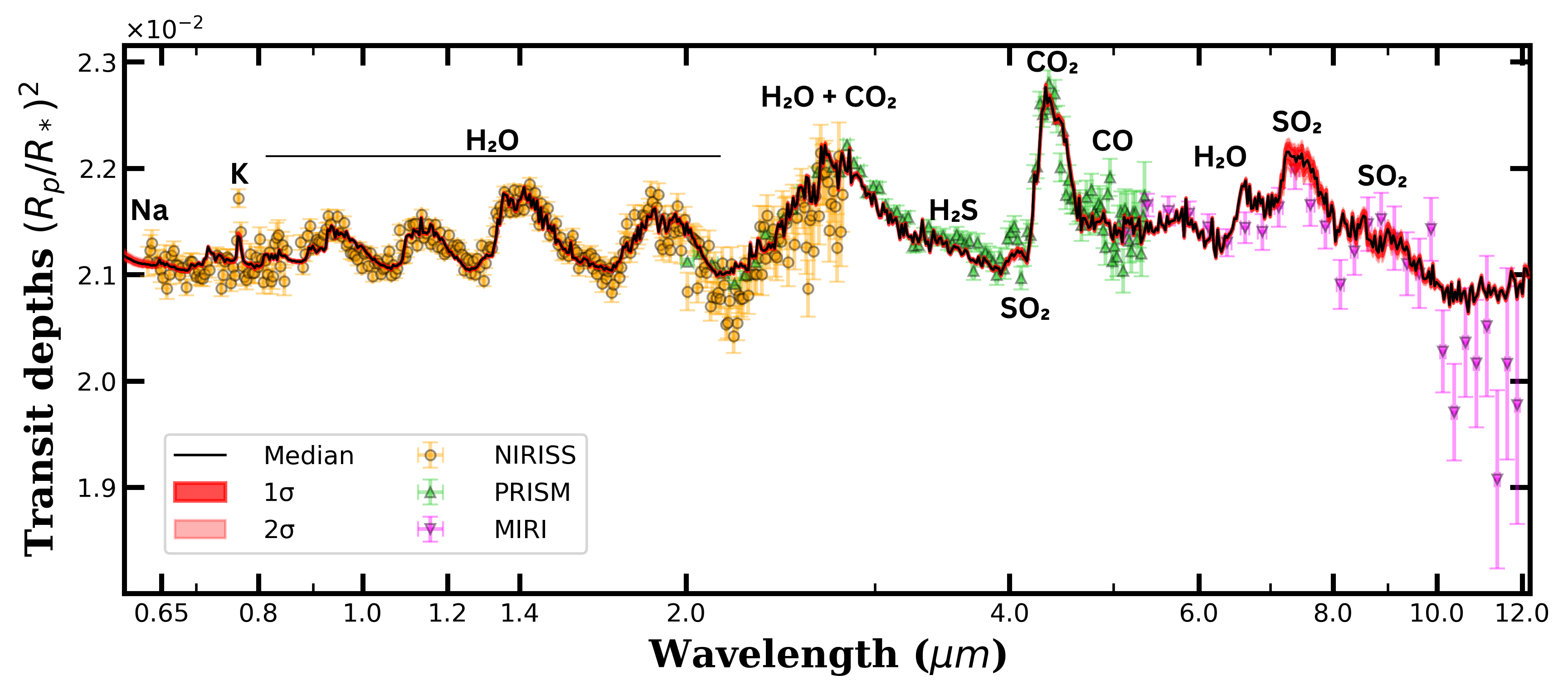}
    \parbox{0.8\linewidth}{\centering (a) Retrieved Spectrum of WASP-39 b using \texttt{PyMultiNest}.}
    \label{fig:hybrid_bayesian_spectra}
    \includegraphics[width=0.85\linewidth]{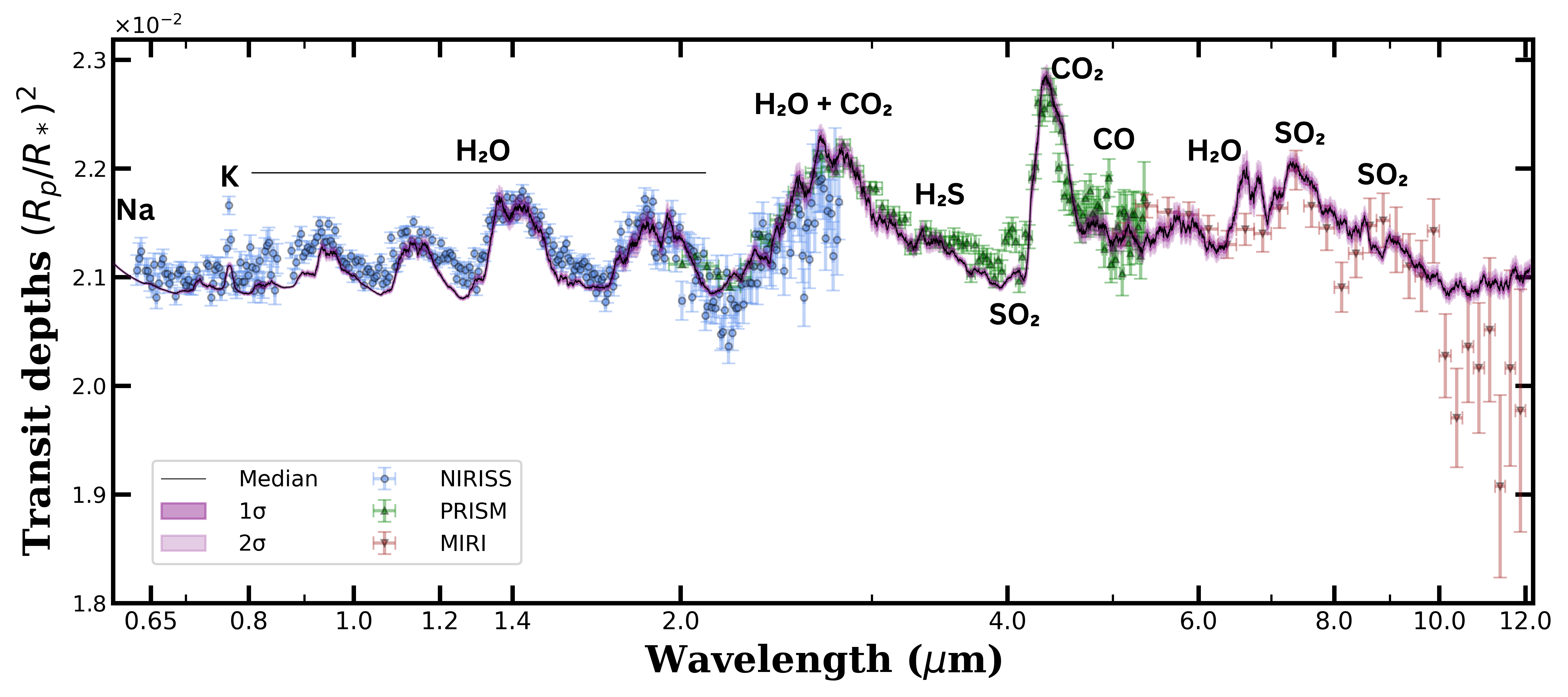}
    \parbox{0.8\linewidth}{\centering (b) Retrieved Spectrum of WASP-39 b using Machine learning (\texttt{Stacking Regressor}).}
    \label{fig:hybrid_ML_spectra}
    \caption{
        The retrieved spectra using the hybrid equilibrium chemistry model and machine learning approaches are shown. Observations from JWST instruments are illustrated with different colored error bars as indicated in the legend.
    }
    \label{fig: hybrid_retrieved_spec}
\end{figure*}
\subsubsection{\textbf{Retrieved Elemental Abundances}}
From the retrieved chemical abundances, the retrievals also inferred elemental ratios for all the chemistry models, as shown in Table \ref{tab:elemental_abundances}.

In the case of free chemistry and Mie scattering aerosol retrieval, we retrieve a super-solar C/O ratio of 0.89 compared to a solar value of 0.59 \citep{asplund2021653} .The retrieved log(O/H), log(C/H), and log(S/H) values are $-1.90^{+0.12}_{-0.11}$, $-1.95^{+0.13}_{-0.12}$, and $-3.91^{+0.13}_{-0.13}$, respectively. These values of O/H, C/H, and S/H correspond to 25.70, 38.90, and 9.33 $\times$ solar, respectively. This suggests a composition where both oxygen and carbon are readily available, likely in stable molecules such as CO, CO$_2$, and H$_2$O, with an atmospheric inventory slightly biased toward oxygen-rich compounds.

In the equilibrium chemistry retrievals, we obtained a high abundance of log(O/H) and log(S/H) with values of $-1.61^{+0.05}_{-0.06}$ and $-3.29^{+0.06}_{-0.06}$, respectively. These enrichment suggest a predominantly oxygen- and sulfur-rich atmosphere. The lower C/O ratio of $0.23^{+0.02}_{-0.02}$ suggests oxygen-dominated chemistry, potentially governed by the high abundance of water vapor as seen in Table \ref{tab:species_abundances}. This may also reflect the influence of high-temperature thermochemical processes, as indicated by the retrieved high-temperature P-T profile shown in Figure \ref{fig:chemistry_vmr_part1}(d). Na/H and K/H abundances are comparatively low, possibly due to their affinity for forming condensed species at lower altitudes, indicating minimal gas-phase presence in this model. However, caution should be exercised when implementing equilibrium chemistry assumptions, as SO$_2$ is observed to have very low abundance in the photospheric region, insufficient to produce the absorption features seen in the observations. This suggests the presence of significant disequilibrium processes in the atmosphere of WASP-39b.

The hybrid equilibrium model provides a balanced view between free chemistry and strict equilibrium, with an enhanced C/O ratio of $0.80^{+0.03}_{-0.01}$ compared to solar and elevated log(O/H), log(C/H), and log(S/H) values of $-2.16^{+0.08}_{-0.06}$, $-2.21^{+0.09}_{-0.07}$, and $-4.15^{+0.06}_{-0.05}$, respectively, with log(S/H) remaining lower than the equilibrium value. This model suggests a slight tendency toward oxygen dominance, although carbon availability remains sufficient to allow for a stable mixture of C- and O-bearing molecules. We obtained a log(S/O) ratio of $-1.99^{+0.07}_{-0.09}$, which is slightly lower than the solar value of $-1.57$ \citep{asplund2021653}, indicating that sulfur abundance is slightly lower than in the equilibrium model, potentially due to sequestration in refractory species like ZnS. The hybrid model reflects a well-mixed atmosphere where equilibrium and non-equilibrium processes jointly shape the atmospheric composition.

For the equilibrium offset retrieval, the C/O ratio approaching $0.89^{+0.01}_{-0.02}$ indicates a relative enhancement of carbon, potentially influenced by dynamic processes that disrupt strict equilibrium. This elevated C/O, alongside high log(O/H) value of $-1.72^{+0.06}_{-0.08}$, implies an atmosphere that is both O- and C-rich, allowing for the potential formation of diverse carbon-oxygen compounds. The super-solar value of S/H suggests that sulfur plays a prominent role, with sulfur-bearing species potentially being key atmospheric constituents, as evidenced by the presence of SO$_2$ and H$_2$S. We once again retrieved sub-solar Na and K abundance ratios, highlighting the limited presence of these alkali metals in the observable photosphere, which may be sequestered in refractories such as silicates or other compounds.

\begin{figure*}
    \centering
    \begin{minipage}{0.48\textwidth}
        \centering
        \includegraphics[width=\textwidth]{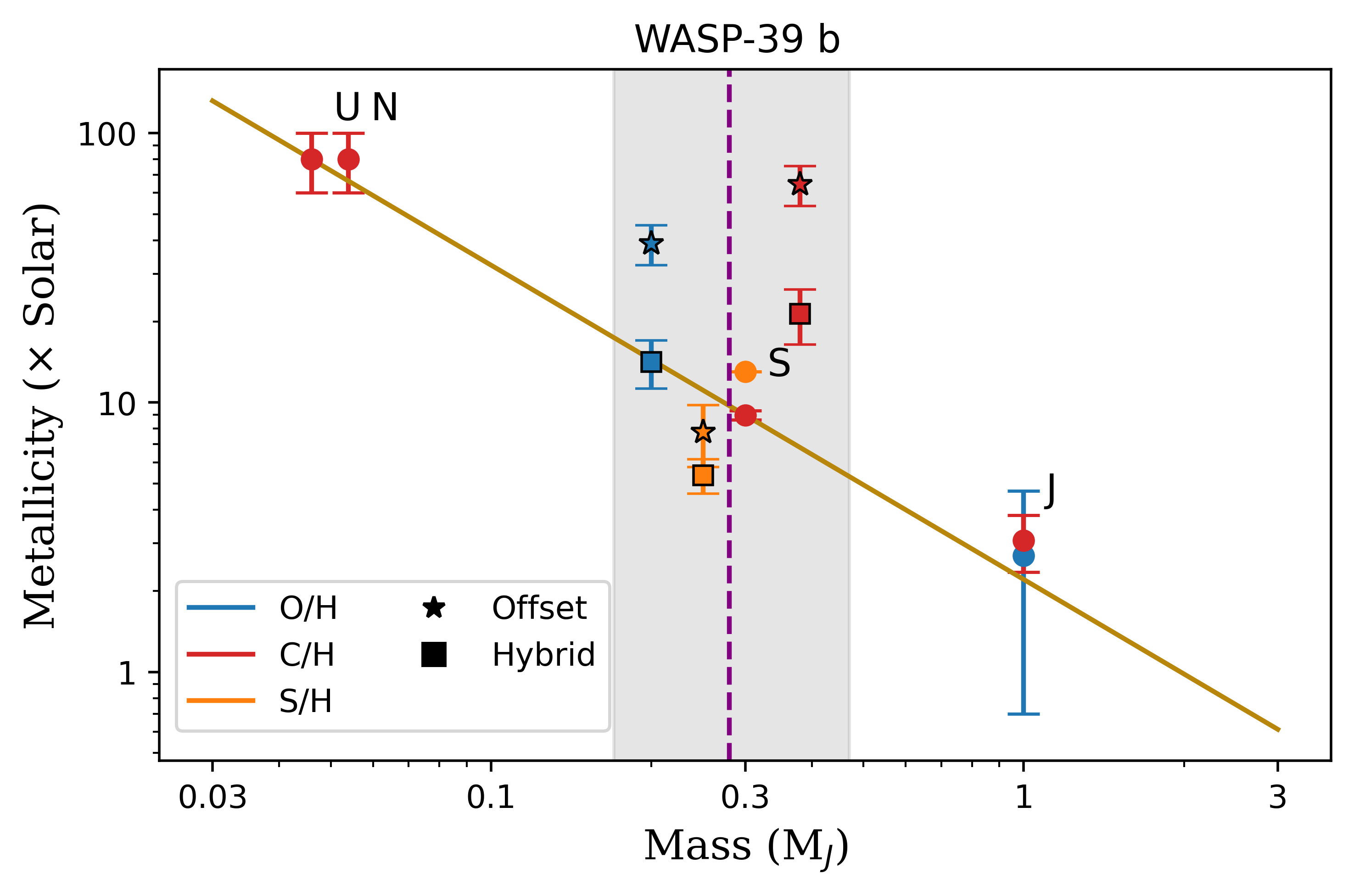}
        \parbox{0.8\linewidth}{\centering (a)}
        \label{fig:abun_Bayes}
    \end{minipage}
    \hfill
    \begin{minipage}{0.48\textwidth}
        \centering
        \includegraphics[width=\textwidth]{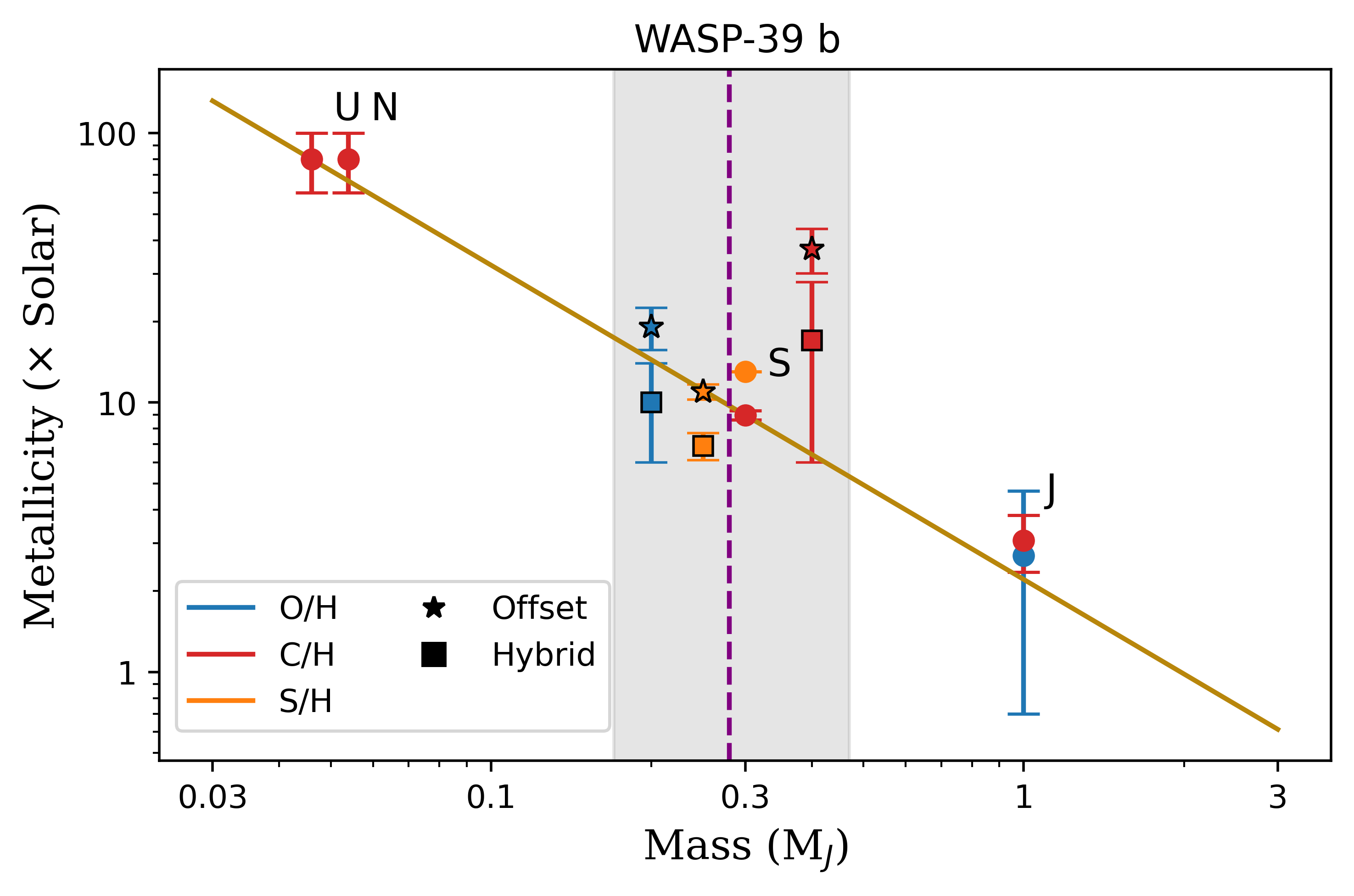}
        \parbox{0.8\linewidth}{\centering (b)}
        \label{abun_ML}
    \end{minipage}
    \caption{The retrieved elemental abundance ratios for oxygen (O), carbon (C), and sulfur (S) corresponding to the best-fit models are displayed. Figures (a) and (b) illustrate values retrieved using Bayesian inference and machine learning, respectively. Additionally, the mass-metallicity trend for the solar system's giant planets—Jupiter (J), Saturn (S), Uranus (U), and Neptune (N) (\cite{wong2004updated, fletcher2009methane})—is also shown, highlighting a clear linear relationship between planetary masses and elemental abundances. The vertical dotted line indicates the mass of WASP-39b (0.28 M$_J$) for reference. Note that the data points for WASP-39b within the shaded region have been slightly shifted horizontally to improve visual clarity.}
\label{fig:elem_abundance}
\end{figure*}

Figures \ref{fig:elem_abundance}(a) and \ref{fig:elem_abundance}(b) present the inferred abundances of O, C, and S for WASP-39b obtained with the hybrid and equilibrium offset retrievals, in comparison to those observed in giant planets within our solar system. Overall, the elemental abundances demonstrate an enhancement in metallicity compared to solar values, providing important insights into the formation and evolution of giant exoplanets \citep{Madhusudhan_2019}.

\begin{figure*}[]
    \centering
    \includegraphics[width=0.90\linewidth]{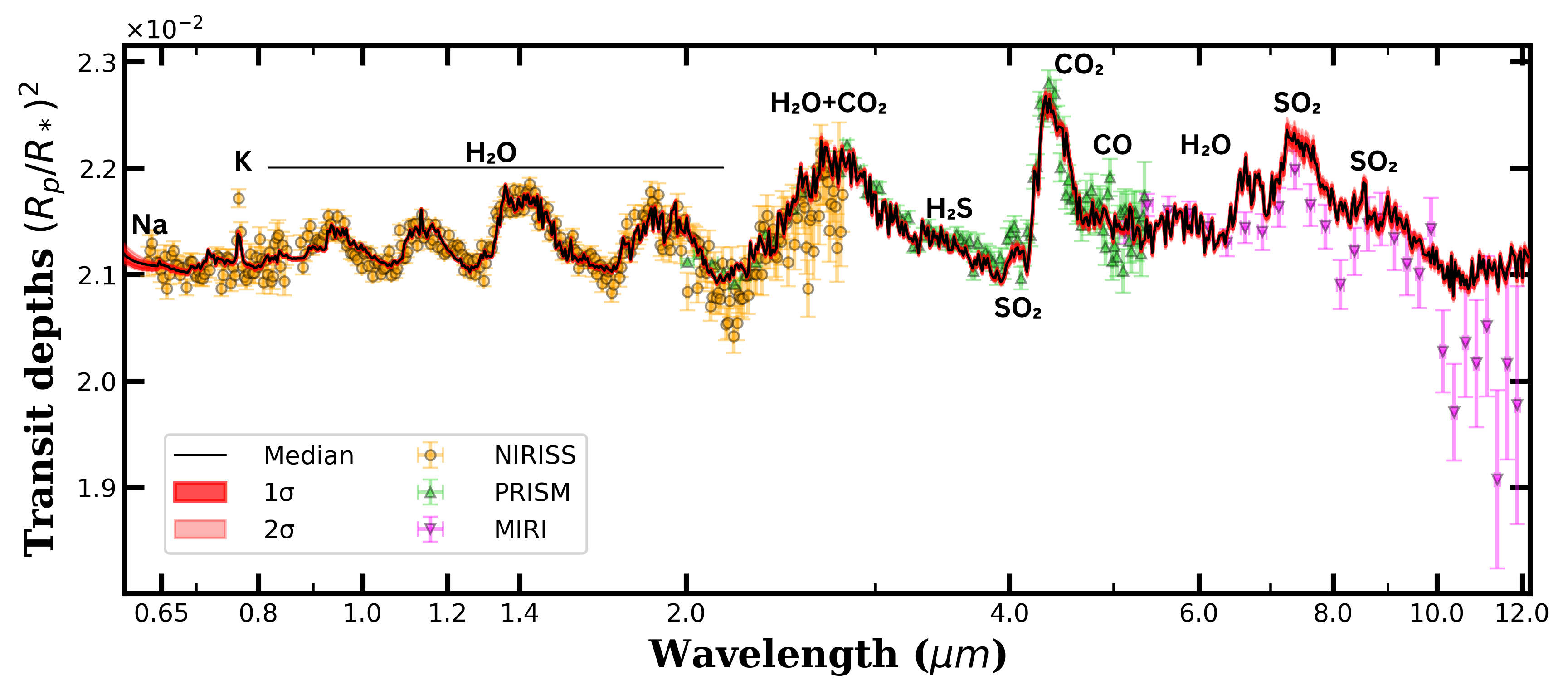}
    \parbox{0.8\linewidth}{\centering (a) Retrieved Spectrum of WASP-39 b using \texttt{PyMultiNest}.}
    \label{fig:chemoffset_bayesian_spectra}
    \includegraphics[width=0.89\linewidth]{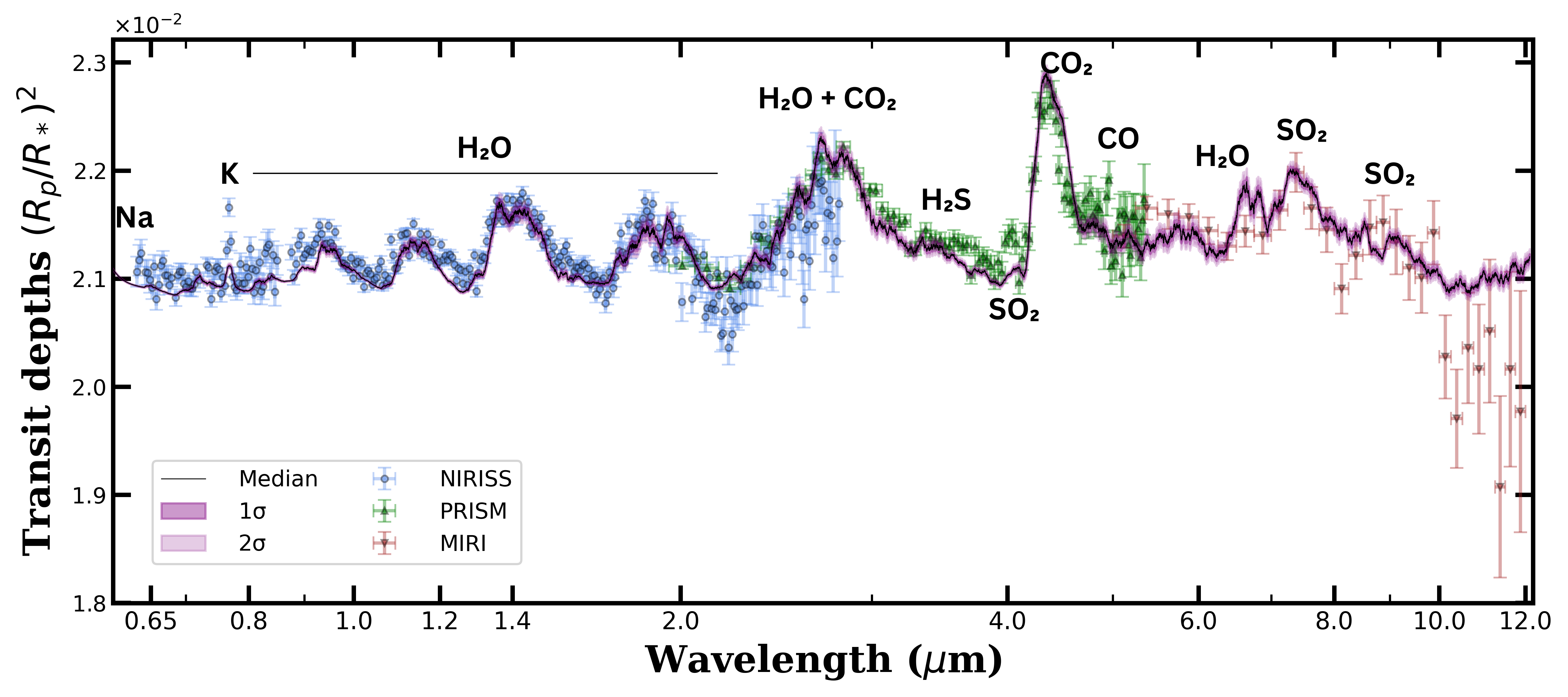}
    \parbox{0.8\linewidth}{\centering (b) Retrieved Spectrum of WASP-39 b using Machine learning (\texttt{Stacking Regressor}).}
    \label{fig:ML_spectra}
    \caption{
        The retrieved spectrum using the equilibrium offset chemistry model is shown, with the black line representing the median fit. The orange contour marks the corresponding 1$\sigma$ uncertainty interval. Observations from JWST instruments are illustrated with different colored error bars as indicated in the legend.
    }
    \label{fig:offset_retrieved_spec}
\end{figure*}

\section{\textbf{Discussions}}

Retrievals conducted using both Bayesian inference and machine learning methodologies yielded robust constraints on the physical and chemical properties of WASP-39b’s atmosphere. Here, we discuss the statistical significance of the best-fit models and examine how these statistical metrics compare across the individual datasets.

\subsection{\textbf{The Best-fit Model}} \label{sec:dis}

In the following sections, we present a discussion of the models that best explain the combined spectroscopic data from JWST NIRISS, NIRSpec PRISM, and MIRI. We focus on the statistically favored Bayesian model, as well as the best-performing machine learning model, providing a comprehensive evaluation of their respective fit qualities.

\subsubsection{\textbf{Bayesian Model}}

Having performed all the retrievals, we conclude that the hybrid chemistry model provides the best statistical fit. To identify the best-fit model, we used the reduced chi-squared ($\chi^{2}$) criterion, a statistical measure that quantifies the goodness-of-fit by comparing observed data with model predictions, normalized by the number of data points and model parameters. The hybrid chemistry retrieval yielded a log Bayesian evidence (ln(Z)) value of 3148.86 with a reduced $\chi^{2}$ value of 2.97. The best-fit retrieved spectrum for the combined NIRISS, PRISM, and MIRI observations is shown in Figure \ref{fig: hybrid_retrieved_spec}(a), along with the obtained elemental abundances in Figure \ref{fig:elem_abundance}(a), compared with those of the Solar System's giant planets.

As discussed in Section \ref{subsection : hybd_result}, the results suggest that the hybrid approach, which combines equilibrium chemistry with free chemistry to account for disequilibrium processes, is well-suited to explain the observed spectrum. This finding underscores the importance of incorporating flexible chemistry profiles in exoplanetary retrievals.

However, since exoplanetary atmospheres are typically not in chemical equilibrium, the equilibrium offset model also emerges as a strong candidate when greater flexibility is required to approximate volume mixing ratios under disequilibrium conditions. For this model, we retrieved a log Bayesian evidence value of 3163.09 with a reduced $\chi^{2}$ value of 2.98, making it the second-best model statistically. The corresponding best-fit retrieved spectrum is shown in Figure \ref{fig:offset_retrieved_spec}(a), along with the obtained elemental abundances in Figure \ref{fig:elem_abundance}(a).

\subsubsection{\textbf{Machine Learning Model}}

We show the spectra retrieved from the hybrid and equilibrium offset chemistry models in Figure \ref{fig: hybrid_retrieved_spec}(b) and Figure \ref{fig:offset_retrieved_spec}(b), respectively, along with the obtained elemental abundances in Figure \ref{fig:elem_abundance}(b) using the \texttt{Stacking Regressor} algorithm, the default machine learning (ML) model in \texttt{NEXOTRANS}. For the equilibrium offset chemistry model, all the machine learning models were run as discussed in Section \ref{subsection : machine learning}. Table \ref{tab:ml-models-parameters} in Appendix \ref{subsec:base} provides a comparative analysis of all the models. 

We find that the \texttt{Stacking Regressor} fits the observational dataset better than the other models (\texttt{Random Forest}, \texttt{Gradient Boosting}, and \texttt{k-Nearest Neighbour}). The average $R^{2}$ score for the \texttt{Stacking Regressor} (0.76) is the highest, compared to the average $R^{2}$ scores of \texttt{Random Forest} (0.67), \texttt{Gradient Boosting} (0.52), and \texttt{k-Nearest Neighbour} (0.65). A higher $R^{2}$ score indicates a better model prediction. Further details about the $R^{2}$ score are discussed in Appendix \ref{comparision}.


The median estimates obtained from both Bayesian and Machine learning (ML) methodologies exhibit strong agreement. However, the posterior distributions derived using ML have more constrained uncertainty intervals, as shown in Figures \ref{fig:corner_eq_ml} and \ref{fig:corner_hybrid_ml}, compared to the broader dispersion observed in Bayesian-derived posteriors (shown in Figures \ref{fig:corner_eq_ultra} and \ref{fig:corner_hybrid_ultra}). The Bayesian method thoroughly explores the entire parameter space, whereas the ML-based perturbation method, as discussed in Section \ref{subsubsection:training}, primarily captures local sensitivity around the median of the predicted data. This localized approach limits its ability to account for global correlations, leading to the narrower uncertainty intervals observed in the ML-derived corner plots.


\subsection{\textbf{Assessment of Bayesian Model Fit Across the Full 0.6–12 $\mu$m Dataset}}

The retrieval analysis on the full 0.6–12 µm dataset results in a reduced $\chi^2$ of 2.97 for the best-fit hybrid equilibrium model, indicating a minute model-data mismatch when compared to the lower reduced $\chi^2$ value for the MIRI only retrieval. To assess whether this discrepancy is driven by a specific instrument or represents a broad wavelength-dependent effect, we examined the reduced $\chi^2$ values for individual datasets by performing retrievals with the global best-fit hybrid equilibrium model with aerosols (ZnS, MgSiO$_3$). The retrieval on MIRI (5–12 $\mu$m) yields reduced $\chi^2$ value of 2.14, suggesting a good fit, while NIRISS (0.6–2.8 $\mu$m) produces a higher reduced $\chi^2$ of 2.67. The largest deviation occurs in NIRSpec PRISM (2–5.3 $\mu$m, excluding saturated data below 2 $\mu$m), where the reduced $\chi^2$ is 3.19, indicating comparative underfitting and driving the overall higher global reduced $\chi^2$ value. The best-fit retrieved spectra for these individual cases are presented in Figure \ref{fig:individual_spectrum} in the appendix \ref{individual_section}. It is important to note here that the global best-fit retrieval model combining data from NIRISS, PRISM, and MIRI may not be equally compatible with PRISM, which might lead to increased reduced chi-squared values. This underscores the interplay between model complexity and wavelength coverage, suggesting that certain models may be more suited to specific spectral regions than others. Consequently, the joint fitting of multi-instrument data presents significant challenges and emphasizes the need for further model exploration.

\section{\textbf{Conclusion}}\label{sec:conc}

In this paper, we demonstrated the capabilities of a new exoplanet atmospheric retrieval framework \texttt{NEXOTRANS} by combining three datasets obtained with JWST, namely, NIRISS, NIRSpec PRISM, and MIRI, which span a vast wavelength range from 0.6 to 12.0 $\mu$m. The availability of data for such wide wavelength coverage provided a unique and much better view into the atmospheric properties and processes of WASP-39 b in comparison to HST observations or individual JWST instruments. The key findings and conclusions from the retrievals performed are summarized below:

\begin{enumerate}
    \renewcommand{\labelenumi}{\textit{\roman{enumi})}}
    
    \item Retrievals on the combined transmission spectra helped us constrain significant spectral contributions from H$_2$O, CO$_2$, CO, H$_2$S, and SO$_2$ across the entire 0.6 - 12 $\mu$m wavelength range. These findings are consistent with previously retrieved values, with only minor differences due to the extended wavelength range.

    \item Taking into account the model with the least reduced $\chi^2$ i.e., the modified hybrid equilibrium with $\chi^2_{red}$ = 2.97, which represents the best fit, we obtained the volume mixing ratios (VMRs) for the major O-, C-, and S-bearing molecules. We retrieved a supersolar C/O ratio of $0.80^{+0.03}_{-0.01}$. Additionally, we obtained supersolar abundances for O/H, C/H, and S/H, corresponding to log values of $-2.16^{+0.08}_{-0.06}$, $-2.21^{+0.09}_{-0.07}$, and $-4.15^{+0.06}_{-0.05}$, respectively. These elemental abundances represent enhancements of $14.12^{+2.86}_{-1.82} \times$ solar, $21.37^{+4.93}_{-3.18} \times$ solar  and $5.37^{+0.79}_{-0.65}$ $\times$ the solar values.

    \item Given the evidence of chemical disequilibrium processes in the atmosphere of WASP-39 b, we also adopted a flexible chemical framework known as the modified equilibrium offset chemistry model. In this approach, the equilibrium mixing ratio profiles from the hybrid equilibrium model are allowed to deviate via an offset factor, providing greater flexibility. Within this retrieval configuration, the best-fit volume mixing ratio (VMR) profiles indicate a slight depletion of CO$_2$ with an offset of 0.33, an enhancement of CH$_4$ with a multiplicative offset of 1.73, and no significant offsets for H$_2$O, CO, or H$_2$S. This model yielded a reduced $\chi^2$ value of 2.98, representing the second-least value among the tested configurations.

    \item The Patchy Cloud Deck and Haze Model inferred the presence of small-grained haze particles with large scattering cross sections. This enabled the inclusion of contributions from Mie scattering aerosols in the transmission spectrum modeling, which also helped us constrain the properties of high-altitude aerosols such as ZnS and MgSiO$_3$, as suggested previously by \citet{Constantinou, Constantinou_2023}.

    \item The retrieval performed by assuming equilibrium chemistry alone obtains the highest reduced $\chi^2$, with a value of 3.35. The retrieved SO$_2$ abundance in the photospheric region is not sufficient to show the minute spectral absorption feature observed in the JWST data. This confirms the presence of disequilibrium processes, such as photochemistry, in the atmosphere of WASP-39 b.

   \item The retrieved temperature profiles show variations among the assumed chemical models. We obtained a gradient temperature profile in the case of the free-chemistry retrievals, ranging between $\sim$950 and 1200 K. For both the hybrid and equilibrium offset retrievals, the obtained profiles are broadly consistent with each other with temperatures between 900-950 K.
   The equilibrium retrieval comparatively obtains a much larger temperature range, between $\sim$960 and 1000 K.
 
   \item The inclusion of MIRI data in the 5.0 - 12.0 $\mu$m wavelength range enabled the most robust constraints on SO$_2$. We recover a range of median abundances for log(SO$_2$) between -6.25 and -5.73 for the different chemical models in the Bayesian retrievals, excluding the equilibrium chemistry case.

   \item Retrievals on individual datasets from NIRISS, NIRSpec PRISM, and MIRI revealed that the goodness of fit of a model depends on the wavelength region being analyzed. Using the best-fit hybrid equilibrium chemistry model with aerosols, we obtained reduced $\chi^2$ values of 2.67, 3.19, and 2.14 for NIRISS, NIRSpec PRISM, and MIRI, respectively. In contrast, for the combined dataset, we obtained a value of 2.97. This suggests potential challenges in jointly fitting multi-instrument data with a particular model.


    \item We also demonstrated the robustness of our machine learning-based retrieval method, which employs a supervised ensemble learning approach utilizing a Stacking Regressor model. This method leverages the strengths of three distinct algorithms--Random Forest, Gradient Boosting, and k-Nearest Neighbors, effectively combining their predictive capabilities. All these algorithms provide reasonably close constraints individually on the atmospheric parameters; however, the Stacking Regressor showed the best R$^2$ value. The machine learning retrievals also showed consistency with the results obtained using Bayesian methods, demonstrating possibilities for performing comparative exoplanetology in computationally efficient ways.

\end{enumerate}


\begin{acknowledgments}
L.M. acknowledges financial support from DAE and DST-SERB research grant (MTR/2021/000864) from the Government of India for this work. T.D. thanks Mr. Aniket Nath of NISER for his contributions and assistance with Bayesian analysis in the early stages of the \texttt{NEXOTRANS} project. D.D. extends gratitude to Mr. Rahul Arora of NISER for sharing his experience with equilibrium chemistry code development, which has been highly beneficial for the \texttt{NEXOCHEM} project. We would like to thank the anonymous
referee for constructive comments that helped improve the manuscript.

\end{acknowledgments}

\vspace{5mm}
\facilities{JWST}


\software{Python \citep{10.5555/1593511}, numba \citep{lam2015numba}, matplotlib \citep{Hunter:2007}, mpi4py \citep{dalcin2005mpi, dalcin2008mpi, dalcin2019mpi, dalcin2021mpi4py, rogowski2022mpi4py}, Sci-kit learn \citep{scikit-learn}.}
%

\clearpage
\section{\textbf{APPENDIX}}\label{sec:apdx}

\subsection{\textbf{BUILT-IN EQUILIBRIUM CHEMISTRY SOLVER: \texttt{NEXOCHEM}}} \label{sec: nexochem_exp} 
 
\texttt{NEXOTRANS} features a built-in Python-based equilibrium chemistry solver called \texttt{NEXOCHEM} (Next-generation EXOplanet equilibrium CHEmistry Model), which is both computationally efficient and fast. The code is based on the methodologies described by \citet{white1958chemical} and \citet{Eriksson_Holm_Welch_Prelesnik_Zupancic_Ehrenberg_1971}. It employs Gibbs free energy minimization using an iterative Lagrangian optimization method with a Lambda correction algorithm. Additionally, we have implemented parallelization using the Message Passing Interface (MPI) \citep{article} and utilized the \texttt{Numba-Jit} \citep{lam2015numba} compiler to further enhance the code's performance. Our code has also been benchmarked against \texttt{FastChem} \citep{stock(2018)}. 

\begin{figure*}[htbp]
    \centering
    \includegraphics[width=1\linewidth]{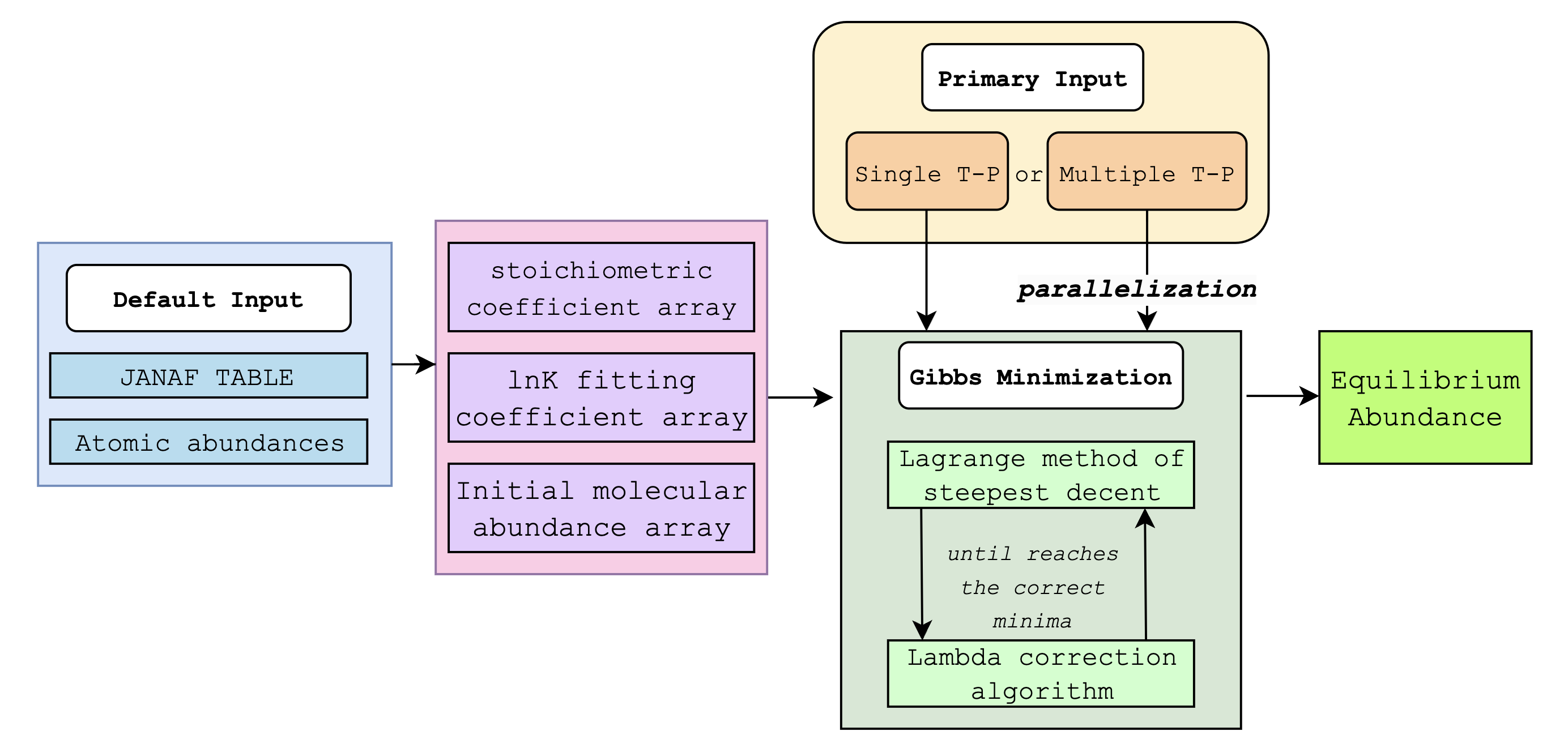}
    \caption{Schematic representation of the computational workflow of \texttt{NEXOCHEM}. 
    The architecture comprises three main components: 
    (1)~Default inputs module, which generates thermodynamic parameters and abundances of chemical species; 
    (2)~Primary input module, responsible for pressure-temperature profile acquisition; 
    (3)~Solver module, which performs free energy minimization of the system. 
    This integrated approach ensures robust chemical equilibrium calculations for exoplanetary atmospheric conditions.}
    \label{nexochem_flowchart}
\end{figure*}

\subsubsection{\textbf{Method}}

The calculation of equilibrium abundances of chemical species can be achieved by performing Gibbs free energy ($G$) minimization on the system at a given pressure-temperature ($P$-$T$) condition to determine the equilibrium abundances. In Figure \ref{nexochem_flowchart}, we illustrate the Gibbs minimization approach followed in \texttt{NEXOCHEM}.

We have adopted the expressions from \citet{Eriksson_Holm_Welch_Prelesnik_Zupancic_Ehrenberg_1971}:
\begin{equation}
\frac{G_{sys(T)}}{RT} = \sum_{i=1}^{n}x_i\left[\frac{g_i^0(T)}{RT} + \ln P + \ln\frac{x_i}{N}\right] 
\label{gibbs_energy}
\end{equation}

\begin{equation}
\frac{g_i^0(T)}{RT} = \frac{1}{R}\left[\frac{G_i^0 - H_{298}^0}{T}\right] + \frac{\Delta_f H_{298}^0 \cdot 1000}{RT}
\label{standard_gibbs}
\end{equation}

In this calculation, we consider only a gaseous atmosphere and neglect the condensation term. Here, $G_{sys(T)}$ is the total Gibbs free energy of the system, $x_i$ denotes the number of moles of $x$ of the $i^{th}$species, and $g_i^0(T)$ represents the chemical potential at standard state, given in J/mol. The gas constant is $R = 8.3144621$ J/K/mol, $H_{298}^0$ is the enthalpy in the thermodynamic standard state at a reference temperature of 298.15 K, and $G_i^0$ is the Gibbs free energy in J/mol. The term $(G_i^0 - H_{298}^0)/T$ is the free energy function in J/K/mol, and $\Delta_f H_{298}^0$ is the heat of formation at 298.15 K in kJ/mol. The values of $(G_i^0 - H_{298}^0)/T$ and $H_{298}^0$ can be obtained from the sixth and fourth columns of the \texttt{NIST-JANAF} tables\footnote{\url{https://janaf.nist.gov}} \citep{10.1063/1.555666, alma99776233502341}. For temperatures not listed in the table, the free energy values are calculated using spline interpolation. 




To conserve the elemental abundances of each species, the mass balance equation must be satisfied at every layer of the atmosphere:

\begin{equation}
\sum_{i=1}^n a_{ij} x_i = b_j \quad (j = 1, 2, ..., m)   
\end{equation}

Here, the stoichiometric coefficient $a_{ij}$ represents the number of atoms of the element $j$ in the $i^{\text{th}}$ species (e.g., for H$_2$O, the stoichiometric coefficient of H is 2 and for O it is 1), and $b_j$ denotes the total number of moles of the $j^{\text{th}}$ element initially present in the mixture.

Finally, to minimize Equation \ref{gibbs_energy} at a specific pressure and temperature, we employ a technique that minimizes a multivariate function under a mass constraint using the Lagrangian steepest-descent method. This methodology, derived in \citet{white1958chemical}, is also adopted in the \texttt{TEA} code as described by \citet{Blecic_2016}. Our approach for minimization is identical to that used in the \texttt{TEA} code. Specifically, we apply the Lagrange method with Lagrange multipliers and use the Lambda correction algorithm to determine the equilibrium abundances of all species, as detailed by \citet{Blecic_2016}.


To save computational time, we fit $\frac{g_i^0(T)}{RT}$ for each chemical species using the following expression:  
\begin{equation}
\ln K = \frac{a_0}{T} + a_1 \ln T + a_2 + a_3 T + a_4 T^2, 
\label{lnk}
\end{equation}
where  
\begin{equation}
\frac{g_i^0(T)}{RT} = -\ln K.
\end{equation}  

\begin{figure*}
    \centering
    \begin{minipage}{0.49\textwidth}
        \centering
        \includegraphics[width=\textwidth]{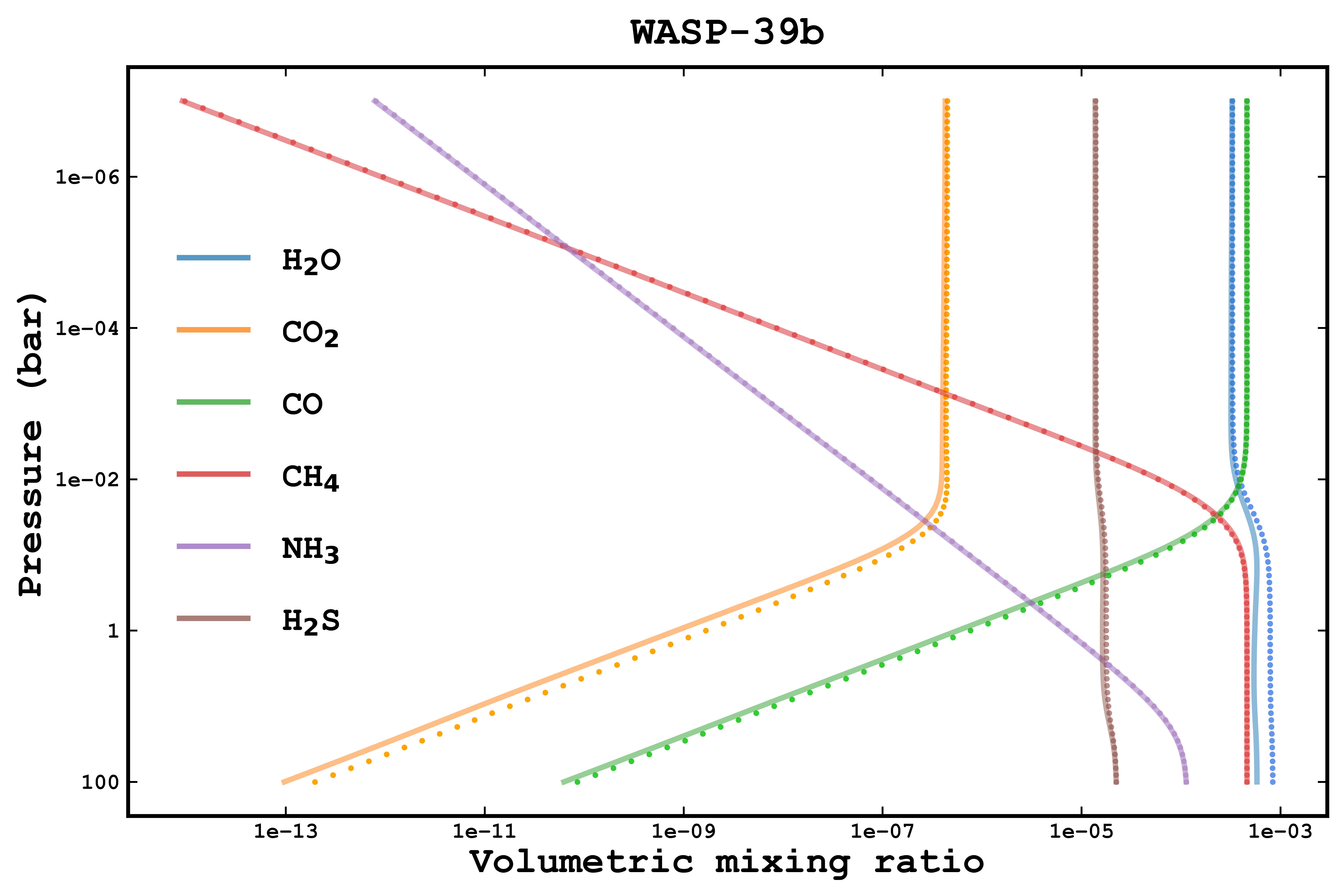}
        
    \end{minipage}
    \hfill
    \begin{minipage}{0.49\textwidth}
        \centering
        \includegraphics[width=\textwidth]{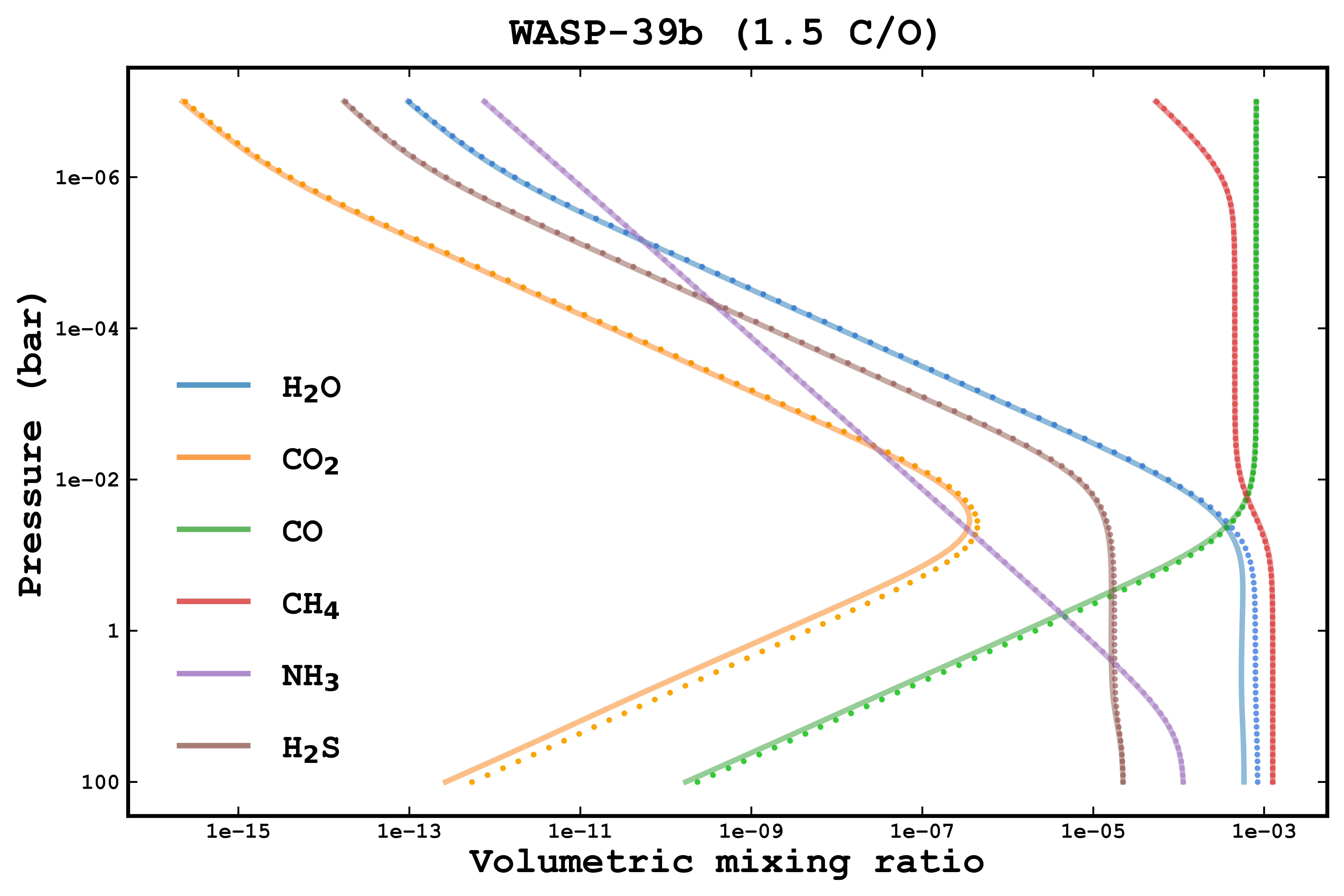}
    \end{minipage}
    \vspace{0.5cm}
    \begin{minipage}{0.49\textwidth}
        \centering
        \includegraphics[width=\textwidth]{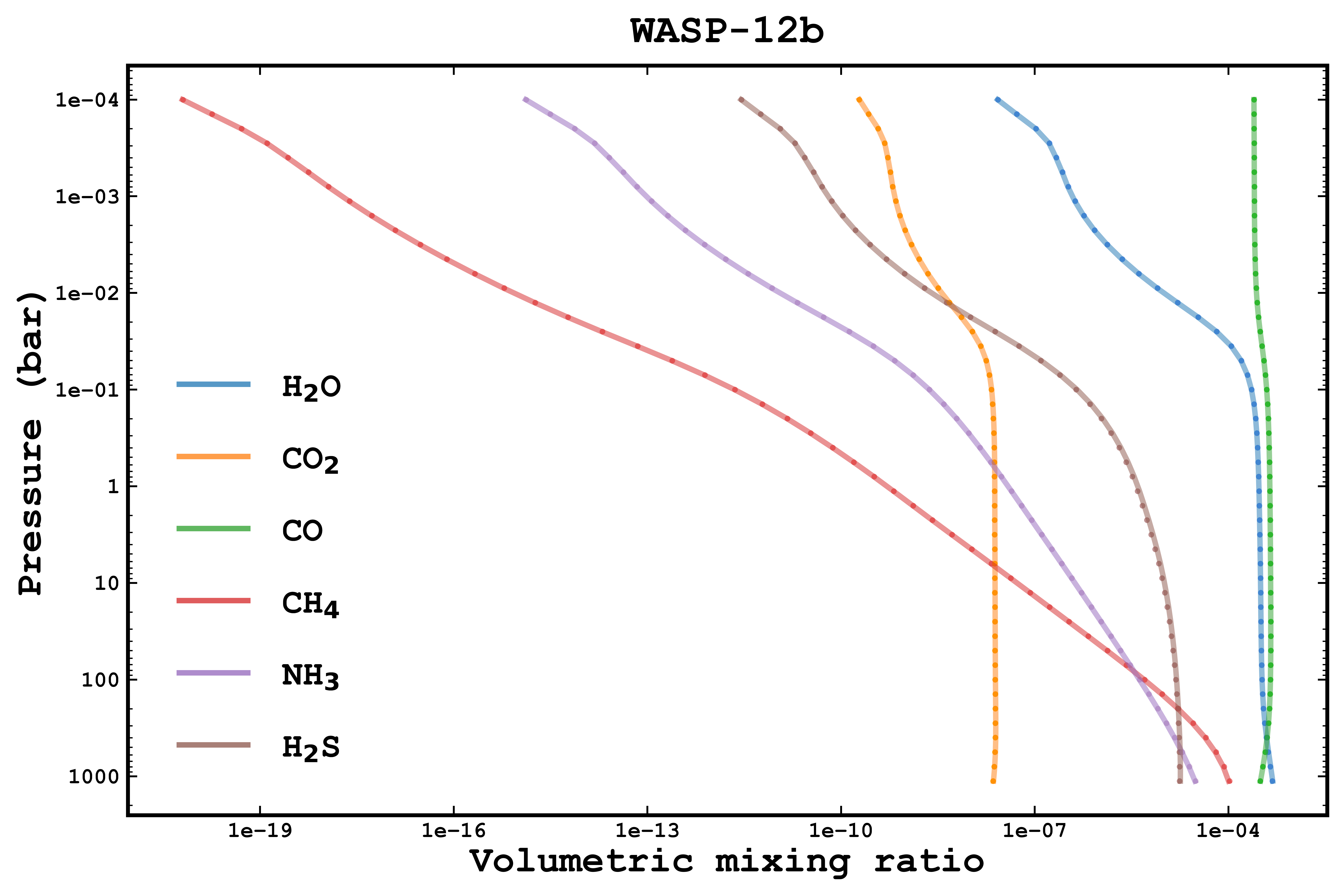}
    \end{minipage}
    \hfill
    \begin{minipage}{0.49\textwidth}
        \centering
        \includegraphics[width=\textwidth]{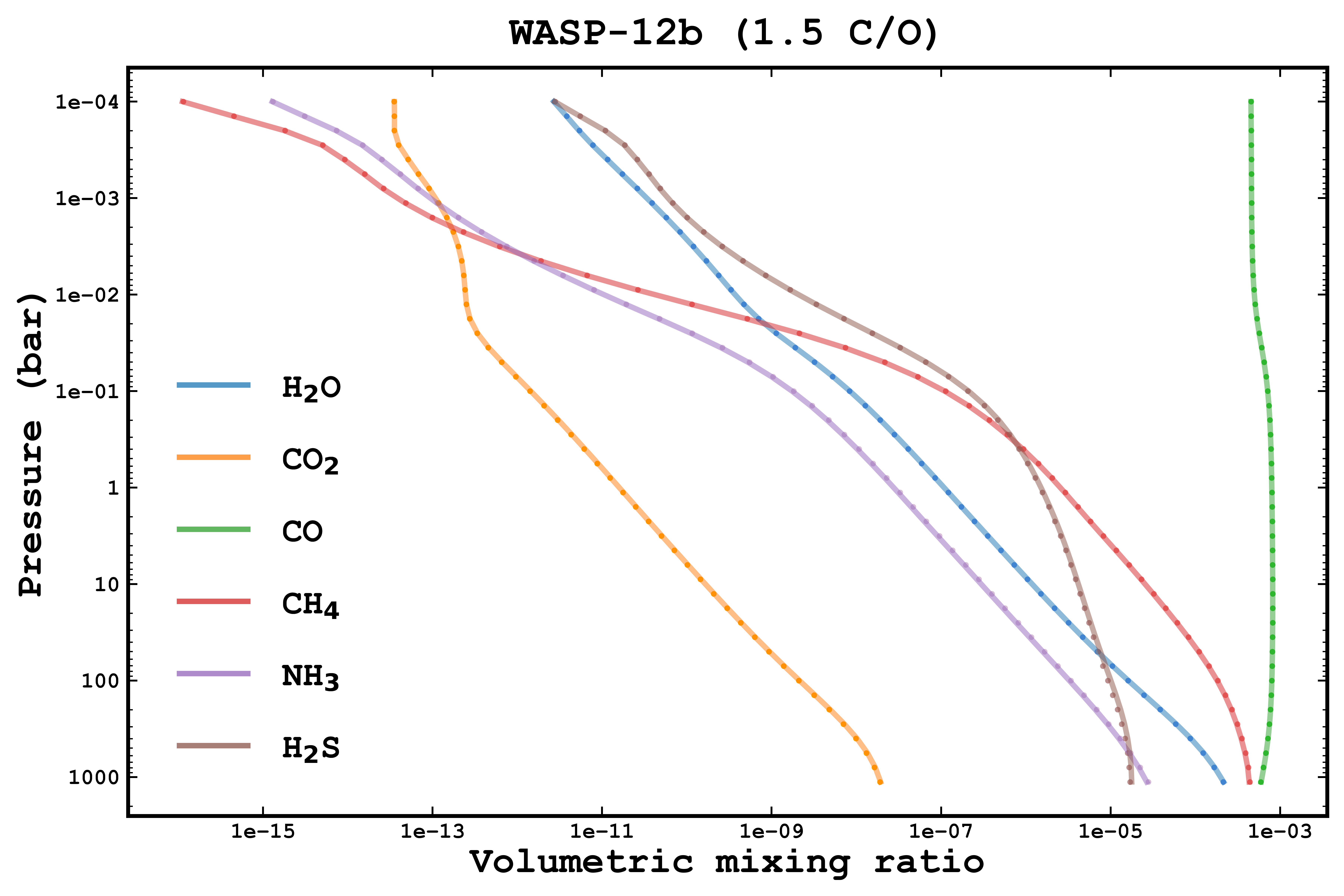}
    \end{minipage}
    
    \caption{Comparison between \texttt{NEXOCHEM} (dotted points) and \texttt{FastChem} (solid lines) for Wasp-39 b (P-T profile is taken from \ref{fig:pt_offset} ) and Wasp-12 b (P-T profile is taken from \cite{stevenson2014deciphering}) with different C/O values}
    \label{nexochem_benchmark}
\end{figure*}

Here, $K$ denotes the equilibrium constant, and $a_i$ are the fitting coefficients. A similar approach has been adopted by \citet{stock(2018)} and \citet{woitke2018equilibrium}.

\subsubsection{\textbf{Grid Generation and Validation}}\label{nexochem_grid}


To incorporate the effects of equilibrium chemistry at each layer of the atmosphere and to initialize our radiative transfer calculations across a broad parameter space, we generated a grid for the atmosphere using both the \texttt{FastChem} \citep{stock(2018)} and \texttt{NEXOCHEM} thermochemical equilibrium codes. We benchmarked the results of our \texttt{NEXOCHEM} grid against those of \texttt{FastChem}. Figure \ref{nexochem_benchmark} presents the comparison plots for the planets WASP-39 b and WASP-12 b. Our grid consists of more than 500,000 data points, spanning a parameter space that includes temperature, pressure, metallicity, and C/O ratios. Specifically, the tested ranges are as follows: 
$T = 300$--$4000$ K, $P = 10^{-7}$--$10^{2}$ bar, $\mathrm{C/O} = 0.2$--$2$, and $[\mathrm{Fe/H}] = 10^{-1}$--$10^{3}$ times solar \citep{Asplund2009}. 
We found excellent agreement between the mixing ratios of all the considered species in the \texttt{NEXOCHEM} and \texttt{FastChem} grids, as shown in Figure \ref{nexochem_benchmark}.

\subsection{\textbf{BASE MODELS USED IN MACHINE LEARNING MODEL}}\label{subsec:base}

\subsubsection{\textbf{Random Forest Regressor}}

\texttt{Random Forest Regressor} (RF) works by creating multiple decision trees during training and combining their predictions to produce a final result, which is the average prediction of its trees \citep{liaw2002classification}.

Given the task of predicting a parameter \( y \) from a set of features \( X \), RF builds several regression trees \( h_b(X) \), where \( b \in \{1, 2, \dots, B\} \), and averages their outputs, yielding the final prediction \( \hat{y} \):

\begin{equation}
    \hat{y} = \frac{1}{B} \sum_{b=1}^{B} h_b(X).
    \label{fg}
\end{equation}

The RF loss function can be optimized by reducing the mean squared error (MSE) between the predicted value and the true value. The MSE of each tree is given by the following equation:

\begin{equation}
\label{qwe}
    \text{MSE} = \frac{1}{n} \sum_{i=1}^{n} (y_i - \hat{y_i})^2,
\end{equation}

where \( y_i \) is the true value, and \( \hat{y_i} \) is the predicted value. \\

To reduce overfitting, RF introduces randomness by selecting a random subset of features through bootstrapping, where each tree is trained on a bootstrapped dataset. This randomness leads to generalization of the RF model \citep{breiman2001random}.

\subsubsection{\textbf{k-Nearest Neighbors Regressor}}

The \texttt{k-Nearest-Neighbor} (kNN) regressor is a non-parametric technique that makes predictions based on interactions in the local neighborhood of a dataset \citep{altman1992introduction}. Unlike parametric models, \texttt{k-Nearest-Neighbor} avoids assumptions about the data distribution and instead follows the principle that similar inputs should produce similar results \citep{peterson2009k}.

To predict the value for a new data point \( x \), the \texttt{k-Nearest-Neighbor} method (kNN) uses a distance measure (usually Euclidean distance) to find the nearest \( k \) points. The formula for calculating the distance between \( x \) and any point \( x_i \) is given as:

\begin{equation}
d(x, x_i) = \sqrt{\sum_{j=1}^p (x_j - x_{ij})^2},
\label{eee}
\end{equation}
where \( x_i \) is any training data point, and \( p \) represents the number of features in the dataset.

The target prediction \( \hat{y} \) of \( x \) is then calculated as the weighted average of the target values of the \( k \) neighbors. This is given by:  
\begin{equation}
\hat{y} = \frac{\sum_{i \in N_k(x)} w_i y_i}{\sum_{i \in N_k(x)} w_i}
\label{zz}
\end{equation}

In this formula, \( N_k(x) \) represents the set of \( k \) nearest neighbors. Since most of the neighbors are close to \( x \), the target value of each neighbor is weighted by \( w_i \) and is generally adjusted such that the neighbor pair prediction value decreases with distance. Testing the parameter value can lead to more accurate predictions by taking into account the influence of the nearest neighbors on the query \citep{dudani1976distance}.

Importantly, a small value of \( k \) allows the model to accommodate local changes that may introduce variation; a larger value of \( k \) reduces the variance but introduces a slight bias \citep{hastie2009elements}.
\\
\begin{table*}[htbp]
\centering
\caption{Comparision of retrieved parameters between different Machine learning models and \texttt{PyMultiNest} for the full wavelength range.}
\resizebox{0.98\textwidth}{!}{
\hspace{-1.8cm}
\begin{tabular}{lccccccccc}
\toprule
\hline
\multirow{2}{*}{ } & \multicolumn{1}{c}{log(Na)} & \multicolumn{1}{c}{log(K)} & \multicolumn{1}{c}{log(H$_2$O)} & \multicolumn{1}{c}{log(CO$_2$)} & \multicolumn{1}{c}{log(SO$_2$)} & \multicolumn{1}{c}{log(H$_2$S)} & \multicolumn{1}{c}{log(CO)} & 
\multicolumn{1}{c}{R$^2$} \\

\cmidrule{1-9}

 & \multicolumn{8}{c}{\textit{\textbf{Machine Learning Models}}} \\
[0.2cm]
\texttt{Random Forest}& $-6.62^{+0.78}_{-0.19}$ & $-8.30^{+0.01}_{-0.01}$ & $-2.51^{+0.05}_{-0.04}$ & $-4.15^{+0.04}_{-0.01}$ & $-6.00^{+1.00}_{-0.01}$ & $-3.55^{+0.02}_{-0.02}$ & $-1.89^{+0.04}_{-0.02}$ & 
0.67 \\
[0.2cm]
\texttt{Gradient Boosting} & $-6.59^{+0.79}_{-0.21}$ &  $-8.30^{+0.01}_{-0.01}$ & $-2.51^{+0.06}_{-0.01}$ & $-4.12^{+0.01}_{-0.01}$ & $-6.00^{+1.33}_{-0.01}$ & $-3.56^{+0.04}_{-0.03}$ & $-1.90^{+0.08}_{-0.04}$ &
0.52 \\
[0.2cm]
\texttt{K-Nearest Neighbour} & $-6.80^{+0.01}_{-0.01}$ & $-8.40^{+0.40}_{-0.20}$ & $-2.51^{+0.03}_{-0.06}$ & $-4.08^{+0.06}_{-0.06}$ & $-6.00^{+1.00}_{-0.01}$ & $-3.55^{+0.01}_{-0.01}$ & 
$-1.89^{+0.02}_{-0.01}$ &
0.65 \\
[0.2cm]
\texttt{Stacking Regressor} & $-6.80^{+1.00}_{-0.01}$ & $-8.30^{+0.01}_{-0.01}$ & $-2.57^{+0.08}_{-0.25}$ & $-4.17^{+0.07}_{-0.70}$ & $-6.00^{+1.00}_{-0.01}$ & $-3.61^{+0.03}_{-0.26}$ & $-1.89^{+0.07}_{-0.53}$ & 
0.76 \\
[0.2cm]
 & \multicolumn{8}{c}{\textit{\textbf{Nested Sampling}}} \\
[0.2cm]
\texttt{PyMultiNest} & $-7.21^{+0.38}_{-0.68}$ & $-8.35^{+0.16}_{-0.18}$ & $-2.84^{+0.04}_{-0.06}$ & $-4.64^{+0.09}_{-0.06}$ & $-5.73^{+0.15}_{-0.16}$ & $-3.77^{+0.08}_{-0.08}$ & $-1.51^{+0.03}_{-0.05}$ 
& \\
\hline

\bottomrule
\end{tabular}}
\label{tab:ml-models-parameters}
\end{table*}
\subsubsection{\textbf{Gradient Boosting Regressor}} \label{grad_boosttt}

\texttt{Gradient Boosting} (GB) is an ensemble technique that combines weak learners, usually decision trees, to build robust predictive models \citep{friedman2001greedy}. By correcting the errors made by previous decision trees, GB tries to predict a target variable \( y \) from features \( X \), improving the prediction. In the case of regression, GB aims to predict a target variable from features \( X \), by iteratively improving the prediction \( \hat{y} \) in each step. At step \( m \), the prediction is updated as:
\begin{equation}
\label{play}
\hat{y}^{(m)} = \hat{y}^{(m-1)} + \eta \cdot h_m(X),
\end{equation}

where \( \eta \) is the learning rate, and \( h_m(X) \) is the decision tree that fits the residuals of the past prediction \( \hat{y}^{(m-1)} \).

The objective of GB in regression is to minimize a differentiable loss function, typically the mean squared error (MSE), between the predicted values \( \hat{y} \) and the true values \( y \). At each step, the algorithm calculates the negative gradient of the loss with respect to the estimated value based on the residual from the next tree,
\begin{equation}
\label{swastik}
r_i^{(m)} = -\frac{\partial L(y_i, \hat{y_i}^{(m)})}{\partial \hat{y_i}^{(m)}}.
\end{equation}

The effectiveness of GB stems from its ability to focus on complex examples and relational patterns \citep{natekin2013gradient}. Regularization techniques such as downsampling and subsampling help reduce overfitting \citep{buhlmann2007boosting}.

\subsubsection{\textbf{Comparison of Machine Learning models}}\label{comparision}
Among the evaluated models, the \texttt{Stacking Regressor}, which combines \texttt{Random Forest}, \texttt{Gradient Boosting}, and \texttt{k-Nearest Neighbour}, demonstrated superior predictive performance with the highest ($R^2$) score of 0.855, as shown in Table \ref{tab:ml-models-parameters}. The $R^2$ score is a measure of how well the model is performing by comparing the predicted value with the actual value. It ranges from 0 to 1, with $R^2$ = 1 indicating perfect prediction. The $R^2$ score is given by the equation \ref{r2 eqtn}, where $S_{res}$ is the sum of squares of residual error and $S_{tot}$ is the total sum of errors.

\begin{equation} \label{r2 eqtn}
R^{2} = 1 - \frac{S_{res}}{S_{tot}}
\end{equation}

The enhanced performance of the \texttt{Stacking Regressor} suggests that leveraging the complementary strengths of these base models through ensemble learning effectively captures both linear and non-linear relationships in the data, resulting in more robust predictions compared to individual models. The comparison of retrieved chemical abundances for equilibrium offset chemistry using different machine learning methods is shown in Table \ref{tab:ml-models-parameters}. The results are also comparable with the \texttt{PyMultiNest} retrieval, as shown in Table \ref{tab:ml-models-parameters}. The corner plot for retrievals done using hybrid chemistry and equilibrium offset chemistry is also shown for both \texttt{PyMultiNest} (Figures \ref{fig:corner_eq_ultra} and \ref{fig:corner_hybrid_ultra}, respectively) and machine learning (\texttt{Stacking Regressor}) in Figures \ref{fig:corner_eq_ml} and \ref{fig:corner_hybrid_ml}, respectively.

\newpage
\subsection{\textbf{\texttt{NEXOTRANS} Benchmark Table}} \label{retrieval_benchmark_results}

The retrieved parameters, along with the corresponding model assumptions and statistical significance for each case in the \texttt{NEXOTRANS} benchmarking dataset, are summarized in Table~\ref{benchmark_table}. This table also includes results from eight other retrieval algorithms, helping validate the retrieval framework implemented in \texttt{NEXOTRANS}.
\begin{table*}[h] 
\centering 
\caption{The table below summarizes the results for the free retrievals performed on the MIRI dataset, including the goodness of fit for each individual retrieval. The cloud model employed by each retrieval code is also mentioned. Notably, the \texttt{NEXOTRANS} retrieval results are in agreement with those from the other eight retrieval codes \citep{powell2024sulfur}.
}
\vspace{1.0pt}
\renewcommand{\arraystretch}{1.2} 

\begin{tabular}{|c|c|c|c|c|}
\hline & log(H$_2$O)  & log(SO$_2$) & Red $\chi^2$ & Cloud model \\

\hline & \multicolumn{4}{|c|}{\textbf{Eureka!}} \\

\hline ARCiS & $-1.5_{-0.6}^{+0.3}$ & $-5.5_{-0.5}^{+0.4}$  & 1.54 & grey, patchy \\
\hline Aurora & $-3.9_{-3.5}^{+2.3}$ & $-5.4_{-0.9}^{+0.8}$  & 1.06 & haze + grey cloud, patchy \\
\hline CHIMERA & ${ -1.9}_{-0.5}^{+0.4}$  & $-5.8_{-0.5}^{+0.4}$ & 1.24 & haze + grey cloud, patchy \\
\hline Helios-r2 & $-1.6_{-0.5}^{+0.3}$ & $-5.7_{-0.5}^{+0.4}$  & 1.77 & grey \\
\hline NEMESIS & $-1.6_{-0.6}^{+0.3}$  & $-5.6_{-0.5}^{+0.4}$  & 1.54 & grey, patchy \\
\hline PyratBay & $-1.5_{-0.6}^{+0.3}$  & $-5.5_{-0.6}^{+0.5}$ & 1.50 & grey, patchy \\
\hline TauREx & $-1.6_{-0.5}^{+0.3}$  & $-5.5_{-0.5}^{+0.4}$  & 1.53 & grey \\
\hline POSEIDON & $-1.6_{-0.7}^{+0.3}$  & $-5.6_{-0.3}^{+0.3}$  & 1.55 & haze + grey cloud, patchy \\
\hline NEXOTRANS & $-1.4_{-0.2}^{+0.3}$  & $-5.8_{-0.4}^{+0.4}$  & 1.35 & grey \\
\hline NEXOTRANS & $-1.4_{-0.2}^{+0.4}$  & $-5.7_{-0.4}^{+0.5}$  & 1.41 & grey, patchy \\
\hline NEXOTRANS & $-1.4_{-0.2}^{+0.4}$  & $-5.7_{-0.4}^{+0.4}$  & 1.15 & haze + grey cloud, patchy \\

\hline & \multicolumn{4}{|c|}{\textbf{Tiberius}} \\

\hline ARCiS & $-2.0_{-0.9}^{+0.5}$  & $-5.6_{-0.7}^{+0.6}$  & 1.10 & \\
\hline Aurora & ${ -1.5}_{-0.5}^{+0.4}$  & $-5.3_{-0.6}^{+0.5}$  & 1.14 & \\
\hline CHIMERA & $-2.3_{-0.7}^{+0.6}$  & $-5.9_{-0.5}^{+0.4}$  & 1.73 & \\
\hline Helios-r2 & ${-2.0 }_{-0.8}^{+0.5}$  & $-5.7_{-0.6}^{+0.5}$  & 1.37 &  \\
\hline NEMESIS & ${-2.1 }_{-0.9}^{+0.5}$  & $-5.7_{-0.6}^{+0.6}$  & 1.07 & \\
\hline PyratBay & $-1.9_{-1.2}^{+0.9}$  & $-5.5_{-0.9}^{+0.6}$  & 1.12 & same as above\\
\hline TauREx & $-1.8_{-0.9}^{+0.5}$  & $-5.3_{-0.8}^{+0.6}$  & 1.08 & \\
\hline POSEIDON & $-1.9_{-0.6}^{+0.4}$  & $-5.6_{-0.5}^{+0.4}$  & 1.54 & \\
\hline NEXOTRANS & $-1.6_{-0.3}^{+0.5}$  & $-5.5_{-0.4}^{+0.5}$  & 0.98 & \\
\hline NEXOTRANS & $-1.6_{-0.3}^{+0.6}$  & $-5.5_{-0.4}^{+0.5}$  & 1.04 &  \\
\hline NEXOTRANS & $-1.6_{-0.3}^{+0.5}$  & $-5.5_{-0.4}^{+0.5}$  & 1.14 &  \\

\hline & \multicolumn{4}{|c|}{\textbf{SPARTA}} \\

\hline ARCiS & $-1.3_{-0.6}^{+0.2}$  & $-5.3_{-0.6}^{+0.4}$  & 1.15 & \\
\hline Aurora & $-1.1_{-0.8}^{+0.2}$  & $-5.0_{-0.6}^{+0.4}$  & 0.95 & \\
\hline CHIMERA & $-1.8_{-0.4}^{+0.7}$  & $-5.6_{-0.4}^{+0.6}$  & 1.29 & \\
\hline Helios-r2 & $-1.4_{-0.7}^{+0.3}$  & $-5.3_{-0.5}^{+0.4}$  & 1.36 &  \\
\hline NEMESIS & $-1.7_{-0.9}^{+0.4}$  & $-5.6_{-0.6}^{+0.5}$  & 1.20 & \\
\hline PyratBay & $-1.4_{-1.0}^{+0.4}$  & $-5.3_{-0.8}^{+0.5}$ & 1.16 & same as above \\
\hline TauREx & $-1.5_{-0.9}^{+0.3}$  & $-5.3_{-0.7}^{+0.4}$  & 1.15 & \\
\hline POSEIDON & $-1.2_{-0.4}^{+0.2}$  & $-5.3_{-0.3}^{+0.3}$ & 1.36 & \\
\hline NEXOTRANS & $-1.2_{-0.2}^{+0.1}$  & $-5.5_{-0.3}^{+0.3}$ & 1.12 & \\
\hline NEXOTRANS & $-1.3_{-0.2}^{+0.4}$  & $-5.6_{-0.4}^{+0.5}$  & 1.17 & \\
\hline NEXOTRANS & $-1.2_{-0.2}^{+0.3}$ & $-5.5_{-0.3}^{+0.3}$  & 1.13 & \\

\hline
\end{tabular}
\label{benchmark_table}
\end{table*}
 
 \subsection{\textbf{Individual Retrieved Spectra}}\label{individual_section}

\begin{figure*}[]
    \centering
    \hspace{0.2cm}
    \includegraphics[width=0.812\linewidth]{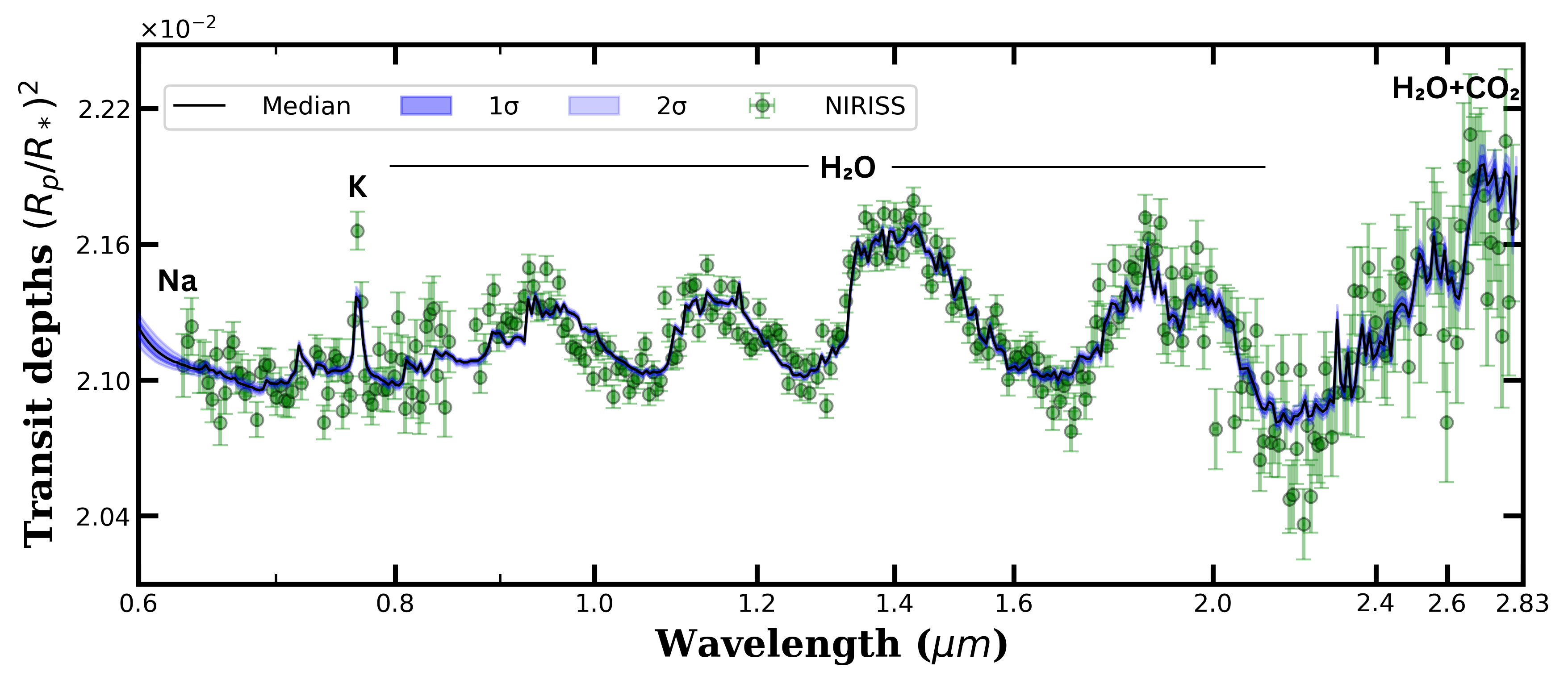}
    \parbox{0.8\linewidth}{\centering (a) Best-fit spectrum on NIRISS dataset.}
    \label{fig:chemoffset_bayesian_spectra}
    \includegraphics[width=0.8\linewidth]{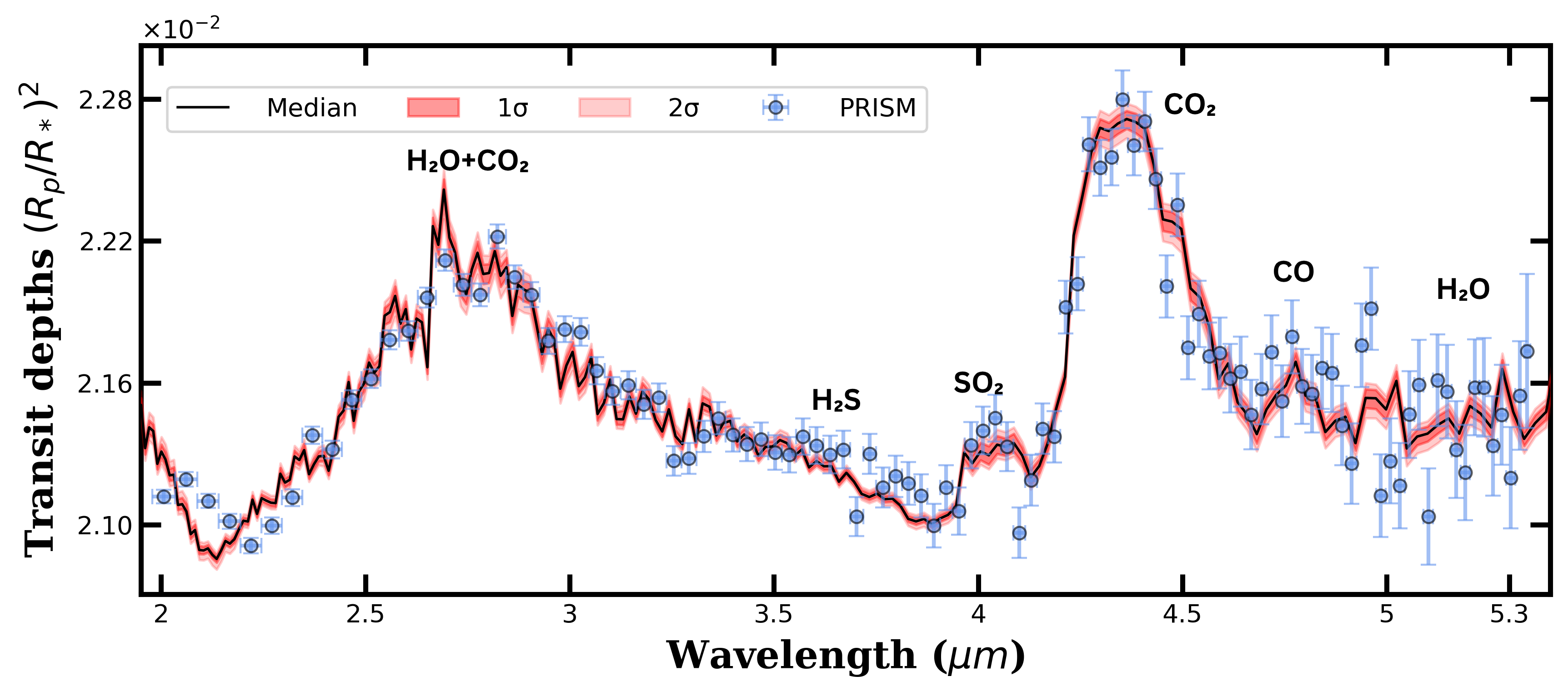}
    \parbox{0.8\linewidth}{\centering (b) Best-fit spectrum on PRISM dataset.}
    \label{fig:ML_spectra}
    \includegraphics[width=0.8\linewidth]{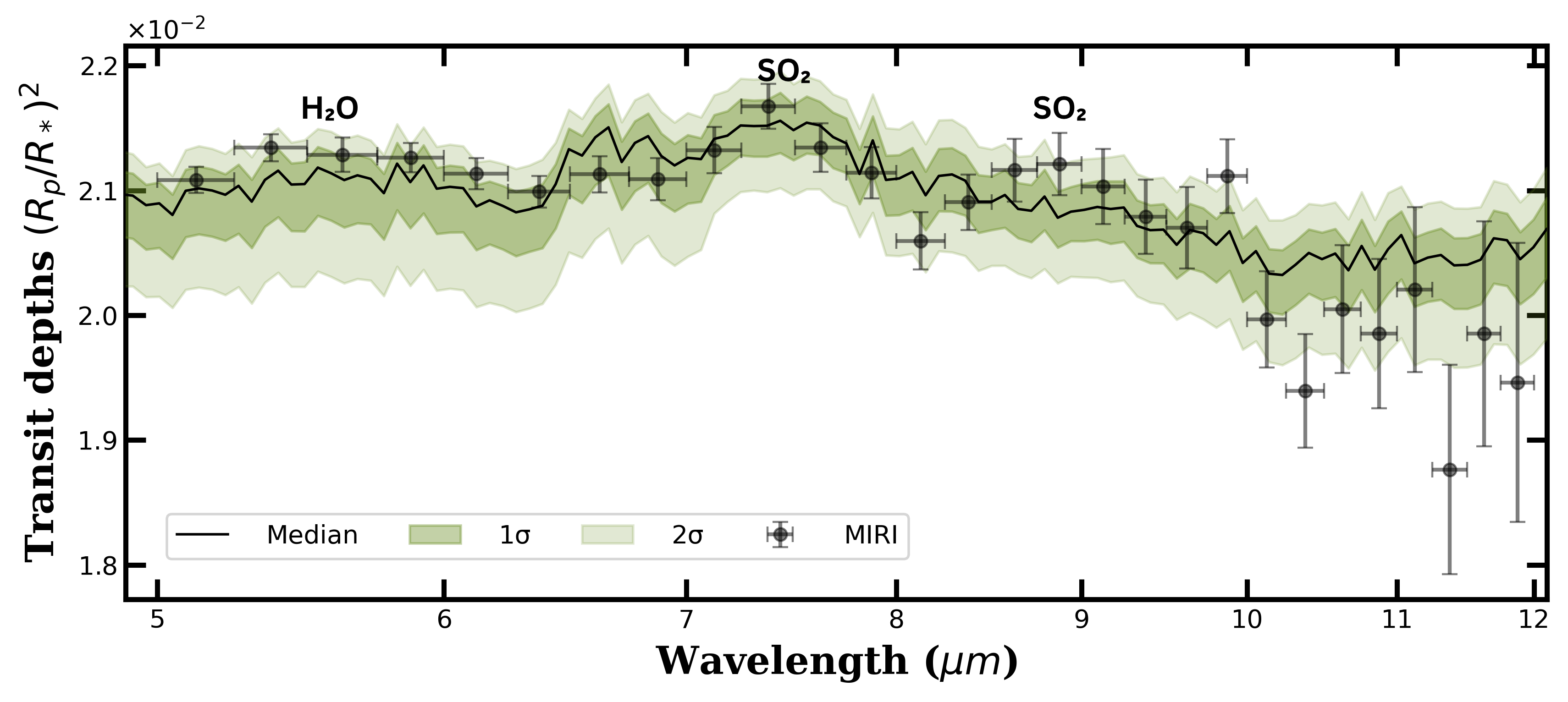}
    \parbox{0.8\linewidth}{\centering (c) Best-fit spectrum on MIRI dataset}
    \label{fig:ML_spectra}
    \caption{
       Retrieved best-fit individual spectra for NIRISS, NIRSpec PRISM, and MIRI using the global best-fit hybrid equilibrium model with aerosols (ZnS, MgSiO$_3$) using \texttt{NEXOTRANS}. The reduced $\chi^2$ values for NIRISS, NIRSpec PRISM and MIRI retrievals are 2.67, 3.19 and 2.14 respectively.
    }
    \label{fig:individual_spectrum}
\end{figure*}

The best-fit retrieved spectra for each JWST dataset--NIRISS, PRISM, and MIRI--are shown in Figure~\ref{fig:individual_spectrum}. These retrievals were performed using the global best-fit hybrid equilibrium model to evaluate the reduced $\chi^2$ values contributed by each individual dataset.

 \clearpage
\newpage

\subsection{\textbf{Retrieved corner plots}} \label{corner_plots}

\begin{figure*}[h]
     \includegraphics[width=\linewidth]{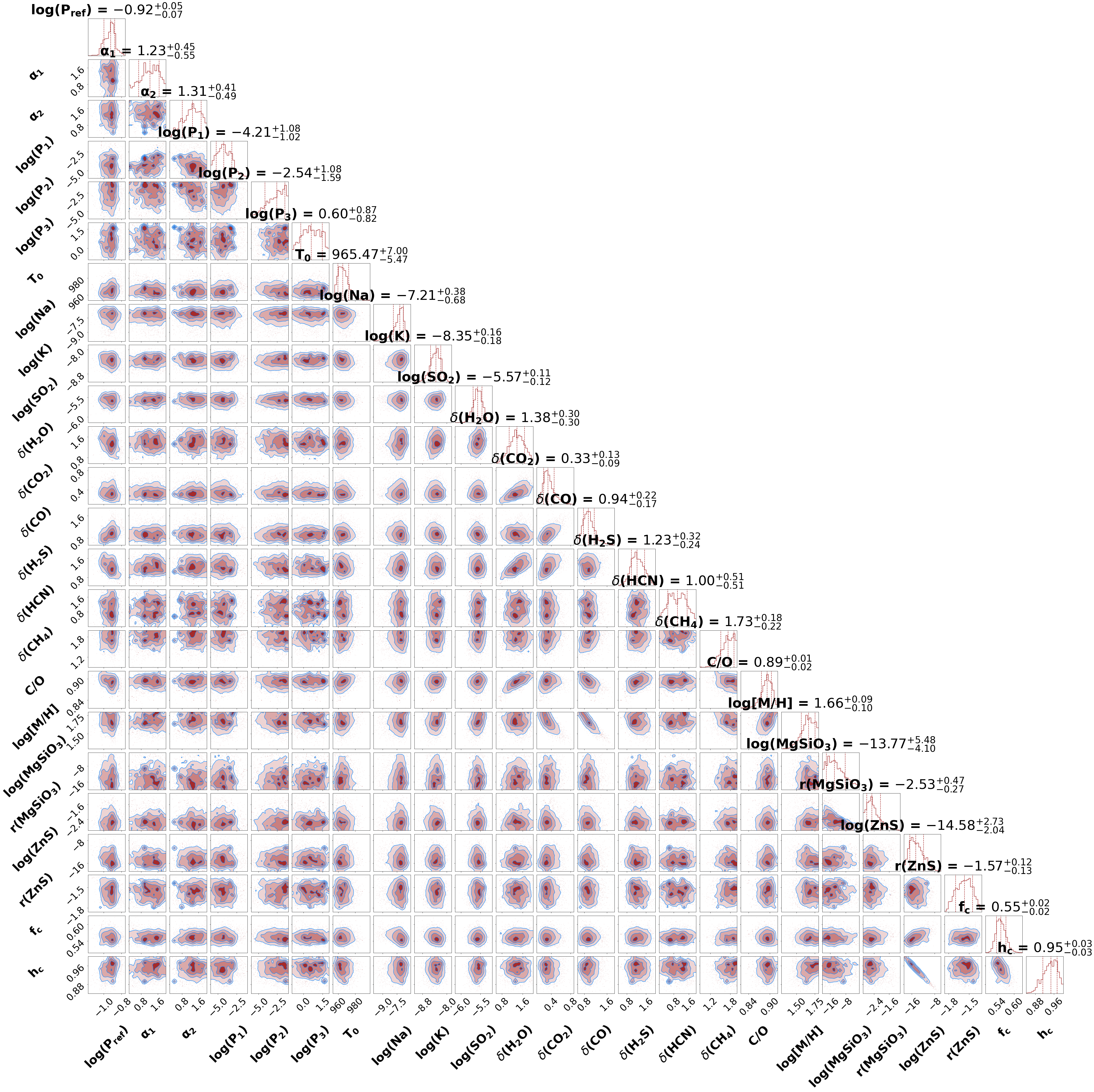}
      \caption{PyMultiNest retrieved Posterior distribution for the mie scattering aerosol model assuming equilibrium offset chemistry. Here, log(P$_{ref}$) = reference pressure; $\alpha$1 and $\alpha$2 are slopes of the PT profile; log(P$_1$), log(P$_2$) and log(P$_3$) are the pressures at different layers; T$_0$ = temperature at reference pressure; $\delta$ represents the offsets multiplied to the vmr of the chemical species; C/O = Carbon-Oxygen ratio; M/H = metallicity; r(species) are the modal particle size of the aerosol particles; f$_c$ = aerosol coverage fraction; h$_c$ = slope of the aerosol vertical profiles.}
      \label{fig:corner_eq_ultra}
\end{figure*}

\begin{figure*}[h]
     \includegraphics[width=\linewidth]{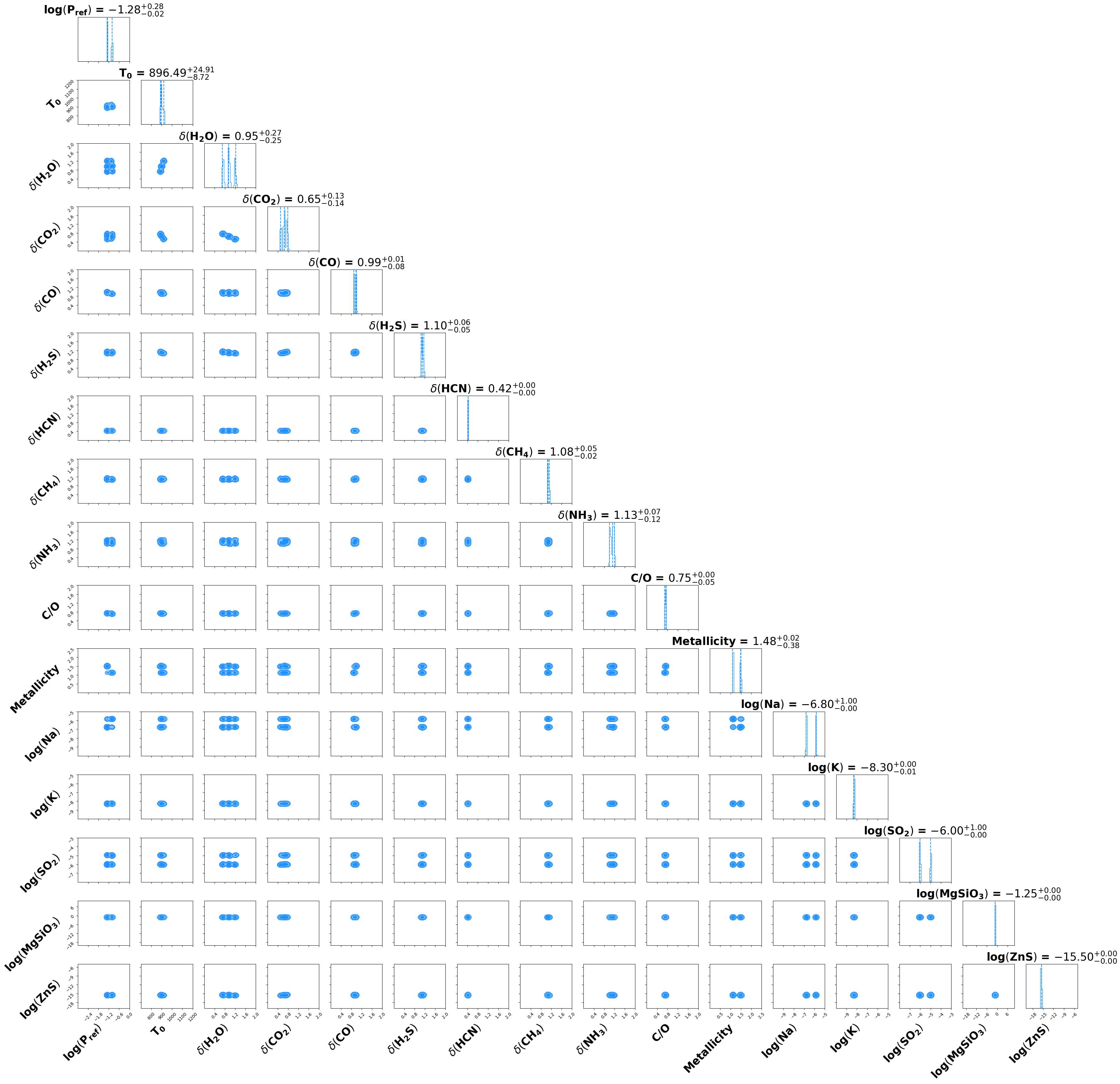}
      \caption{Posterior distribution for the mie scattering aerosol model assuming equilibrium offset chemistry using Machine learning (\texttt{Stacking Regressor}).}
      \label{fig:corner_eq_ml}
\end{figure*}

\begin{figure*}[h]
     \includegraphics[width=\linewidth]{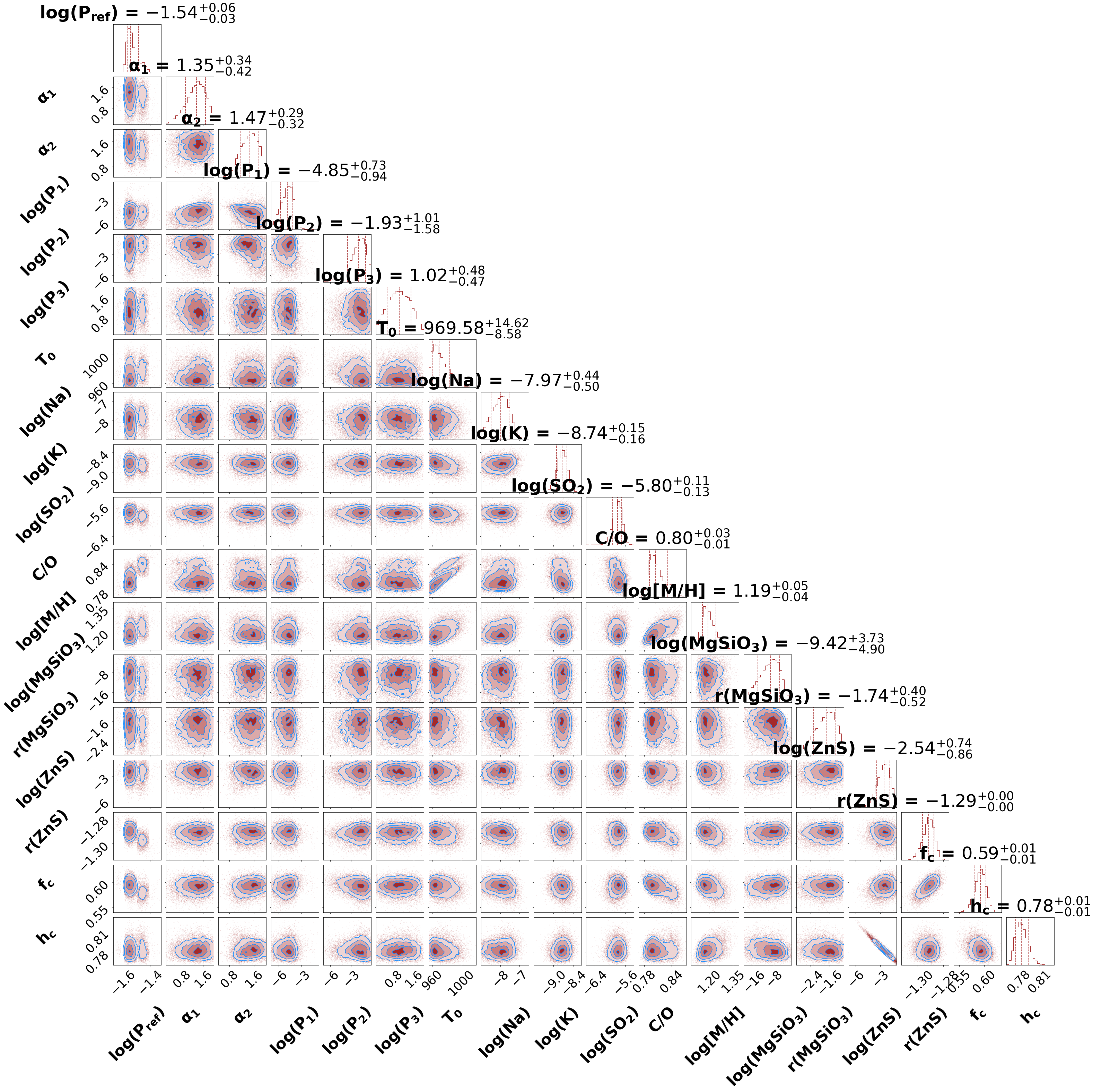}
      \caption{PyMultiNest retrieved full posterior distribution for the mie scattering aerosol model assuming hybrid equilibrium chemistry.}
      \label{fig:corner_hybrid_ultra}
\end{figure*}

\begin{figure*}[h]
     \includegraphics[width=\linewidth]{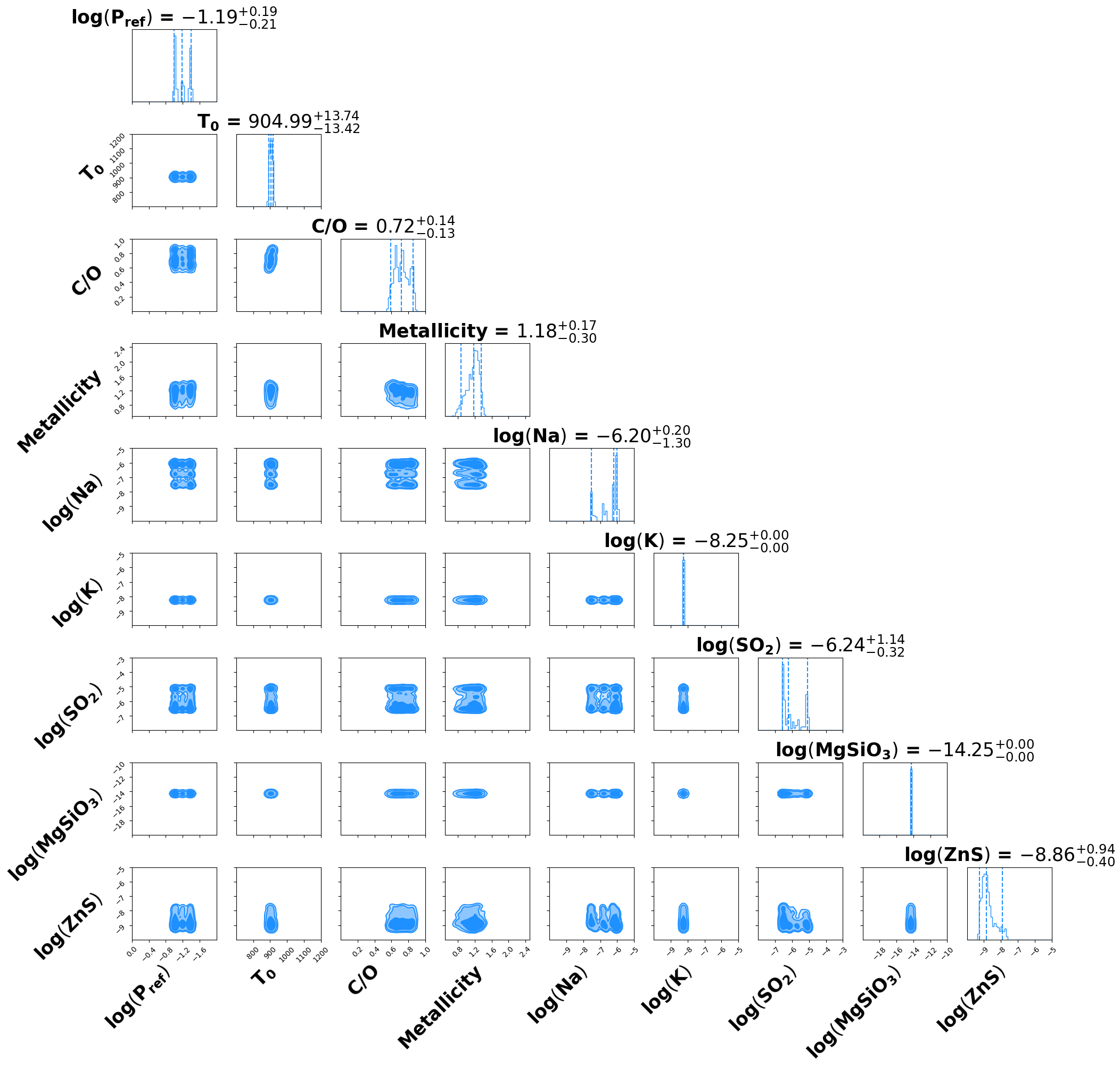}
      \caption{Posterior distribution for the mie scattering aerosol model assuming hybrid chemistry using Machine learning (\texttt{Stacking Regressor}).}
      \label{fig:corner_hybrid_ml}
\end{figure*}

The corner plots for the best-fit models obtained using \texttt{PyMultiNest} and machine learning (\texttt{Stacking Regressor} are presented below. These plots correspond to the retrieval configuration incorporating hybrid equilibrium chemistry, equilibrium offset chemistry and aerosols. Figures \ref{fig:corner_eq_ultra} and \ref{fig:corner_eq_ml} present the corner plots of the retrieved parameters using the modified equilibrium offset chemistry model, derived from the Bayesian and machine learning methods, respectively. Figures \ref{fig:corner_hybrid_ultra} and \ref{fig:corner_hybrid_ml} shows the corresponding corner plots obtained from the modified hybrid equilibrium chemistry model.

\clearpage
\bibliography{main}{}
\bibliographystyle{aasjournal}



\end{document}